%% file: main.tex
\RequirePackage{fix-cm}

\documentclass[smallextended]{svjour3}       %
\smartqed  
\usepackage{natbib}
\usepackage{amsmath}                            
\usepackage{multirow}                                
\usepackage{indentfirst}                           
\usepackage[colorlinks,linkcolor=blue,anchorcolor=blue,citecolor=blue,urlcolor=blue]{hyperref}    
\usepackage{enumitem}
\usepackage{comment}
\usepackage{url}
\usepackage{enumitem}
\usepackage[titletoc,title]{appendix}
\usepackage{booktabs}
\usepackage{adjustbox,lipsum}
\usepackage[figuresright]{rotating}
\usepackage{todonotes}
\usepackage{listings}
\usepackage{xcolor}
\usepackage{outlines}
\usepackage{tabularx}

\definecolor{codeblue}{rgb}{0.3,0.3,0.9}
\definecolor{codegreen}{rgb}{0,0.6,0}
\definecolor{codegray}{rgb}{0.5,0.5,0.5}
\definecolor{codepurple}{rgb}{0.58,0,0.82}

\lstdefinestyle{javastyle}{
    language=Java,
    basicstyle=\ttfamily\scriptsize,
    keywordstyle=\color{codeblue},
    stringstyle=\color{codegreen},
    commentstyle=\color{codegray},
    morecomment=[l][\color{codepurple}]{\#},
    numbers=left,
    xleftmargin=1em,
    numberstyle=\tiny\color{codegray},
    stepnumber=1,
    numbersep=10pt,
    tabsize=1,
    showspaces=false,
    showstringspaces=false,
    breaklines=true,
    breakatwhitespace=true,
    escapeinside={(*@}{@*)}
}

\usepackage[T1]{fontenc}
\usepackage{lmodern}
\usepackage{textcomp}

\definecolor{light-cyan}{gray}{0.80}
\usepackage{dcolumn}
\usepackage{amsmath}
\usepackage{amssymb}
\usepackage{caption}
\usepackage[ruled]{algorithm2e}

\usepackage{subcaption}

\usepackage[abs]{overpic}

\definecolor{OliveGreen}{rgb}{0,0.6,0}

\newcommand{\pa}[1]{\noindent\textbf{#1}}

\begin{document}

\title{A Multi-Language Perspective on the Robustness of LLM Code Generation}


\author{Fazle Rabbi\and Zishuo Ding \and
        Jinqiu Yang
}


\institute{
Fazle Rabbi \at
Concordia University\\
\email{fazle.rabbi@mail.concordia.ca}
\and
Zishuo Ding\at
The Hong Kong University of Science and Technology \\
\email{zishuo.ding@uwaterloo.ca}
\and
Jinqiu Yang\at
Concordia University \\
\email{jinqiu.yang@concordia.ca}
}

\date{Received: date / Accepted: date}

\maketitle

\begin{abstract}
Large language models have gained significant traction and popularity in recent times, extending their usage to code-generation tasks. While this field has garnered considerable attention, the exploration of testing and evaluating the robustness of code generation models remains an ongoing endeavor. Previous studies have primarily focused on code generation models specifically for the Python language, overlooking other widely-used programming languages. In this work, we conduct a comprehensive comparative analysis to assess the robustness performance of several prominent code generation models and investigate whether robustness can be improved by repairing perturbed docstrings using an LLM. Furthermore, we investigate how their performance varies across different programming languages. To accomplish this, we introduce perturbations in four key areas of the prompt: DocString, function name, syntax, and format. We have compiled and released a dedicated dataset for this purpose. {Our results show that all models consistently degrade under perturbations across all three languages, but vary in magnitude depending on the language and perturbation type. Larger model size does not reliably improve robustness, and semantic perturbations prove at least as disruptive as syntactic ones. Our LLM-based docString repair yields only marginal gains for simple perturbations and can degrade performance for semantic ones, highlighting the limits of prompt-level mitigation.} 
\end{abstract}


\input{introduction}
\input{background_related}
\input{methodology}

\input{experiments}

\input{refactoring}
\input{lessons_learned}
\input{threats}
\input{conclusion}
\input{funding}


%
%

\input{main.bbl}
\end{document}

%% file: introduction.tex
\section{Introduction}
\label{intro}
\noindent In the last few years, Large language models (LLMs) have shown remarkable performance in a wide range of natural language processing (NLP) tasks, including machine translation, text generation, and summarization. Inspired by the success of LLMs in NLP, researchers recently proposed to apply LLMs to solve code-related tasks, such as 
generating code from a natural language (NL) prompt and completing code from a program context (e.g., code and NL comment).

\begin{figure}
\centering
\includegraphics[width=.75\textwidth]{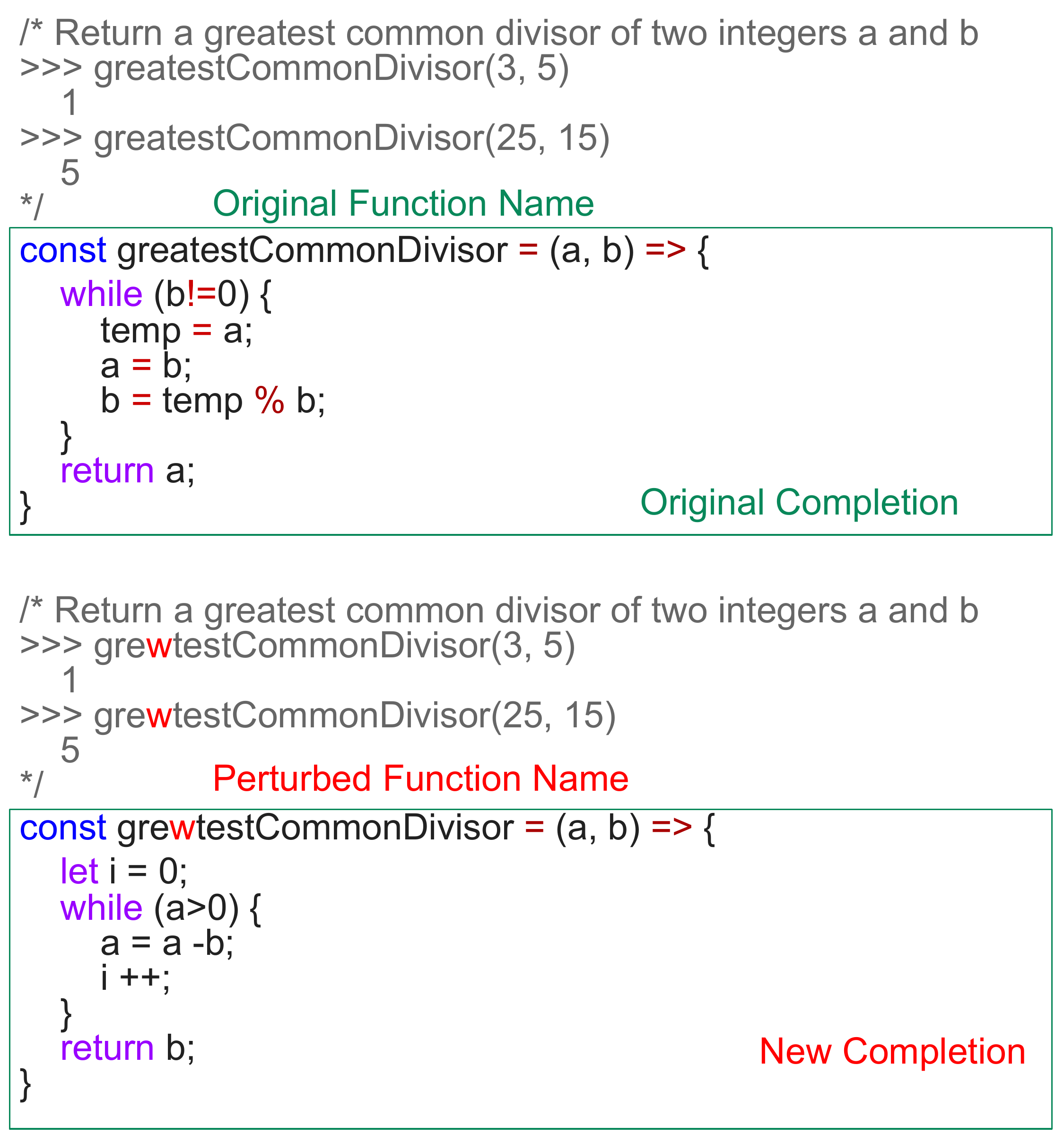}
\caption{A modification in the function name causes incorrect code generation by CodeGen-6B-Multi model}
\label{fig:perturbed_gc}
\end{figure}

Despite these advancements, the robustness of the LLMs on code generation tasks has not been extensively explored. LLMs are known to suffer from the issue of robustness in NLP tasks, i.e., studies\citep{samanta2017towards,gao2018black,li2018textbugger,jin2020bert,li2020bert} in NLP find that a slight change in the input can cause the model to generate a different prediction from the original result. Meanwhile, some recent studies ~\citep{Wang2023, copilot_robustness} focusing on LLM for code generation find that the Large Language Code generation model may also be sensitive to small perturbations on the prompt. For instance, Figure \ref{fig:perturbed_gc} demonstrates that even a minor function name modification can cause the model to produce incorrect code, although the non-perturbed prompt produced code that passed the test cases. Therefore, to test the robustness of code generation models,~\citet{Wang2023} propose the first evaluation benchmark, ReCode, by customizing 28 semantic-preserved perturbations on the prompts (e.g., docstrings, function names), and then they evaluate the robustness of three code generation models (i.e., CodeGen, InCoder, and GPT-J).

\begin{figure}
\centering
\includegraphics[width=.75\textwidth]{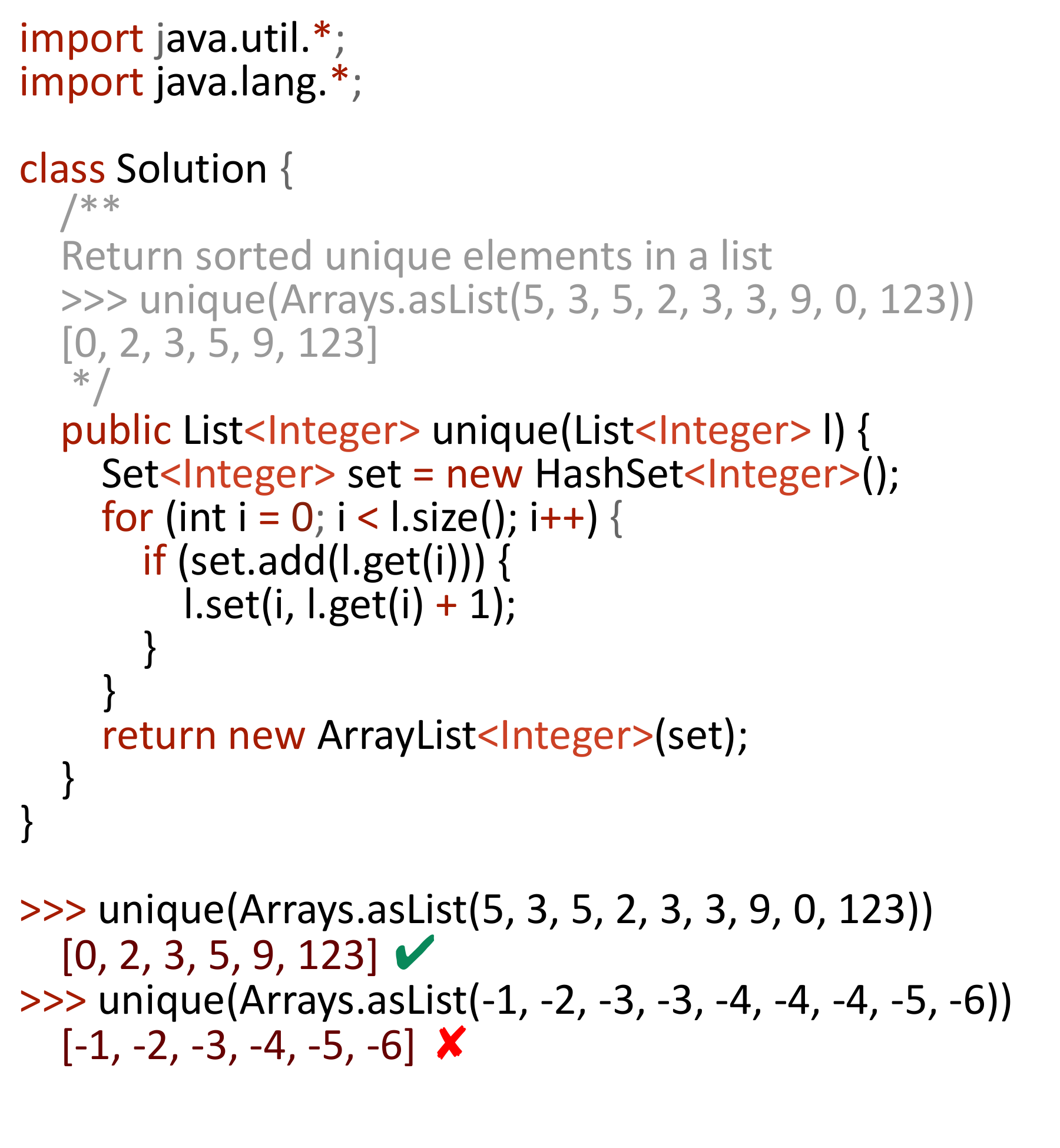}
\caption{A generated code by CodeGen-6B-Multi model which passes the test cases of HumanEval-X but fails on the test cases from EvalPlus.}
\label{fig:humanevalnotenough}
\end{figure}

Previous robustness evaluations of LLMs for code generation~\citep{Wang2023} have focused primarily on Python, leaving open the question of whether these findings generalize to other programming languages with distinct syntax, semantics, and uneven training resources~\citep{austin2021program}. The benchmark used in prior work lacks sufficient test coverage, which can mask failures that appear under more rigorous evaluation suites~\citep{liu2023your}, for instance, as shown in Figure~\ref{fig:humanevalnotenough}, a Java solution may pass all HumanEval-X tests yet fail under the stronger EvalPlus suite. Another line of work has explored input smoothing techniques~\citep{ye2020safer,ji2024defending} to improve LLM performance on security-related instructions, but these methods are tailored to security-specific scenarios and do not address robustness issues arising from natural perturbations in code prompts. The broader problem of repairing or normalizing perturbed prompts, particularly perturbed docstrings to restore performance in code generation remains largely unexplored. This motivates our investigation into multilingual robustness, stronger evaluation settings, and post-perturbation prompt repair strategies for LLMs in code generation.


In this paper, we expand the current robustness assessment and revisit the findings from prior research to understand whether they still hold with different programming languages and under more rigorous testing. 
We aim to provide more insights to guide future research on improving the robustness of LLM-powered code generation.
Specifically, we extend ReCode~\citep{wang2022recode} from two aspects: First, in addition to Python, we evaluated the LLMs on more programming languages, including Java, C++, and JavaScript. 
Second, knowing that the test cases in HumanEval are inadequate, we construct an enhanced multi-lingual dataset of test cases based on EvalPlus~\citep{liu2023your}. EvalPlus~\citep{liu2023your}, an enhancement of the HumanEval~\citep{chen2021evaluating} benchmark, introduces additional corner cases and test cases, leading to a more rigorous evaluation that expands a set of challenges, resulting in a higher failure rate for generated code by LLMs. Since EvalPlus test cases and solutions are in Python, we convert them to Java, C++, and JavaScript and manually update the solutions accordingly. Besides this expanded robustness evaluation, we investigate whether robustness losses caused by prompt perturbations, particularly perturbed docstrings can be mitigated by automatically repairing the docstring using an LLM. The source code and benchmark dataset are publicly available. \\

Our primary contributions are as follows.
\begin{enumerate}
    \item Our work involves the expansion of ReCode, a robustness testing framework initially developed to evaluate the performance of code generation models using Python datasets. In our research, we extend the capabilities of ReCode to encompass additional programming languages such as Java, C++, and JavaScript. 
    \item We significantly extend EvalPlus to EvalPlus-X to include a multi-language dataset. Particularly, we started with the HumanEval-X benchmark, and then we translated the EvalPlus test cases and modified the canonical solutions of HumanEval-X in Java, C++, and JavaScript to ensure correctness and compatibility;
    \item We conduct a comparative analysis to evaluate the performance of code generation models when exposed to perturbed datasets in Java, C++, and JavaScript languages. To the best of our knowledge, no previous research has conducted a comparative study of this nature, evaluating code generation models across multiple languages using perturbed datasets.
    \item We release robustness evaluation datasets for Java, C++, and JavaScript, which are derived from our curated dataset based on HumanEval-X and EvalPlus test cases. These perturbed datasets provide valuable resources for future research and related endeavors.
    \item We reveal prevalent linguistic and structural factors that drive robust drops in LLM code generation, showing how different programming languages and perturbation types amplify performance drops.
    \item We experimented with an automatic mitigation strategy that uses LLMs to fix perturbed docstrings and demonstrate that it partially recovers robustness across models and languages.
\end{enumerate}

{Our evaluation of six LLMs across Java, C++, and JavaScript reveals several noteworthy findings. First, robustness drops are consistent across all three languages, but their magnitude varies depending on the language and perturbation type, indicating that robustness is a language-dependent property that cannot be assessed from a single language alone. Second, semantic perturbations prove at least as disruptive as syntactic ones, underscoring the need for diverse perturbation types in robustness benchmarks. Third, model size does not reliably predict robustness. In several cases, larger models are more brittle than smaller ones under semantics-preserving perturbations. Finally, our prompt-repair strategy, using an LLM to reconstruct perturbed docstrings before code generation, yields only marginal gains for simple perturbations and can occasionally degrade performance for semantic ones, suggesting that lightweight prompt-level repairs alone are insufficient to restore robustness.}\\

{ To guide our investigation, 
this study addresses the following three research questions:
\begin{enumerate}
    \item \pa{RQ1: How robust are LLMs across different programming languages?} \\We evaluate how the performance of six LLMs degrades under semantics-preserving 
    perturbations across Java, C++, and JavaScript, using the EvalPlus-X test suites.
    \item \pa{RQ2: What are the major factors that contribute to the difference 
    in the robustness of LLMs on different programming languages?} \\We investigate 
    which input and output features are most strongly associated with robustness 
    failures, and whether these factors differ across languages.
    \item \pa{RQ3: Can the robustness drop be mitigated by fixing the perturbed 
    prompts?} \\We explore whether using an LLM to automatically repair perturbed 
    docstrings before code generation can recover the lost robustness.
\end{enumerate}
}
\noindent\textbf{Artifact.}
We share the data and code in this replication package\footnote{\url{https://github.com/frabbisw/robustextended}}.

%% file: background_related.tex
\section{Background and Related Work}
\label{background}
\noindent In this section, we discuss the necessary background information to enhance the clarity of our work. And then we present the recent work that is related to our research.

\subsection{Background}
\noindent In this part, we introduce two fundamental concepts essential to this work: (1) LLM code generation, which is the task of generating source code based on the given natural language (NL) description and other information, and  (2) robustness testing on LLM code generation, which aims to measure the ability of a system to function correctly despite the perturbation of the input.

\begin{figure}
\centering
\includegraphics[width=.75\textwidth]{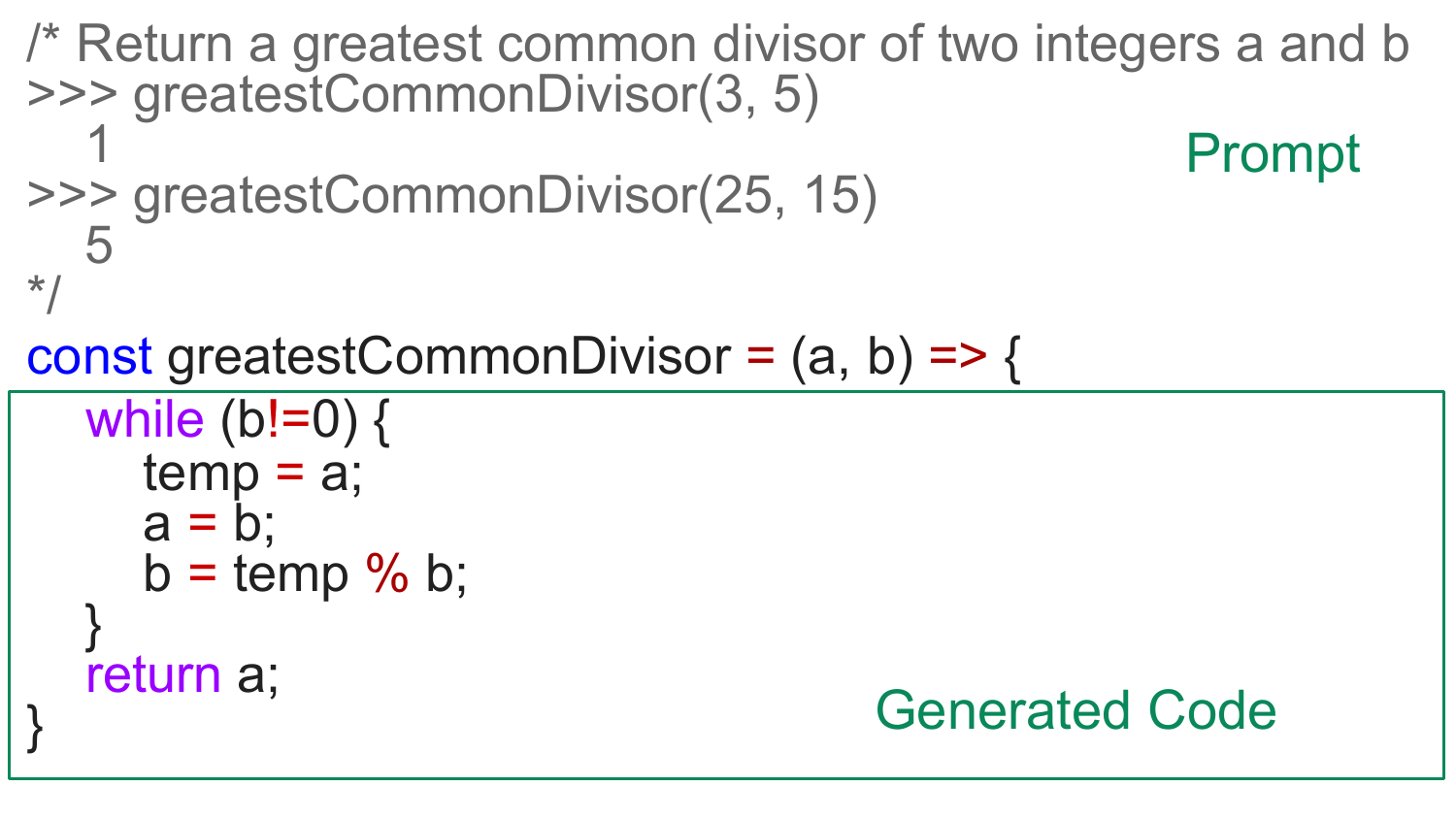}
\caption{An example of a generated code by CodeGen-6B-Multi snippet from a sample prompt}
\label{fig:sample_gc}
\end{figure}

LLM code generation techniques leverage LLMs to generate a code snippet based on a given prompt, which usually contains information about the method’s signature, an NL description of the code’s functionality, and example test cases in the prompt. An LLM is required to generate a complete code snippet based on the given information. Often, the performance of LLM code generation is measured by benchmarks, such as HumanEval~\citep{zheng2023codegeex}, EvalPlus~\citep{liu2023your}, etc. Such benchmarks utilize carefully crafted test cases to evaluate the generated code by LLMs. The example in Figure~\ref{fig:sample_gc}, shows cases that an LLM generates the code from the prompt that contains a docString describing the method’s functionality and a method signature.

\label{rtocgm}
AI robustness is a known issue, and using LLM for code generation is no exception to the robustness challenge. In the context of LLM code generation, robustness testing is formed as techniques to assess an LLM’s performance degradation when a prompt is slightly perturbed while keeping the same semantics. During testing, a negligible perturbation is performed on the original prompt, and then the modified prompt is fed into the model to generate a new code snippet, which is later compared with the original prediction.

In this work, we focus on evaluating the robustness of code generation models. During robustness testing, we apply small, semantics-preserving modifications to the original prompt, for example, changing a single character in a function name, and then ask the model to generate code from this perturbed input. After generation, the correctness of the model's output is assessed using the same test cases as in the nominal (unperturbed) setting. Following prior work, ReCode~\citep{wang2022recode}, which studies robustness only in Python, we adapt and extend its 29 perturbation techniques to support C++, Java, and JavaScript. 

Formally, let $\mathcal{F\left ( \cdot  \right )}$ be the code generation model, that takes in a prompt $\mathbf{p} =\left (p_1, p_2, \dots, p_N\right )$, and generates a complete function, $\mathbf{y}=\left (y_1, y_2, \dots, y_M\right )$, of which the correctness is tested against the set of $K$ test cases \(T = \{t_1, t_2,...,t_K\}\). To test the robustness of the model $\mathcal{F\left ( \cdot  \right )}$, a new prompt $\mathbf{p}_{adv}$ is generated by slightly perturbing the original prompt (i.e., $\mathbf{p}$), which should follow the following requirements:
\begin{equation}
    Sim(\mathbf{p}_{adv}, \mathbf{p}) \geq \epsilon 
\end{equation}
Where \(Sim(.) \rightarrow (0,1)\) is the similarity between the original and the perturbed prompt. To preserve the semantic meaning of the original prompt, this similarity value must be larger than the threshold. And the robustness of the model $\mathcal{F\left ( \cdot  \right )}$ is measured by executing the newly generated function $\mathcal{F}\left ( \mathbf{p}_{adv}\right )$ against the same set of test cases $T$.
%

Robustness testing assesses the ability of a component or system to function correctly despite invalid inputs or stressful conditions. It evaluates how well a system handles unexpected scenarios and aims to identify failures caused by errors or unusual inputs. The goal is to ensure the system remains stable, avoids crashes, and maintains its functionality. Robustness testing enhances software reliability, reduces the risk of failure, and optimizes performance in abnormal circumstances. Generally, two types of robustness testing are common in Machine Learning, which are as follows:

\textbf{Black-Box Testing:} Black box testing involves limited access to a model's inputs and the dataset for labels or confidence intervals. It examines the external behavior of software or machine learning models, emphasizing functionality, user interface usability, and compliance with external standards. Operating without knowledge of internal workings, this testing method validates the desired output. To assess the robustness of code generation models, black box testing is employed. This ensures the models can handle unexpected scenarios and typographical errors, bolstering their reliability and performance.

\textbf{White-Box Testing:} White box robustness testing offers testers extensive access to a model's output, architecture, gradients, and training data, facilitating a thorough examination of code, algorithms, and data structures. While our research primarily focuses on black box testing, it is important to acknowledge that white box testing provides valuable insights into the system's inner workings and remains essential for evaluating software and machine learning models. This methodology enables the identification of vulnerabilities and weaknesses, but it requires expertise due to its in-depth analysis of the system's internals.

This work does not target adversarial manipulation of inputs. Instead, our focus is on robustness evaluation, where the goal is to examine whether code generation models can still perform reliably when nominal prompts are altered by natural or accidental variations. While adversarial attacks are designed to intentionally mislead models, the perturbations used in this study are semantically preserving and randomly generated, reflecting realistic user mistakes rather than crafted exploits. Our evaluation measures how model performance shifts under these natural perturbations using standard test pass metrics, allowing us to assess robustness degradation without adversarial intent.

Our exploration of related works is organized into four distinct sections, each providing a comprehensive overview of the following areas.
\label{related}
\subsection{Related Work} 
\noindent Researchers have proposed various approaches for the task of code generation, and recently, the LLM-based code generation models have achieved impressive performance. 

\citet{nijkamp2022codegen} introduces CodeGen, a series of transformer-based code generation models with various parameters for multiple languages, where the 16 billion parameter variant performs the best. CodeGeex by \citet{zheng2023codegeex} is another multi-language-based series of models for code generation, where the largest model consists of 13 billion parameters. Other notable works in the field of code generation include Codex by~\citet{chen2021evaluating}, Incoder by \citet{fried2022incoder}, and GPT-J by \citet{gpt-j}. Besides, LLMs such as GPT-3 by \citet{brown2020language} have shown potential for code-generation tasks. Furthermore, works such as CodeBERT by \citet{feng2020codebert} and PLBART by \citet{ahmad2021unified} focus on training LLMs specifically on code. 

Meanwhile, to evaluate the performance of code generation models, prior work has proposed several benchmarks. The HumanEval dataset by \citet{chen2021evaluating} consists of handcrafted prompts and test cases designed for Python, making it suitable for evaluating generated Python code. An expansion of this dataset is HumanEval-X\citep{zheng2023codegeex}, which retains the same prompts but includes test cases for multiple programming languages. Additionally, Evalplus by~\citet{liu2023your}, an enhanced version of Humaneval, featuring a broader range of test cases and updated canonical solutions was proposed to effectively identify incorrect code, especially in challenging corner cases.

To our knowledge, the first work of generating adversarial examples was done by~\citet{szegedy2013intriguing}, where they apply a specific barely noticeable perturbation that is discovered by maximizing the network's prediction error, causing the image classifier network to incorrectly classify an image. Some other similar works \citep{goodfellow2014explaining} \citep{moosavi2016deepfool} \citep{kurakin2018adversarial} \citep{papernot2016limitations} \citep{chib1995understanding} have been found to generate adversarial examples with different approaches.

Our research has uncovered various attacks on Natural Language Models that are relevant to our work. ~\citet{papernot2016crafting} pioneer the crafting of Adversarial Input Sequences for Recurrent Neural Networks in Natural Language Processing. Similarly, \citet{samanta2017towards} propose an approach involving modifications to original text samples, such as the deletion or replacement of important words and the introduction of new words. \citet{gao2018black} develop a black-box attack technique that alters characters to generate small perturbations, deceiving deep-learning classifiers. \citet{li2018textbugger} focus on character-level and word-level attacks, considering word importance ranking and employing perturbations on NLP classifier models. \citet{jin2020bert} introduce TextFooler, a semantic black-box attack challenging the robustness of the BERT model. 

Additionally, BERT-ATTACK by \citet{li2020bert} targets vulnerable words in input sequences, generating substitutes using BERT. \citet{iyyer2018adversarial} design SCPN, a strategy for sentence paraphrasing without compromising quality. \citet{naik2018stress} propose an evaluation methodology using artificially generated stress tests to assess inferential judgments. In software engineering, \citet{zhou2022adversarial} develop ACCENT, testing the robustness of code comment generation by replacing identifiers while preserving functionality and proposing a new training method. 

These studies collectively contribute to the understanding of attacks in Natural Language Processing and software engineering domains, advancing techniques for generating adversarial samples and evaluating model robustness.

Although many attacks have been proposed to challenge the robustness of the Machine Learning Models, few are concerned with the code generation model. 

ReCode, introduced by \citet{wang2022recode}, stands as a comprehensive testing framework designed to evaluate the robustness of code generation models specifically tailored for Python datasets. In our study, we closely align with and build upon the foundations established by ReCode. This framework puts forth a diverse set of 28 perturbations encompassing function names, docstrings, and code. The approach outlined in ReCode is evaluated against various popular code generation models, providing valuable insights into their performance and vulnerabilities. By following the methodology outlined in ReCode, we aim to extend the understanding of code generation model robustness in our research.\\
An additional notable study we encountered is CodeAttack by \citet{jha2022codeattack}, which presents a black-box attack methodology specifically tailored to exploit the vulnerabilities of Programming language models through code-specific adversarial attacks. This approach leverages the inherent structure of code to generate adversarial code samples that are both efficient and effective while remaining undetectable. The authors identify vulnerable tokens by masking words and inputting them into the models. CodeAttack primarily emphasizes three areas: code-to-code translation, code-to-code repair, and code-to-comment generation, highlighting the diverse applications of their approach within the realm of programming. 

In their recent work, \citet{zhuo2023robustness} conduct an empirical study focusing on the prompt-based adversarial robustness of CODEX. The authors uncover the vulnerability of the model to meticulously crafted perturbations. Moreover, they propose some methods to mitigate this issue and enhance the model's robustness.\\
\citet{liu2023your} extend the HumanEval dataset with more test cases using ChatGPT and mutation, and their findings revealed that a few test cases alone are inadequate to assess the correctness of a function. The inclusion of additional generated test cases led to a higher occurrence of failures, highlighting the importance of employing a diverse set of test cases to comprehensively evaluate function accuracy.\\
Besides, recent studies have explored various aspects of code generation. Such as, \citet{li2026exploratory} and \citet{li2024fine} have focused on secure code generation, while \citet{ling2025bias} and  \citet{rabbi2026socialbias} investigate social bias in code generation. \citet{lin2024soen} examine code generation through the lens of emulating the software development process, and \citet{saha2024specification} propose methods for generating code using the source code from other programming languages. More recently, \citet{cheng2025cfceval} evaluate security aspects in LLM-generated code, \citet{li2025secure, li2025prompt} investigate automated pipelines for secure code generation, and \citet{rabbi2025babelcoder} propose an agentic approach to code translation with specification alignment.

{ The studies above leave several important  questions unaddressed. First, existing robustness evaluations of LLMs for code generation~\citep{Wang2023, copilot_robustness} have focused exclusively on Python, leaving it unclear whether their findings generalize to other widely-used programming languages with distinct syntax and semantics. Second, the benchmarks used in prior robustness work rely on HumanEval, whose test cases have been shown to be insufficient~\citep{liu2023your}, potentially masking failures that would be exposed under more rigorous evaluation. Third, while prompt-repair and input-smoothing techniques have been explored in NLP~\citep{ye2020safer,ji2024defending}, their  applicability to code generation robustness remains unexplored. This work directly targets these gaps by extending robustness evaluation to Java, C++, and JavaScript, adopting a stronger test suite based on EvalPlus, and investigating whether LLM-based docstring repair can recover lost robustness.}

%% file: methodology.tex
\section{Methodology}
\label{methodology}
\noindent We applied semantic-preserving perturbations to two types of coding tasks: (1) code generation given a Docstring and method signature, and (2) code completion given a Docstring and an incomplete method with its first few lines.

For code generation task, we include the perturbations to revise the docstrings and method signatures. and in the code completion task, from docstrings and incomplete methods. We applied two types of function and Docstring perturbations for the code generation task, and format and syntax perturbations for the code completion task.

As our study focuses on comparisons across multiple programming languages, we prepared a robustness framework that can be applied to multiple PLs, including Java, C++, and JavaScript. In this section, we will discuss the methodology step by step.

\subsection{Study Overview}
\begin{figure}
    \centering
    \includegraphics[width=0.95\linewidth]{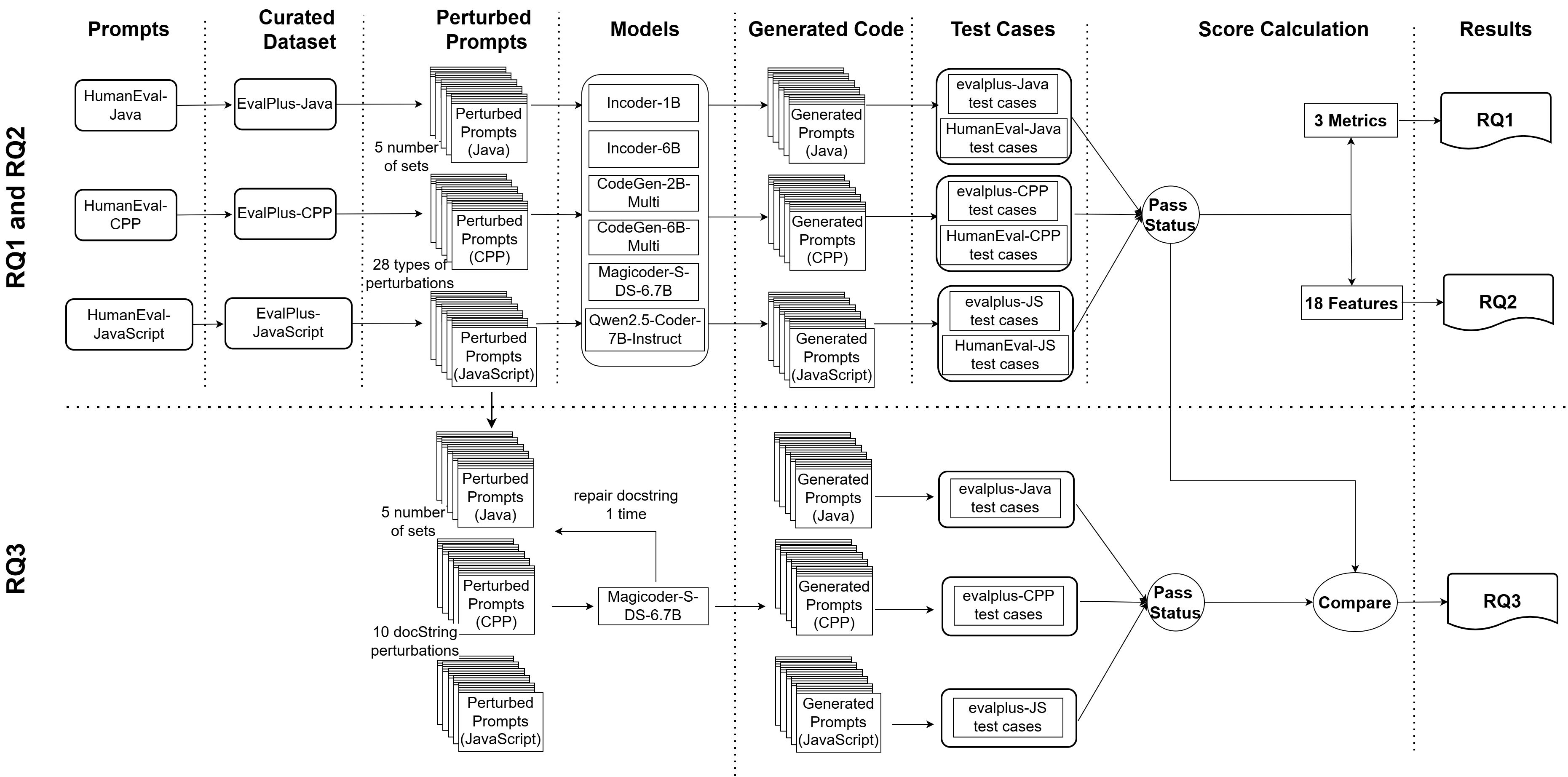}
    \caption{Overview of the study}
    \label{fig:overview}
\end{figure}

\noindent This work focuses on the end-to-end code generation task under a black box setting, where it has access only to the model's predictions with corresponding inputs. As depicted in Figure~\ref{fig:overview}, the experimental pipeline proceeds through distinct stages following the prompt construction. First, the nominal and perturbed prompts for Java, C++, and JavaScript are processed by six distinct Large Language Models (LLMs): Incoder (1B and 6B), CodeGen-Multi (2B and 6B), Magicoder-S-DS-6.7B, and Qwen2.5-Coder-7B-Instruct. 

The models generate code as output, which is then subjected to evaluation. To ensure a thorough assessment, the generated solutions are validated against both the original HumanEval test cases and the extended EvalPlus test cases. For RQ1, performance is quantified using three robustness metrics: Robust Pass (RPs@1), Robust Drop (RDs@1), and Robust Relative (RRs@1). These metrics are computed to produce a comparative report on model robustness.

For RQ2, we use 18 predefined features described in Section 5.2. These feature scores are computed from the generated code for both perturbed and nominal prompts, as well as from the prompts themselves. Using the pass/fail status and the feature values, we calculate impact scores for each language and each perturbation type. The results are then summarized to address the research question. Further details are provided in Section 5.2.

For RQ3, we focus on a single model, MagiCoder-S-DS-6.7B, selected from six evaluated models for its ability to repair docstrings. Only docstring perturbed datasets are considered. First, the model is prompted to repair the docstring within each perturbed prompt. The repaired docstring is then reintegrated into the prompt, and these repaired prompts are used to generate code following the same procedure as before. The generated code is then compared against code produced from the original perturbed docstrings. The results of this comparison are analyzed and visualized in RQ3.

The rest of this section discusses the dataset preparation, and the four types of perturbations, which consider DocString, function name, format, and syntax to generate examples, as well as the metrics used for robustness evaluation.

\subsection{Dataset Preparation} \label{evalplusprep} 
\noindent In this part, we describe the steps taken to prepare the datasets used for evaluating our experiments. Our modifications aim to enhance the robustness evaluation of code generation models by extending the scope of assessment beyond previous benchmarks.

To address the limitations of~\citet{wang2022recode}, which only evaluates code generation robustness in a single programming language (i.e., Python), we extend the evaluation to a multilingual setting by utilizing the HumanEval-X dataset. HumanEval-X, introduced by~\citet{zheng2023codegeex}, serves as a benchmark for assessing the multilingual capabilities of code generation models. The dataset includes Python, C++, Java, JavaScript, and Go. {From the remaining languages, we selected C++, Java, and JavaScript as the most representative and widely used languages in both industry and open-source development, covering diverse language paradigms (compiled vs. interpreted, statically vs. dynamically typed). Go was not selected as it is paradigmatically similar to C++ and Java, also compiled and statically typed, and would therefore contribute limited 
additional diversity to our evaluation.} We also exclude Python from our evaluation since it was already thoroughly studied in prior work, ReCode~\citep{wang2022recode}. We focused on C++, Java, and JavaScript, which represent diverse language paradigms (compiled vs. interpreted, statically vs. dynamically typed) and are widely used in real-world software development. This selection allows for a balanced and representative evaluation across different programming styles.

Prior research~\citep{liu2023your} has highlighted critical shortcomings in the HumanEval dataset used by ReCode~\citep{wang2022recode}, particularly regarding the insufficient number of test cases per coding problem and imprecise natural language (NL) descriptions of expected program behaviors. Our analysis of HumanEval-X aligns with these findings, revealing that certain canonical solutions remain vulnerable to corner cases and that many test cases fail to adequately address such scenarios.

{
To improve evaluation robustness, we extend EvalPlus to EvalPlus-X by incorporating the HumanEval-X prompts as well as the expanded test set from EvalPlus~\citep{liu2023your}. EvalPlus enhances HumanEval by refining natural language descriptions, correcting canonical solutions, and increasing the number of test cases per problem. However, because EvalPlus is limited to Python, we translated its test cases into C++, Java, and JavaScript to enable multilingual evaluation. In addition, we revised the canonical solutions in HumanEval-X to ensure they remain consistent with the additional test cases introduced by EvalPlus. Together, these enhancements make EvalPlus-X a more comprehensive and robust benchmark for evaluating multilingual code generation models.

The translation was carried out by three researchers, each independently responsible for one target language. We extend this dataset to multilingual by manually rewriting the prompts, canonical solutions, and test cases in C++, Java, and JavaScript through the following steps:

\begin{enumerate}
    \item We find it necessary to update certain canonical solutions because they failed to handle corner cases, inputs at the edges of valid ranges or uncommon conditions that reveal latent bugs. Figures \ref{fig:humanevalnotenough} and \ref{fig:humanevalnotenoughrobust} show two representative failures: in Figure \ref{fig:humanevalnotenough} the model produces a wrong answer for a corner input, while in Figure \ref{fig:humanevalnotenoughrobust} the original solution leads to a timeout. Both examples demonstrate that passing the standard HumanEval-X tests does not guarantee robustness. To mitigate this, we follow the EvalPlus methodology to refine the affected canonical solutions so they correctly handle these edge cases. In total, between 30 and 50 canonical solutions were modified for each of the three languages. Each modified solution was validated by executing it against all converted test cases, and only solutions that passed were retained. After applying those fixes, we generate prompts from the revised solutions for subsequent evaluation.

    \item We convert the test cases from EvalPlus, originally written in Python, into Java, C++, and JavaScript. During this process, we ensure that data types are transformed into their equivalent counterparts (e.g., converting lists to \texttt{ArrayList} in Java and \texttt{vector} in C++).

    \item We exclude test cases that rely on Python-specific type flexibility that does not exist in statically typed languages. For example, Python allows heterogeneous lists containing mixed types such as integers and strings, which are not valid in C++ or Java. Similarly, Python supports arbitrarily large integers, whereas Java, C++, and JavaScript are bounded by fixed-size data types. These exclusions are therefore driven by fundamental type system incompatibilities rather than difficulty. The impact of these exclusions is minimal: the original EvalPlus Python benchmark has an average of 764.1 test cases per problem, while our translated versions retain an average of 722.56 (C++), 748.48 (Java), and 709.99 (JavaScript) as stated in ~Table\ref{stat_tc}, a reduction of less than 7\% in all cases. We therefore do not expect these exclusions to materially affect the difficulty or validity of our benchmark. Throughout this process, we strive to maintain the method signature, including return types and parameter types, identical to those found in the original prompts.
\end{enumerate}
}

\subsection{Semantic Preserved Perturbation}
\noindent After preparing our dataset by converting the test cases to multiple programming languages and updating the canonical solutions accordingly, we apply semantically preserved perturbations to the prompts for each language in a distinct manner. For the code generation task, we perturb the prompt by applying 7 function name perturbations and 10 docstring perturbations. For the code completion task, we first construct incomplete code snippets using the prompt and a portion of the canonical solution. We then apply 6 syntax perturbations and 6 format perturbations to these incomplete codes. The specific details of the perturbation techniques are described in the following sections.

\subsubsection{Perturbation on DocString}
\noindent We applied 10 semantic-preserved perturbations on the DocString parts of the prompts. A docString is the human-written natural language to describe the method's functionality. In a prompt, DocString provides the major instruction about how a method will be generated by a code generation model.

In this work, we follow ReCode's implementation, which is based on NL-Augmenter~\citep{dhole2021nl} for natural language transformations of datasets dealing with natural language\footnote{https://github.com/GEM-benchmark/NL-Augmenter}. We follow ten perturbation techniques for DocString from ReCode, which cover character-level, word-level, and sentence-level changes. As DocString is agnostic across languages, we directly apply the perturbations from ReCode to Java, C++, and JavaScript. The explanations with examples are shown in Table \ref{docstringexample}. These examples are collected from our perturbed dataset which we generated from the HumanEval-X dataset\footnote{https://huggingface.co/datasets/THUDM/humaneval-x}. For SynnonymSubstitution in DocString, synonyms by~\citep{miller1995wordnet} are used to replace words in DocString with their synonyms. For BackTranslation, the whole DocString is translated into another language and then retranslated to English again to paraphrase~\citep{sugiyama2019data,li2019improving}. 

\textbf{Adaptation of Docstring Perturbations to multilingual} To perturb a DocString, first, it is separated from the whole prompt, and the perturbation techniques are applied. The DocString may contain keywords of programming languages from code that must be untouched as changing them leads to altering the meaning. To maintain this, a blacklist of different programming languages is created which prevents the system from perturbing them. To separate the keywords and generate the blacklist, we utilize Tree-Sitter-Java, Tree-Sitter-C++, and Tree-Sitter-JavaScript for JavaSc, C++, and JavaScript, respectively. The function names, variable names, and types are extracted from a code snippet by Tree-Sitter. 

\begin{table}
\centering
\caption{{Perturbations Examples for docString}}
\large
\label{docstringexample}
\resizebox{\textwidth}{!} { 
\begin{tabular}{| p{0.35\textwidth} | p{0.4\textwidth} | p{0.6\textwidth} |}
\hline
\textbf{Perturbation} & \textbf{Explanation} & \textbf{Example} \\
\hline
Nominal & Original Prompt & Filter an input list of strings only for ones that start with a given prefix. \\ \hline
Back Translation & Translate to another language and re-translate to English & Filter an input list of strings only for strings starting with a \textbf{certain} prefix. \\ \hline
Butter Fingers Perturbation & Replace random characters with some other characters & Fil\textbf{h}er an input list of strings only for ones that start \textbf{q}ith a given prefix. \\ \hline
Change Char Case & Change case of random characters. & Fil\textbf{T}er a\textbf{N} inpu\textbf{T} list of strings Only f\textbf{O}r one\textbf{S} that st\textbf{A}rt With a Giv\textbf{EN} prefix. \\ \hline
English Inflectional Variation & Change the form of random verbs in a sentence & Filter an input list of strings only for ones that start with a \textbf{gives} prefix. \\ \hline
Swap Characters Perturbation & Swap two adjacent characters with each other & Filt\textbf{re} an input list \textbf{fo} strings only for ones that start with a given prefix. \\ \hline
Synonym Insertion & Insert a word's synonym adjacent to its position & Filter an input \textbf{stimulation} list of strings only for ones that start \textbf{get down} with a given \textbf{impart} prefix. \\ \hline
Synonym Substitution & Replace a word with its synonym & Filter an input list of strings only for ones that start with a \textbf{give} prefix. \\ \hline
Tense Transformation Future & Transform the tense to future & Filter an input list of strings only for ones that will start with a \textbf{will give} prefix. \\ \hline
Tense Transformation Past & Transform the tense of the sentence to past & Filter an input list of strings only for ones that started with a \textbf{gave} prefix. \\ \hline
Whitespace Perturbation & Insert or remove whitespaces at random places & Filter \textbf{aninpu t} list of \textbf{str ings} only  for ones that \textbf{startwith} a given \textbf{pr e fix}. \\
\hline
\end{tabular}}
\end{table} 

\subsubsection{Perturbation on Function Name} \noindent With this scope of perturbation, the function name is changed slightly, alike the DocString. The 7 perturbations used for Function renaming are shown in Table~\ref{tab:funcexm} with examples and short explanations. Among the 7 perturbations, the CamelCase perturbation is applied to C++ and JavaScript function names, where the original functions are written in snake case. Conversely, for Java, the snake case perturbation is applied since function names are written in camel case. 

Among the 7 perturbations, InflectionalVariation and SynonymSubstituition are word-level, and the rest of them are character-level perturbations. CamelCase perturbation is applied to the C++ dataset as the function names in HumanEval-X for C++ are written following the snake case format. On the other hand, SnakeCase is applied to Java and JavaScript function names as they follow the camel case format in function names. 

\textbf{Adaptation of Function Name Perturbations to Multiple Programming Languages:} The semantic-preserved perturbations applied to function names require splitting a function name into multiple words. Different programming languages have different naming conventions. 
To apply semantics-preserving perturbations to function names, we first split the names into individual tokens based on the naming convention used in each language. For Java and JavaScript, we assume camelCase and split accordingly. For C++, we assume snake\_case. If the detected naming style does not match the expected convention, it is first converted to the appropriate form (camelCase or snake\_case) before splitting.
Once split, the function name is treated as a sequence of words, and perturbations are applied to these words. These changes preserve the overall meaning while introducing variation in the name. 
\begin{table}
\centering
\caption{{Perturbations Examples for function name}}
\label{tab:funcexm}
\resizebox{\textwidth}{!}
{ \begin{tabular}{|l|l|l|}
\hline
\textbf{perturbation} & \textbf{explanation} & \textbf{Example} \\
\hline
Original & original function name & getPositive \\
SnakeCase & change to snake cases from snake cases & get\textbf{\_}positive \\
CamelCase & change to camel cases from camel cases & getPositive \\
ButterFinger & change a random character inside a function name  & get\textbf{O}ositive \\
SwapChar & swap two adjacent characters inside a function name & getP\textbf{so}itive \\
ChangeChar & Change Case of random characters & get\textbf{p}osi\textbf{T}ive \\
InflectionalVariation & Change form of a word in a function name & \textbf{gotten}Positive \\ 
SynonymSub & Replace a word in a function name & \textbf{catch}Positive \\
\hline
\end{tabular}}
\end{table} 

\subsubsection{Perturbation on Syntax}
\noindent Following the approach of~\citet{wang2022recode}, we apply syntax-level perturbations to the code portion of prompts in a code completion setting. Unlike code generation, which uses only the docstring and method signature, code completion provides part of the function body to simulate a developer writing or editing code.
Specifically, for each coding problem, we start with the canonical solution and select approximately the first half of its lines, following the strategy used in \citet{wang2022recode}, to create a baseline prompt. This prompt includes the original docstring and method signature, along with the selected code lines. We apply syntax-preserving perturbations only to these added lines. These perturbations introduce surface-level variation without changing the underlying semantics.
To create partial code prompts for all three programming languages, we use the HumanEval-X dataset. For each language, we extract the canonical solution, truncate the function body, and apply the perturbation strategy described above. In this context, partial code refers to a prompt that includes the docstring, method signature, and a perturbed portion of the function body, tailored for code completion tasks.

We use the same six perturbations used by ReCode~\citep{wang2022recode} where three of them: DeadCode Insertion, Operand Swap, and For-While-Switch are generated by NatGen~\citep{chakraborty2022natgen}. The rest three of them: Variable Renaming by CodeBERT \citep{feng2020codebert}, Variable Renaming with NatGen style, and generate random variable names using a combination of alphabetic and numeric characters are prepared by ReCode itself.

\begin{figure}
    \centering
    \includegraphics[width=0.85\linewidth]{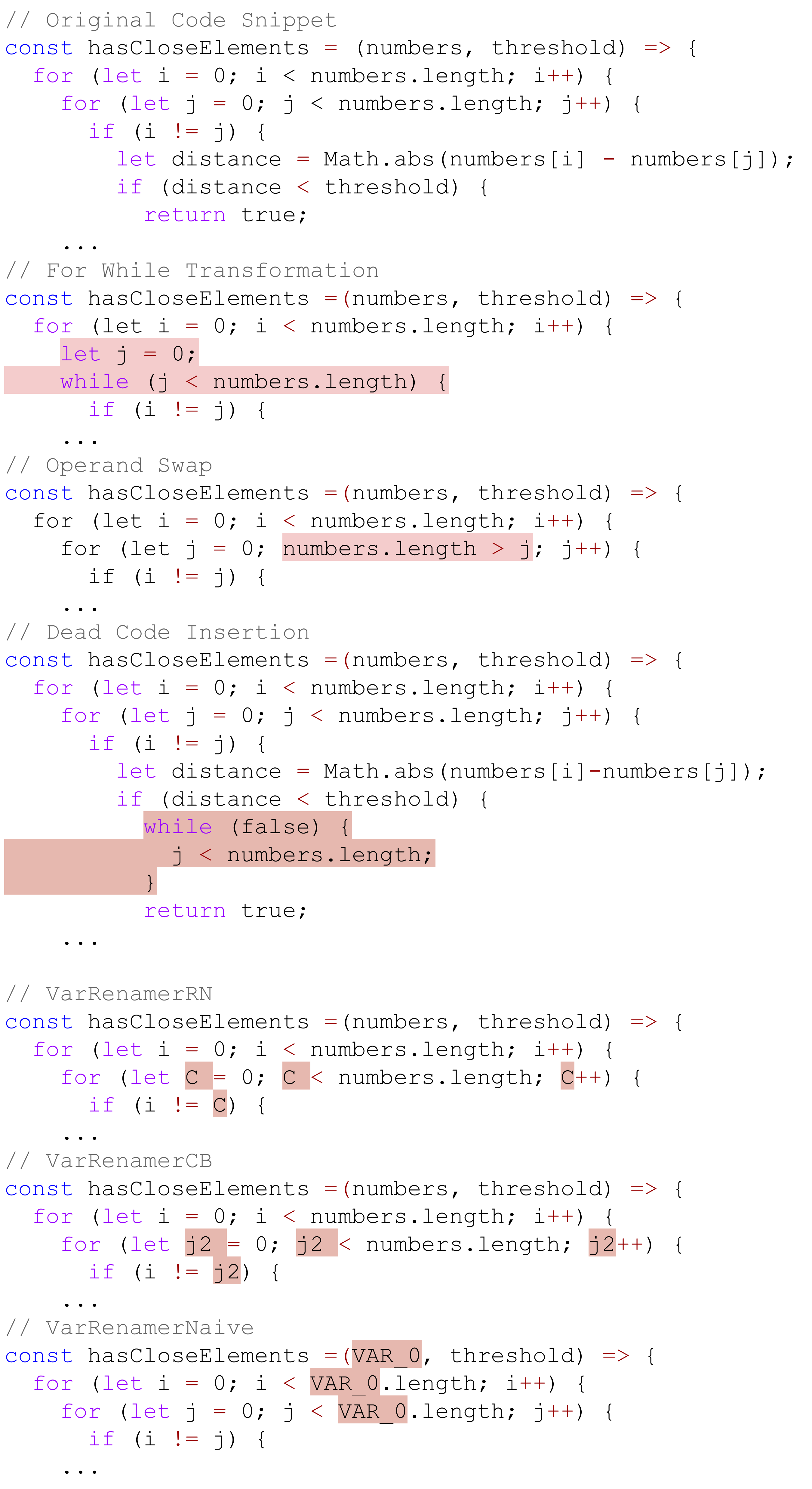}
    \caption{Syntax Perturbations (JavaScript)}
    \label{fig:syntaxexamples}
\end{figure}

The following list explains the syntax perturbations used in this work. Figure \ref{fig:syntaxexamples} illustrates examples of how the perturbations affect the original code snippet. The changes that happen in each of the syntax perturbations are highlighted.
\begin{enumerate}
    \item \textbf{DeadCodeInserter}: Introduces irrelevant and non-executable code into the original codebase by inserting a code block at a random location. The inserted block may take the form of a loop that iterates zero times or an ``if'' statement that always evaluates to false. Its content is randomly selected from nearby code statements, constrained by small Tree-Sitter node sizes. This transformation is implemented using NatGen \citep{chakraborty2022natgen}, which provides mechanisms for programmatically modifying code syntax trees to inject such dead code in a syntactically valid way.
    \item \textbf{ForWhileTransformer}: Select a random for loop and transfer it to an equivalent while loop or vice versa.  
    \item \textbf{OperandSwap}: Randomly pick a binary logical operation from a set of possible ones and switch the positions of the two operands involved in that operation (e.g. $a<b$ to $b>a$). If necessary, the operator is adjusted to ensure that the semantic meaning of the expression remains the same.
    \item \textbf{VarRenamerCB}: This transformation substitutes a new name created by CodeBERT~\citep{feng2020codebert} for the most frequently used variable name in a particular code snippet. A mask token is used to substitute the original variable name. After that, CodeBERT generates a list of alternative names for each occurrence of the mask token along with a probability score. The name that received the highest rating overall gets chosen as the replacement.
    \item \textbf{VarRenamerNaive}: This modification chooses the partial code's most often used variable name and changes it to ``VAR\_0.'' This is the NatGen package's original implementation. The change does not alter the meaning.
    \item \textbf{VarRenamerRN}: Selects the most commonly used variable name from the partial code and replaces it with a randomly generated string composed of a mix of alphabetic and numeric characters. 
\end{enumerate}

\noindent \textbf{Adaptation of Syntax Perturbations to multilingual:} Initially, the task involves generating partial prompts from the nominal prompts. To accomplish this, we make adjustments to ReCode's code and introduce specific rules for creating partial code, effectively separating the header, DocString, body, and other segments for Java, C++, and JavaScript. Then, we leverage the NatGen libraries to generate syntax-level perturbations.

\subsubsection{Perturbation on Format}
\noindent This perturbation is employed to evaluate whether the code generation model can effectively handle a perturbed prompt with identical information to the original one but with a distinct formatting style. It tests the model's ability to handle variations in formatting while generating code. In this work, we use 6 types of formats as follows.

\begin{enumerate}
    \item \textbf{doc2comments}: This perturbation transforms a Docstring (written as a block comment) into multiple lines of inline comments.
    \item \textbf{new\_line\_aftercode}: A new empty line is added after the code.
    \item \textbf{new\_line\_afterdoc}: A new empty line is added after the docString. 
    \item \textbf{new\_lines}: A new line is added at a random position between any two lines in the full prompt.
    \item \textbf{split\_lines}: Split the longest line of code and split it into two lines in the middle.
    \item \textbf{tab\_indent}: random replacements of space indentation with tabs or replace tabs with four spaces. 
    Although languages such as Java, JavaScript, and C++ are not sensitive to indentation, this perturbation is used to check the robustness of code generation models. 
\end{enumerate}

\textbf{Adaptation of Format Perturbations:} Format perturbations use the same partial codes as Syntax perturbations. We extend ReCode's code to apply formatting separately for Java, C++, and JavaScript, considering differences in the header, DocString, and body positions. Unlike Python, Java, C++, and JavaScript allow direct newlines between tokens without requiring a backslash (For Python, "\textbackslash\textbackslash\textbackslash n" is placed for newlines, while for other languages, they are simply "\textbackslash\textbackslash n".)

We applied 29 perturbation strategies in total: 7 perturbations on function name, 10 on docstring, 6 syntax perturbation, and 6 formatting perturbations. For each coding problem, 28 of these strategies were used to alter the original prompt. The remaining strategy depends on the language (camelCase is used for C++/JavaScript, whereas snake\_case is used for Java). Each strategy produces five perturbed versions of the prompt, resulting in 28 × 5 perturbed prompts per problem, and we generate one code sample for each perturbed prompt. Some strategies may occasionally generate identical text, but using a non-zero sampling temperature during generation still produces varied model outputs across runs.

We follow ReCode’s practice of generating five perturbed prompts per strategy to capture the natural variability introduced by stochastic perturbation methods. Multiple samples provide a more stable and representative estimate of each strategy’s robustness, as some techniques exhibit high variability while others are largely deterministic.

For certain strategies, especially when perturbing very short text (e.g., single-word function names), some or all of the five perturbed prompts may become identical. Deterministic methods such as back-translation can even produce identical outputs across all five samples. We treat such duplication as an inherent property of the strategy, reflecting its limited perturbation diversity rather than a methodological issue.

Identical perturbed prompts do not affect our robustness results; they reflect the limited variation that some perturbation methods naturally produce.

\subsection{Evaluation}
\label{eval}
\noindent There are several metrics used by previous works to evaluate the performance of code generation. In our evaluation, we consider both the worst-case passes (i.e., for each coding problem, all perturbed prompts $s$ of a prompt must be passed) and partial-case passes (a partial number $t<s$ of the perturbed prompts from a prompt must be passed). Alike ReCode~\citep{wang2022recode}, we use the following metrics for our evaluation. We choose a greedy approach for all of the metrics, skipping the sampling process to eliminate the randomness effect and enable fair comparisons~\citep{wang2022recode}.

\subsubsection{Robust Pass (\texorpdfstring{$RP_s@k$}))} 
\noindent Pass@k is a commonly used metric for code generation tasks. As we are generating multiple perturbed prompts from a single prompt, we use a modified version of this metric. From the initial prompt, there is $s$ number of perturbed prompts. Additionally, $k$ generations are made from each perturbed prompt, for a total of $s*k$ generations from a single prompt. In this work, we use $s=5$ and $k=1$. We use the $RP_s@k$ metrics as equation~\ref{rps@k}.
\begin{equation}
    \label{rps@k}
    RP_s@k = \frac{RC_s}{N}
\end{equation}
where $RC_s$ denotes the number of correct prompts and $N$ denotes the total number of prompts.

We follow recent evaluation practices that report \textit{pass@1} as the primary metric for code generation~\citep{chen2021evaluating, nijkamp2022codegen}, and therefore use $k=1$ for each prompt. Our goal is to evaluate the model's \textit{robustness to input variation} rather than its sampling variance. Accordingly, instead of generating multiple outputs from a single prompt, we generate five perturbed versions of each sample. Multiple generations from the same prompt would primarily capture decoding randomness, whereas multiple perturbed prompts reveal how sensitive the model is to small input changes. Using five perturbed prompts thus provides a more direct and meaningful robustness assessment.

\subsubsection{Robust Drop (\texorpdfstring{$RD_s@k)$})}
\noindent The Robust Drop metric is used to measure the decrease in robustness of the generated code under perturbed data compared to the original data. This metric states the fall in the performance compared to the nominal data. In equation \ref{rds@k}, this metric is defined.
\begin{equation}
    \label{rds@k}
    RD_s@k = \frac{Pass_{nominal}@k - RP_s@k}{Pass_{nominal}@k}
\end{equation}
where $Pass_{nominal}@k$ denotes the result for the nominal dataset and $RP_s@k$ denotes the robust relative value for the perturbed dataset we get from equation \ref{rps@k}.

\subsubsection{Robust Relative (\texorpdfstring{$RR_s@k)$})} 
\noindent The Robust Drop metric provides an idea of the performance decrease for the perturbed dataset. But it does not highlight the pair-by-pair performance change between nominal and perturbed datasets. To get the relative changes, the Robust Relative metric is used. There can be instances where models produce inaccurate code for the nominal prompts but accurately predict the perturbed ones. This non-robust behavior should also be taken into account when reporting the model's robustness. For this situation, the Robust Relative metric is useful. In equation \ref{rrs@k}, the Robust Relative metric is defined.
\begin{equation}
    \label{rrs@k}
    RR_s@k = \frac{RC_s^{[+]} + RC_s^{[-]}}{N}
\end{equation}
where $RC_s^{[+]}$ denotes the number of correct cases in perturbations that were detected as incorrect cases for the original prompts. $RC_s^{[-]}$ denotes the opposite, the number of incorrect cases in perturbations that were detected as correct cases for the original prompts. This work follows a greedy setting (n=1 sample), and the equation is made upon this setting, and this metric denotes the number of positive or negative changes for a set of prompts.

%% file: experiments.tex
\section{Experiment Setup}
\noindent In this section, we present the experimental design and setup employed to evaluate the robustness of code generation models. By meticulously addressing each sub-section, we provide a comprehensive overview of our experiments, enabling a deeper understanding of our research implementation and subsequent analysis.
\label{exp}
\subsection{Dataset}
\label{data}
\noindent As described in Section~\ref{evalplusprep}, we construct a new dataset by combining, extending, and converting two existing datasets to evaluate the robustness of code generation models. Below, we provide a detailed description of these source datasets and our final curated dataset used in the experiments.

\noindent\textbf{HumanEval-X.} HumanEval-X\footnote{https://huggingface.co/datasets/THUDM/humaneval-x} is an extended version of the HumanEval\footnote{https://github.com/openai/human-eval} dataset, originally designed to evaluate code generation models in Python. HumanEval-X expands this benchmark to support multiple programming languages, including C++, Java, JavaScript, and Go. It consists of 164 prompts, each accompanied by a canonical solution and language-specific test cases. In our study, we leverage HumanEval-X to assess code translation and bug classification across Java, C++, and JavaScript. This dataset provides a total of 820 test cases for 820 problems, with an average of 164 test cases per programming language and approximately 7.7 unit tests per problem.

\noindent\textbf{EvalPlus.} EvalPlus~\citep{liu2023your} is an enhanced version of HumanEval, originally developed for Python. It introduces a substantially larger set of corner cases and refined prompts that are better aligned with the associated test cases. In this work, we adapt the EvalPlus test cases for our three target programming languages, following the curation process described in Section~\ref{evalplusprep}. The prompts are largely consistent with those in HumanEval-X, except in instances where minor modifications were required to accommodate language-specific corner cases.

While EvalPlus offers more comprehensive and rigorous test coverage, many HumanEval-X reference solutions were not directly compatible with its test cases. To address this, we manually implemented canonical solutions for each target language and evaluated them using the converted EvalPlus test cases. This adaptation ensures that our cross-language evaluations are both valid and comparable. The detailed statistics of the resulting test cases are presented in Table~\ref{stat_tc}.

\begin{table}
\centering
\caption{Statistical measures of the number of test cases for different languages}
\label{stat_tc}
\begin{tabular}{lcccccc}
\toprule
\multirow{2}{*}{~} & \multicolumn{2}{c}{Avg} & \multicolumn{2}{c}{Median} & \multicolumn{2}{c}{Std} \\
\cmidrule(lr){2-3} \cmidrule(lr){4-5} \cmidrule(lr){6-7}
Language & HumanEval-X & Ours & HumanEval-X & Ours & HumanEval-X & Ours \\
\midrule
CPP  & 6.96 & 722.56 & 7.0 & 973.5 & 3.73 & 375.95 \\
Java & 6.5  & 748.48  & 6.0 & 980.0 & 4.55 & 359.39 \\
JS   & 7.95 & 709.99 & 8.0 & 972.0 & 3.73 & 379.72 \\
\bottomrule
\end{tabular}
\end{table}

\subsection{Studied Code Language Model}
\label{targetmodel}
\noindent In our experiment, we focus on evaluating six distinct code generation models: Incoder-1b, Incoder-6b, CodeGen-2b-Multi, and CodeGen-6b-Multi Magicoder-7B, and QwenCode-2.5-6.7B as they have been trained in multiple programming languages. These models are specifically chosen due to their varying number of parameters, allowing us to assess our experiments across a range of model complexities. Due to the excessive size and our inability to run them on our machine, we exclude certain large models, such as CodeGen-16B-Multi, StarCoder-13B, etc, from our experiment. 
\subsection{Setup}
\noindent \textbf{Model Setup and Hyperparameter Selection.}
We use HuggingFace’s \texttt{AutoModelForCausalLM} API to load pretrained code generation models for all target programming languages. We did not perform any training or fine-tuning; we directly evaluated them in their pretrained form. Our goal is to assess the performance of existing models under consistent experimental conditions.

The same set of hyperparameters is applied across models, chosen through a small grid search on a held-out subset of the evaluation data. The temperature is set to 0.2 to reduce randomness and increase the likelihood of generating syntactically valid and test-passing code. The nucleus sampling parameter (\texttt{top\_p}) is set to 0.95 to retain diversity while suppressing low-probability outputs. These values are taken from the ReCode~\citep{wang2022recode}. The maximum generation length is set to 1536 tokens to prevent truncation. This value is based on our experience that it provides sufficient output while keeping inference time low.

We follow the same practice of pre-processing, prompting, and post-processing steps from prior works~\citep{nijkamp2022codegen}~\citep{wei2023magicoder}~\citep{hui2024qwen2}. Language and model-specific stopping rules are applied to keep generations consistent and clean. Extra content, such as redundant helper functions or repeated class definitions, is removed. For Java, generation stops at the appearance of a second \texttt{Solution} class. For JavaScript and C++, it stops at the second docstring or comment block.

\subsection{Evaluation of Perturbation Quality} 
\noindent As stated in Section~\ref{rtocgm}, the perturbations should preserve the semantic meaning of the original prompts while appearing natural. To evaluate the perturbed dataset's quality, we employ three measurements for the docString, function name, and code sections described in the following.

\begin{table}[]
    \centering
    \caption{Similarity Values Between Nominal and Perturbed Datasets}
    \label{tab:sim_values}
        \begin{tabular}{|p{0.25\textwidth}|p{0.1\textwidth}|p{0.1\textwidth}|p{0.15\textwidth}|p{0.20\textwidth}|}
        \hline
        Scope & Java & CPP & JavaScript & ReCode (Python) \\
        \hline
        Syntax     & 0.96 & 0.97 & 0.95 & 0.96 \\
        DataFlow   & 0.97 & 0.97 & 0.94 & 0.97 \\
        DocString  & 0.93 & 0.93 & 0.93 & 0.93 \\
        Func Name  & 0.79 & 0.85 & 0.79 & 0.81 \\
        \hline
    \end{tabular}
\end{table}

\subsubsection{Sentence Similarity Measurement} \noindent For the Java, C++, and JavaScript datasets, we assess the semantic similarity between the perturbed and non-perturbed prompts. To do this, we employ the sentence cosine similarity measurement, following the same methodology as ReCode. We calculate the similarities between the perturbed and nominal datasets using the sentence embeddings from MPNet \citep{song2020mpnet}. To be more specific, we utilize MPNet embeddings to measure the cosine similarities for the docString and function name sections. For docstrings, we directly compare them after extraction, while for function names, we split them into lists of words and then compare them. Table \ref{tab:sim_values} shows the summary of similarities, and Table \ref{tab:sentence_sim_all} shows the similarities for individual perturbations. 
These scores range from 0 to 1, where values close to 0 indicate dissimilarity and values close to 1 indicate high similarity. The two tables show that the scores are generally close to 1, indicating that the perturbations only slightly modify the docstrings and function names while largely preserving their content.

\begin{table}
    \centering
    \caption{Sentence Similarities for individual docString and function name perturbations}
    \label{tab:sentence_sim_all}
    \begin{adjustbox}{max width=\textwidth}
    \begin{tabular}{|p{0.1\textwidth}|p{0.4\textwidth}|p{0.1\textwidth}|p{0.1\textwidth}|p{0.1\textwidth}|}
        \hline
        Categories & Perturbations & Java & JavaScript & CPP \\
        \hline
        \multirow{10}{*}{\centering docString} & ChangeCharCase & 1.0 & 1.0 & 1.0 \\
         & TenseTransformationPast & 0.98 & 0.98 & 0.98 \\
         & EnglishInflectionalVariation & 0.95 & 0.95 & 0.95 \\
         & WhitespacePerturbation & 0.89 & 0.9 & 0.89 \\
         & ButterFingersPerturbation & 0.86 & 0.86 & 0.87 \\
         & TenseTransformationFuture & 0.97 & 0.97 & 0.97 \\
         & SynonymSubstitution & 0.88 & 0.88 & 0.88 \\
         & SwapCharactersPerturbation & 0.9 & 0.89 & 0.9 \\
         & SynonymInsertion & 0.9 & 0.89 & 0.91 \\
         & BackTranslation & 0.94 & 0.93 & 0.92 \\
        \hline
        \multirow{6}{*}{\centering Function} & FuncRenameSynonymSub & 0.72 & 0.75 & 0.72 \\
        & FuncRenameButterFinger & 0.56 & 0.62 & 0.57 \\
        & FuncRenameSwapChar & 0.83 & 0.83 & 0.82 \\
        & FuncRenameSnake/CamelCase & 0.89 & 0.89 & 0.89 \\
        & FuncRenameInflectionalVariation & 0.93 & 0.94 & 0.93 \\
        & FuncRenameChangeChar & 0.87 & 0.91 & 0.86 \\
        \hline
    \end{tabular}
    \end{adjustbox}
\end{table}

\subsubsection{Code Similarity Measurement} 
\noindent To assess the similarities between the perturbed and nominal data, we utilize CodeBLEU's Syntax and DataFlow scores \citep{ren2020codebleu}. For our analysis, we adapt the CodeBLEU score calculation method employed by CodeXGlue \citep{lu2021codexglue}. Additionally, to calculate scores for C++, we incorporate NaturalCC \citep{wan2022naturalcc} since the DataFlowGraph for C++ is not available in CodeXGlue, which is essential for computing the DataFlow score. The summary of similarity scores is presented in Table \ref{tab:sim_values}, while Table \ref{tab:codebleu_all} displays the individual similarity scores. These scores range from 0 to 1 as well, where values closer to 1 indicate higher similarity between the perturbed and nominal code. Overall, the scores are relatively high across most perturbations, suggesting that the code semantics and structure are largely preserved. However, the DeadCodeInserter perturbation yields comparatively lower scores, as the insertion of additional code blocks introduces noticeable structural differences that reduce similarity.

\begin{table}
    \centering
    \caption{CodeBLEU Similarities for individual perturbations}
    \label{tab:codebleu_all}
    \begin{adjustbox}{max width=\textwidth}
    \begin{tabular}{|l|l|l l|l l|l l|}
        \hline
        \multirow{2}{*}{\centering Categories} & \multirow{2}{*}{\centering Perturbations} & \multicolumn{2}{|c|}{Java} & \multicolumn{2}{|c|}{CPP} & \multicolumn{2}{|c|}{JavaScript} \\
         &  & Syntax & Dataflow & Syntax & Dataflow & Syntax & Dataflow \\
        \hline
        \multirow{6}{*}{\centering Syntax} & ForWhileTransformer & 0.94 & 0.99 & 0.96 & 1.0 & 0.93 & 0.99 \\
         & OperandSwap & 0.87 & 0.99 & 0.97 & 1.0 & 0.87 & 1.0 \\
         & DeadCodeInserter & 0.76 & 0.75 & 0.76 & 0.64 & 0.74 & 0.72 \\
         & VarRenamerRN & 0.99 & 0.98 & 0.99 & 1.0 & 1.0 & 0.98 \\
         & VarRenamerCB & 0.98 & 0.97 & 1.0 & 1.0 & 0.99 & 0.97 \\
         & VarRenamerNaive & 0.99 & 0.99 & 0.99 & 1.0 & 1.0 & 0.98 \\
        \hline
        \multirow{6}{*}{\centering Format} & newLineAfterdoc & 1.0 & 1.0 & 1.0 & 1.0 & 1.0 & 1.0 \\
         & SplitLines & 0.99 & 1.0 & 0.99 & 1.0 & 0.97 & 1.0 \\
         & Doc2Comments & 1.0 & 1.0 & 0.92 & 0.99 & 0.89 & 0.69 \\
         & NewLineAftercode & 1.0 & 1.0 & 1.0 & 1.0 & 1.0 & 1.0 \\
         & NewLines & 1.0 & 1.0 & 1.0 & 1.0 & 1.0 & 1.0 \\
         & TabIndent & 1.0 & 1.0 & 1.0 & 1.0 & 1.0 & 1.0 \\
        \hline
    \end{tabular}
    \end{adjustbox}
\end{table}

\subsubsection{Human Annotation of Naturalness and Semantic Similarity in Perturbed Prompts} 

\noindent We conducted a random sampling of 300 prompt pairs, comprising 100 Java, 100 C++, and 100 JavaScript examples, each containing a \textbf{perturbed} and \textbf{non-perturbed} version of the same prompt. Each pair was annotated independently by two annotators, who provided three scores: 
\begin{itemize}
    \item the \textit{naturalness} of the perturbed prompt
    \item the \textit{naturalness} of the non-perturbed prompt
    \item the \textit{semantic similarity} between the perturbed and non-perturbed versions.
\end{itemize}

We follow 5-point Likert scale during the annotation process, both annotators provide discrete scores from ${0, 0.25, 0.5, 0.75, 1}$.
\begin{itemize}
    \item 0 (not natural / not similar)
    \item 0.25 (mostly unnatural / mostly dissimilar)
    \item 0.5 (neutral)
    \item 0.75 (mostly natural / mostly similar)
    \item 1 (fully natural / fully similar). 
\end{itemize}
These annotations were performed by two independent annotators to ensure consistency and reliability. The aggregated results of this annotation study are summarized in Table~\ref{tab:annotations}.

We use Cohen’s Kappa to measure inter-annotator agreement across Naturalness (Nominal), Naturalness (Perturbed), and Semantic Similarity. Ratings were discretized into five ordinal bins following 5-point Likert scale (0, 0.25, 0.5, 0.75, 1.0) for categorical consistency. The kappa values show fair to moderate agreement (0.1-0.29), with slightly lower scores for similarity due to its subjective nature. Overall, these results indicate a consistent and reliable annotation process.

Table~\ref{tab:annotations} presents the average annotation scores and inter-annotator agreement for Java, C++, and JavaScript. The results show that nominal prompts exhibit near-perfect naturalness (1.0), while perturbed prompts retain high naturalness (0.88–0.91), confirming that the perturbations remain largely fluent. Semantic similarity scores (0.94–0.96) indicate that meaning is well preserved across languages. The corresponding Cohen’s kappa values suggest fair to moderate agreement among annotators, reinforcing the consistency of the evaluation while reflecting the inherent subjectivity of semantic judgments in code understanding.

\begin{table}
    \centering
    \caption{Result of Human Annotations (Average Scores and Inter-Annotator Agreement) for 300 Random Samples}
    \label{tab:annotations}
    \begin{tabular}{|l|c|c|c|}
        \hline
        \textbf{Scope} & \textbf{Java} & \textbf{C++} & \textbf{JavaScript} \\
        \hline
        Naturalness (Nominal) $\uparrow$ & 0.996 & 0.996 & 1.000 \\
        Naturalness (Perturbed) $\uparrow$ & 0.914 & 0.888 & 0.900 \\
        Semantic Similarity $\uparrow$ & 0.954 & 0.945 & 0.955 \\
        \hline
        Cohen’s $\kappa$ (Nominal) & 0.10 & 0.12 & 0.08 \\
        Cohen’s $\kappa$ (Perturbed) & 0.22 & 0.29 & 0.15 \\
        Cohen’s $\kappa$ (Similarity) & 0.26 & 0.18 & 0.06 \\
        \hline
    \end{tabular}
\end{table}


\section{Experiment Results}
\label{result}
\noindent In this section, we aim to answer three research questions (RQs). For each RQ, we
present the motivation and the approach to
address the RQ, and the corresponding results.

\begin{figure}
    \centering
    \includegraphics[width=\linewidth]{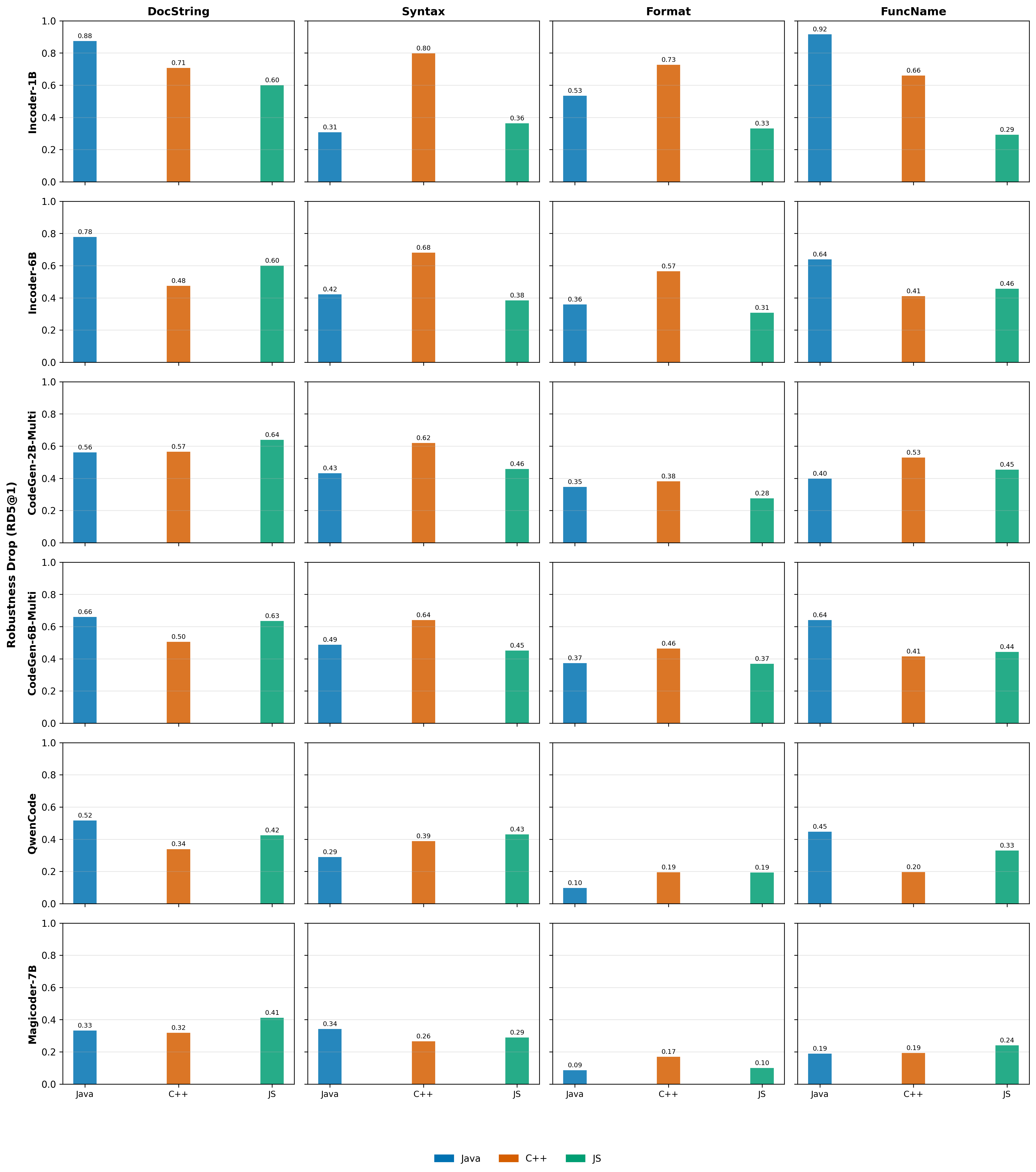}
    \caption{Robust Drop among different languages and models}
    \label{fig:robust_drop}
\end{figure}

\subsection{\textbf{RQ1: How robust are LLMs across different programming languages?}}

\noindent\textbf{Motivation:} Our experiment encompasses three distinct languages and four different models. A prior investigation \citep{wang2022recode} demonstrates the influence of perturbations on the robustness of code generation models. Nonetheless, the question of whether language-specific variations impact robustness remains unanswered. Hence, in this study, we aim to address this gap by exploring discrepancies in LLMs' robustness across diverse programming languages (Java, C++, and JavaScript), shedding light on the extent of disparities among these languages.

\noindent\textbf{Approach:} To investigate this research question, we conduct experiments on Java, C++, and JavaScript using our EvalPlus-X derived from EvalPlus~\citep{liu2023your} and HumanEval-X~\citep{zheng2023codegeex}. We evaluate six LLMs for code generation: Incoder-1B, Incoder-6B, CodeGen-Multi-2B, CodeGen-Multi-6B, Magicoder-7B, and QwenCode-3.5-6.7B. We evaluate the performance using three metrics: $RP_s@k$ measures robustness, $RD_s@k$ assesses the decrease in robustness due to changes, and $RR_s@k$ captures the relative differences between the original and modified datasets. The perturbations related to the Function Name and the DocString are compared with the nominal prompts (prompts that have only DocString and method signature), while the perturbations related to Syntax and Format are compared with the partial prompts (prompts that have some lines of code along with DocString and method signature).

\noindent\textbf{Results:} \textbf{Overall, the performance of models deteriorates after applying the perturbations.} Table~\ref{tab:ressummary_ep} and ~\ref{tab:ressummary_he} present the worst-case performance of six models: Incoder-1B, Incoder-6B, CodeGen-2B-Multi, CodeGen-6B-Multi, Magicoder-7B, and QwenCode-3.5-6.7B under different perturbation types (DocString, Function Signature, Syntax, and Format) across three programming languages (Java, C++, and JavaScript). Similarly, Table~\ref{tab:ressummarypartial_ep} and ~\ref{tab:ressummarypartial_he} show the results in partial-case scenarios, where a model passes at least 3 out of 5 generated samples.

Across all of the tables, we observe a consistent performance drop after applying perturbations. For instance, in Table~\ref{tab:ressummary_ep}, the metric $\text{RD}_{5}@1$, which quantifies the worst-case degradation, shows severe drops up to 0.9 (90\%) in some DocString perturbation cases for Incoder-1B (Java). Even more superficial perturbations, such as Syntax and Format, lead to noticeable performance degradation, albeit to a lesser extent.
Table~\ref{tab:ressummarypartial_ep} highlights that even under relaxed partial-pass criteria, the relative performance drop (RR$_{5}$@1) remains substantial, typically ranging from 5\% to 20\%, depending on the model and language. This suggests that perturbations, whether semantic or syntactic, can significantly impact model behavior and reliability. 

The performance drop of the LLMs is visualized in Figure \ref{fig:robust_drop}. The figure shows the robust drop across four perturbation types: DocString, Syntax, Format, and Function Name, for all 6 models. It is organized into four columns (one per perturbation type) and six rows (one per model). Each plot represents a specific model–perturbation combination (e.g., Magicoder-7B with DocString) across three programming languages. Each plot contains three bars, one for each language. Higher bars indicate a larger robustness drop for that model under the corresponding perturbation.

We observe a slight performance improvement in the partial-case scenario for Java under DocString and Function perturbations. This suggests that in some cases, modifying the function or the NL description can yield better results, hinting at the potential of prompt refinement as a strategy for improving code generation performance.

In general, all models demonstrate performance degradation under perturbations, reaffirming the brittleness of current LLM-based code generation systems. While models such as Magicoder-7B show slightly better robustness, the consistent drop across all settings underscores the need for robustness-aware evaluation and training. Moreover, the variation in impact across programming languages suggests that the models' sensitivity may be language-dependent, warranting further investigation into language-specific robustness strategies.

If we see the language-wise trends, the three languages: Java, C++, and JavaScript exhibit distinct behaviors under input perturbations. Java shows the highest robustness, with notable RP5@1 gains and minimal RD5@1 drops, particularly under Format and Syntax perturbations. Its structured syntax likely helps models infer intent despite surface-level changes.

C++ is the most fragile across the board. It suffers large RD5@1 drops, especially under worst-case perturbations, with only modest improvements under partial conditions. Its complex syntax and variability appear to challenge the models more than the other languages.

JavaScript falls in between. It benefits from Format and Syntax perturbations with decent RP5@1 gains, though it still shows some drops in DocString and Function changes. Its flexible syntax likely helps models recover better than in C++ but not as effectively as in Java.

Overall, Java is the most resilient, followed by JavaScript, while C++ is the most sensitive to input changes.

{These behavioral differences across languages can be partly attributed to their type system characteristics. Statistically typed languages such as Java and C++ enforce strict type checking at compile time, meaning that perturbations, particularly syntax-level changes, are more likely to produce compilation errors, directly amplifying the robustness drop. In contrast, JavaScript's dynamic typing offers more flexibility at runtime, allowing the model to recover from certain surface-level changes without triggering compilation failures. This aligns with our observation that JavaScript tends to show fewer compilation errors and more assertion errors under perturbation, while Java and C++ exhibit the opposite pattern.}

\textbf{Summary of Findings.} The following trends can be observed from the results of RQ1.

\begin{enumerate}
    \item \textbf{A Larger model size does not always guarantee better robustness or pass rate across all programming languages.} As shown in Table~\ref{tab:ressummary_ep} and Table~\ref{tab:ressummarypartial_ep}, in general, we can observe a robustness or pass rate improvement with the increase of model sizes. For instance, for CodeGen in the worst-case scenario, the $RD_5@1$ of JavaScript increases from 0.07 to 0.08 for 2B and 6B, respectively. On the contrary, for Java, we also observe that there is a drop of 0.02 of the $RD_5@1$. The results show that increasing the model size does not always guarantee better robustness and highlight the need for more comprehensive experiments on different programming languages.
    \item \textbf{CodeGen generally performs better than Incoder, while Magicoder and QwenCode further outperform both.} As shown in Tables~\ref{tab:ressummary_ep} and~\ref{tab:ressummarypartial_ep}, CodeGen achieves higher nominal pass@k than Incoder at comparable scales (e.g., 0.16 vs. 0.11 on Java). This is consistent with their training data differences: CodeGen is trained on a much larger token corpus, whereas Incoder, although more multilingual, is heavily skewed toward Python and JavaScript, limiting its cross-language generalization. On the other hand, Magicoder-7B and QwenCode outperform both CodeGen and Incoder, but with different behaviors under perturbations. Magicoder shows higher robustness, maintaining stronger pass@k when prompts are partially truncated or reformatted. This can be attributed to its OSS-INSTRUCT training, which leverages diverse open-source code snippets to reduce synthetic data bias and improve instruction grounding. QwenCode, trained on massive curated multilingual corpora, excels in nominal accuracy but exhibits slightly larger drops under perturbation, suggesting its strong generation ability is more sensitive to prompt variation.
    \item \textbf{Stronger test cases expose more failures across all languages.} From Tables~\ref{tab:ressummary_ep} and \ref{tab:ressummary_he}, it is evident that the stronger test cases from EvalPlus reveal substantially more failures in both nominal and perturbed settings, which aligns with prior findings~\citep{liu2023your}. The same trend holds for the partial cases, as shown in Tables~\ref{tab:ressummarypartial_ep} and \ref{tab:ressummarypartial_he}. However, the robustness drop remains roughly consistent regardless of the absolute performance levels.

\end{enumerate}

\subsection{RQ2: What are the major factors that contribute to the difference in the robustness of LLMs on different programming languages?}

\noindent\textbf{Motivation:} From the insights of RQ1, it becomes apparent that the performance of identical code generation models varies across different languages. This research inquiry delves deeper into uncovering the underlying reasons for such disparities. By doing so, we aim to find out the pivotal factors that contribute significantly to the fluctuations in robustness observed among various programming languages. Through this exploration, we seek to shed light on the influential components that drive the differences in how well models perform on perturbed data across different linguistic contexts.

\begin{figure}
    \centering
    \begin{subfigure}{\textwidth}
        \centering
        \framebox{\parbox{\textwidth}{\centering
        \includegraphics[width=.95\linewidth]{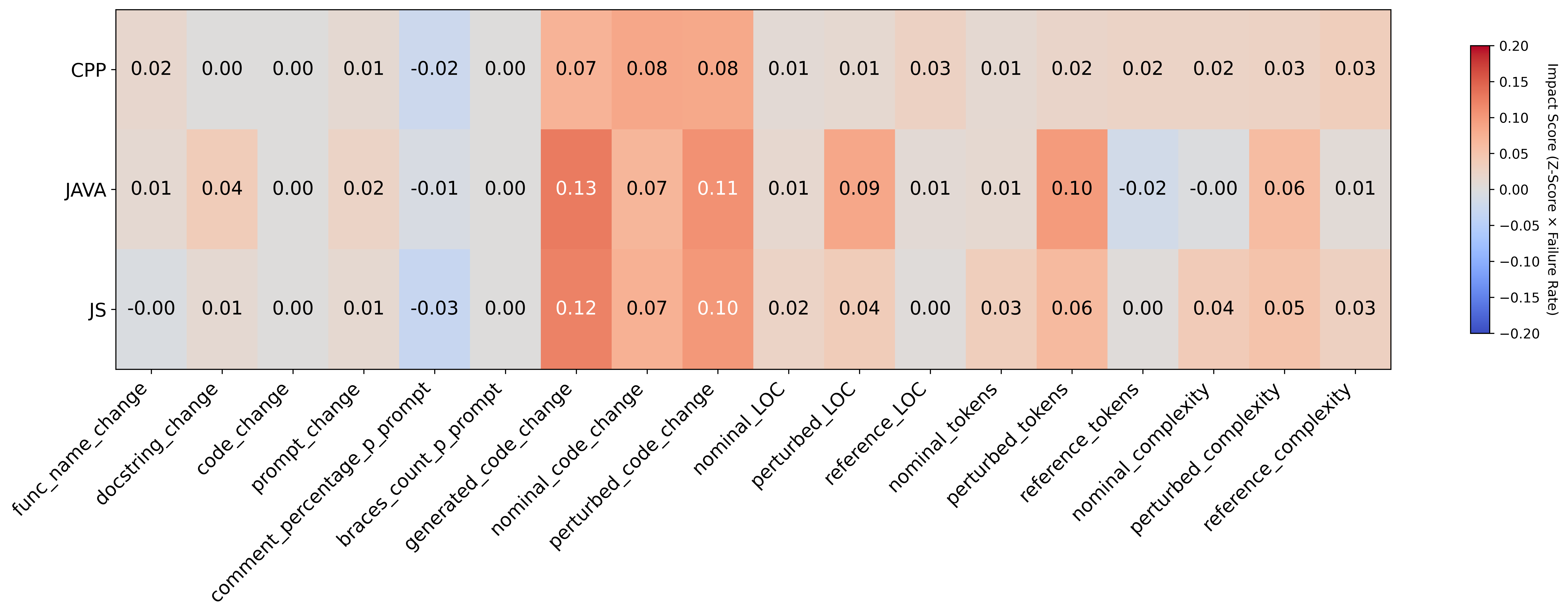}
        }}
        
        \caption{Function Name Perturbation}
        \label{fig:func_name_heatmap}
    \end{subfigure}
    
    \vfill 
    \begin{subfigure}{\textwidth}
        \centering
        
        \framebox{\parbox{\textwidth}{\centering
        \includegraphics[width=.95\linewidth]{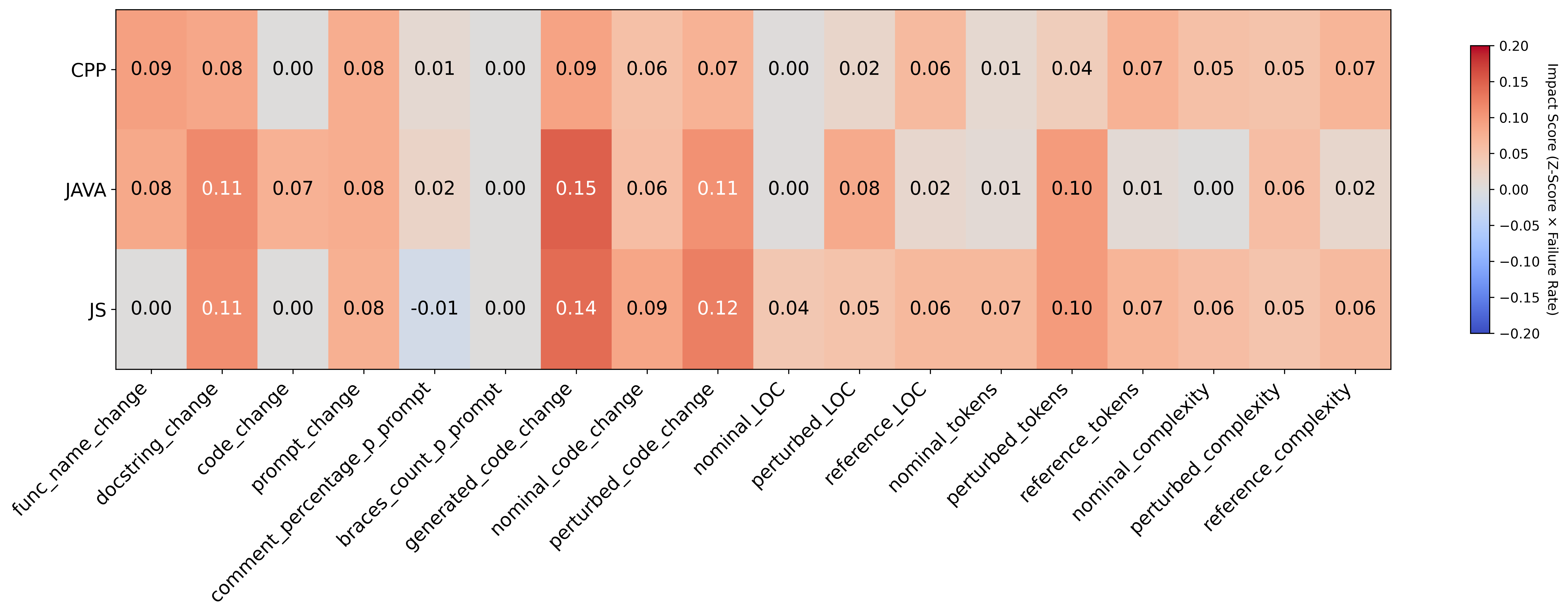}
        }}
        
        \caption{DocString Perturbation}
        \label{fig:nlaugmenter_heatmap}
    \end{subfigure}

    \vfill
    
    \begin{subfigure}{\textwidth}
        \centering
        
        \framebox{\parbox{\textwidth}{\centering
        \includegraphics[width=.95\linewidth]{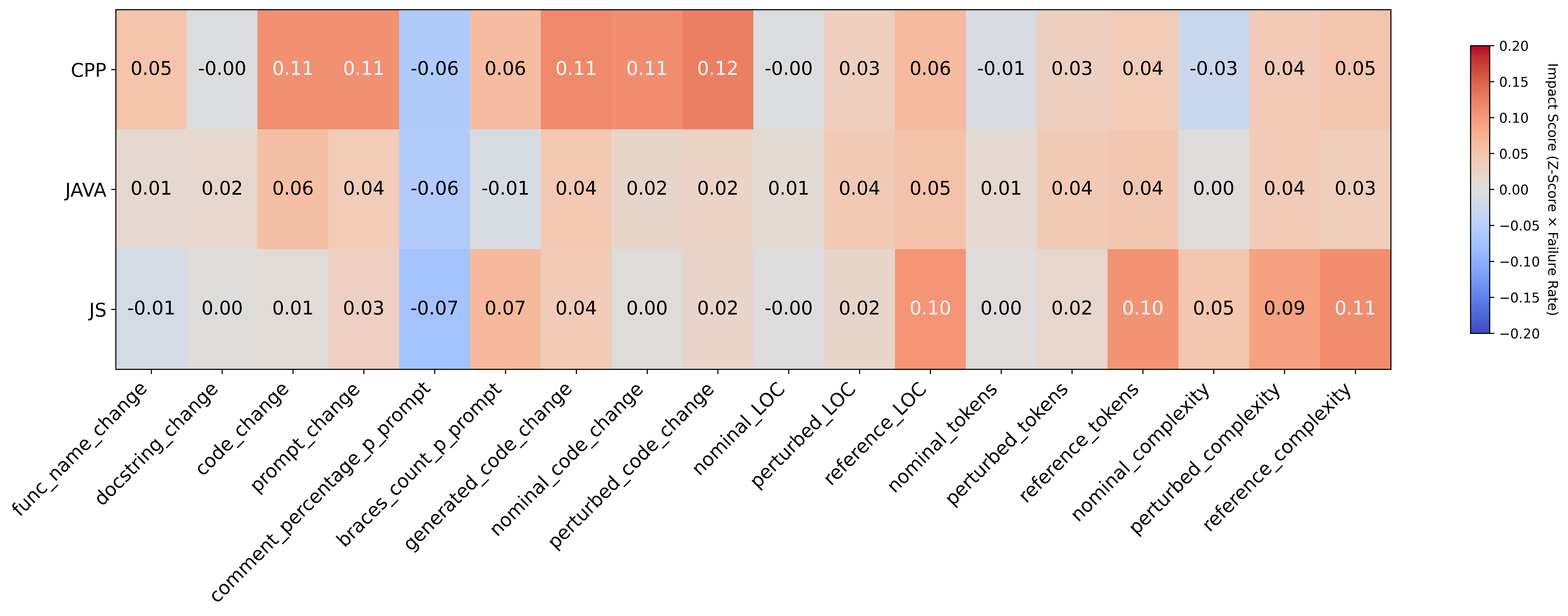}
        }}
        
        \caption{Syntax Perturbation}
        \label{fig:syntax_heatmap}
    \end{subfigure}

    \vfill
    \begin{subfigure}{\textwidth}
        \centering
        \framebox{\parbox{\textwidth}{\centering
        \includegraphics[width=.95\linewidth]{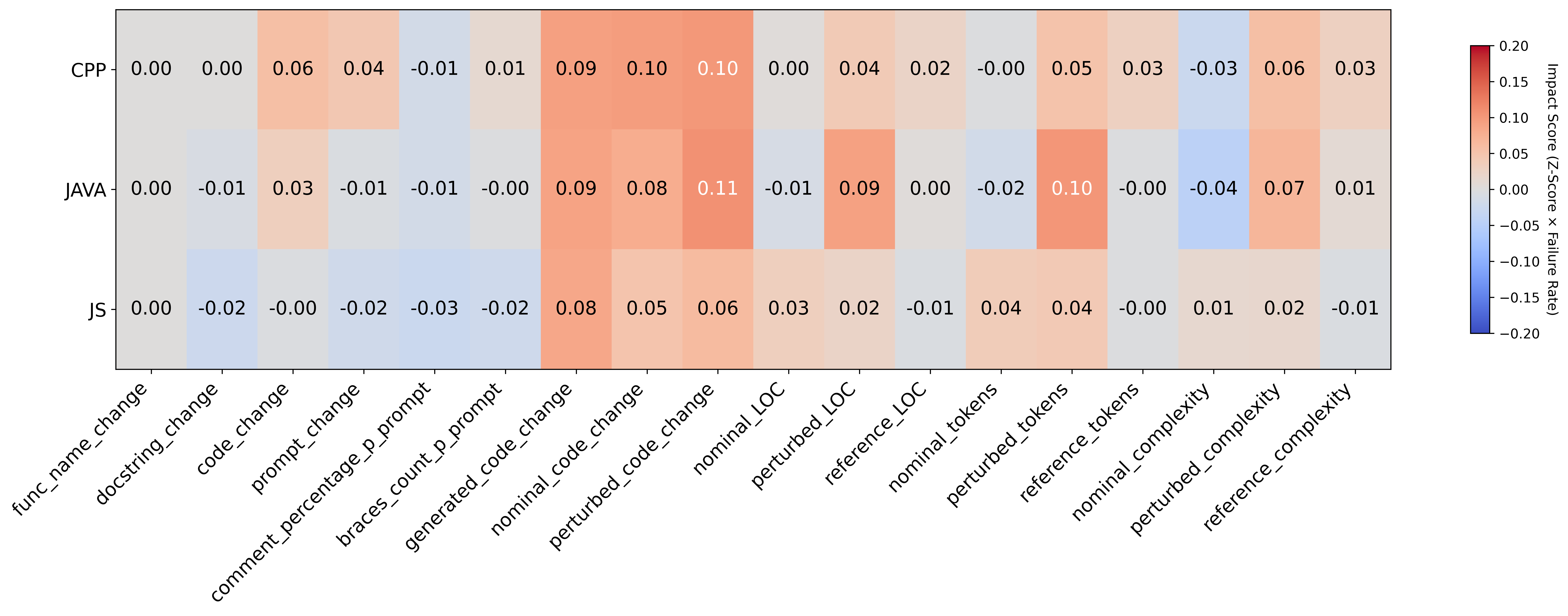}
        }}
        \caption{Format Perturbation}
        \label{fig:format_heatmap}
    \end{subfigure}

    \caption{The heatmaps show the impact score for each feature, calculated relative to the overall average.}
    \label{fig:combined_perturbation_heatmaps}
\end{figure}

\noindent \textbf{Approach:} To diagnose the underlying causes for the robustness disparities observed in RQ1, we employ a quantitative, feature-level quantitative analysis. Our approach is designed to identify which specific characteristics of the input prompts or the generated code are most strongly associated with failure in each language. This is achieved by calculating a weighted impact score for each feature and visualizing these scores in a comparative heatmap. The approach is executed in four sequential steps:

\noindent\textbf{Step 1. Data Selection and Preparation.}
We aim to identify features correlated with the Robust Drop, which represents the performance degradation when models generate code from perturbed versus nominal prompts. Therefore, the dataset for this analysis is scoped to include only samples that were passed on the nominal (unperturbed) prompt. This scoping decision isolates failures induced by a perturbation, distinguishing them from problems the model failed on initially. Then our goal is to identify the features most strongly associated with the samples that subsequently fail on the perturbed prompt.

\noindent\textbf{Feature Calculation.} For each sample, we compute a comprehensive set of 18 features. These features are grouped into two categories:

\textbf{Category 1. Input Perturbation Features:}
Quantify the magnitude of change in prompts.

\begin{enumerate}
    \item[] \textbf{Feature 1-4. Token Dissimilarity (\texttt{func\_name\_change}, \texttt{docstring\_change}, \texttt{code\_change}, \texttt{prompt\_change}):}
    These features compute the percentage of token-level change between two code snippets (a nominal component $P_n$ and a perturbed component $P_p$). We first tokenize both snippets. We then calculate a similarity ratio by finding the total number of matching tokens within all matching subsequences, and relate this to the total number of tokens in both snippets, as defined in Equation \ref{eq:dissim}.
    \begin{equation}
      \label{eq:dissim}
      \text{Dissimilarity \%}(P_n, P_p) = \left( 1 - \frac{2 \cdot |\text{Matches}(P_n, P_p)|}{|P_n| + |P_p|} \right) \times 100
    \end{equation}
    Here, $P_n$ and $P_p$ are the token sequences, and $|\text{Matches}|$ is the count of matching tokens from the sequence matcher.

    \item[] \textbf{Feature 5. Comment Percentage (\texttt{comment\_percentage\_p\_prompt}):}
    Measures the ratio of tokens inside comments to the total number of tokens in the perturbed prompt.
    \begin{equation}
      \text{Comment \%} = \frac{|\text{Tokens in Comments}_{p}|}{|\text{Total Tokens in Prompt}_{p}|}
    \end{equation}

    \item[] \textbf{Feature 6. Scope Delimiter Count (\texttt{scope\_delimiter\_count}):}
    Measures the quantity of syntax elements used to \textbf{open} code blocks and define scope (e.g., counting the opening brace \texttt{\{} in C++/Java). This metric serves as a simple proxy for the code's structural complexity, which may correlate to the model's fragility.
\end{enumerate}

\noindent \textbf{2. Output and Problem Features:}
Quantify the characteristics of the canonical solutions and models generated output, which represents the complexity of the problem.

\begin{enumerate}
    \item[] \textbf{Feature 7-9. Output Dissimilarity (\texttt{generated\_code\_change}, \texttt{nominal\_ code\_change}, \texttt{perturbed\_code\_change}):}
    Measures the token dissimilarity (using Equation \ref{eq:dissim}) between different versions of code to quantify changes in output structure and solution correctness.
    \begin{itemize}
        \item generated\_code\_change: Dissimilarity between the model-generated code from nominal\_prompt and perturbed\_prompt. A high value indicates that the perturbation caused the model to generate a structurally different solution.
        \item nominal\_code\_change: Dissimilarity between the human-written reference\_code and the LLM's generated code from nominal\_prompt. This quantifies the changes in the nominal code generated by the LLM.
        \item perturbed\_code\_change: Dissimilarity between the human-written reference\_code and the generated code by the LLM from the perturbed prompt. This quantifies the changes in the perturbed code generated by the LLM.
    \end{itemize}
    
    \item[] \textbf{Feature 10-12. Lines of Code (LOC) (\texttt{nominal\_LOC}, \texttt{perturbed\_LOC}, \texttt{reference\_LOC}):}
    A count of the total lines of code for the nominal generated code, perturbed generated code, and reference solution, respectively.
    
    \item[] \textbf{Feature 13-15. Token Count (\texttt{nominal\_tokens}, \texttt{perturbed\_tokens}, \texttt{reference\_tokens}):}
    A count of the total tokens for the nominal generated code, perturbed generated code, and reference solution, respectively.
    
    \item[] \textbf{Feature 16-18. Cyclomatic Complexity (\texttt{nominal\_complexity}, \texttt{pertur- bed\_complexity}, \texttt{reference\_complexity}):}
    Measures the structural complexity of the code (e.g., McCabe's complexity) by counting linearly independent paths. This is calculated for the nominal generated code, perturbed generated code, and reference solution.
\end{enumerate}

\noindent \textbf{Statistical Calculation of the Impact Score}
To create a fair, language-to-language comparison, we calculate an Impact Score for every feature. This score quantifies the relationship between each feature and the robustness failures, addressing a key challenge: a feature that is highly associated with failure in a language that less-frequently fails is less impactful overall than a feature associated with failure in a language that frequently fails. The Impact Score ($I$) for a given feature ($F$) and language ($L$) is the product of the feature's Z-Score ($Z$) and the language's Failure Rate ($\text{FR}$).

The calculation follows these steps for each language $L$:

\begin{enumerate}
    \item \textbf{Calculate Failure Rate ($\text{FR}_L$):} The proportion of samples that failed on the perturbed prompt.
    \begin{equation}
      \text{FR}_L = \frac{|\text{Failed Samples}_L|}{|\text{Total Samples}_L|}
    \end{equation}

    \item \textbf{Establish Language-Specific Statistical Norms:} For each feature $F$, we calculate the mean ($\mu_{F,L}$) and standard deviation ($\sigma_{F,L}$) using all samples (passed and failed) for that language $L$.

    \item \textbf{Profile Failed Runs:} We calculate the mean of the failed samples ($\mu_{F,L}^{\text{fail}}$) for that feature $F$ and language $L$.

    \item \textbf{Calculate Z-Score ($Z_{F,L}$):} We standardize the failed-run average against its language-specific statistical norms. This use of Z-score normalization is a standard method for data preprocessing. It is necessary here to transform features with heterogeneous units and scales (e.g., \texttt{nominal\_LOC} versus \texttt{func\_name\_change}) onto a common, dimensionless scale. This allows us to compare the relative unusualness of each feature's value, as the resulting Z-score represents the deviation from the language-specific mean in units of standard deviations. A high positive Z-Score means failed runs are strongly associated with high values of that feature.
            
    \begin{equation}
        \label{eq:z}
      Z_{F,L} = \frac{\mu_{F,L}^{\text{fail}} - \mu_{F,L}}{\sigma_{F,L}}
    \end{equation}

    \item \textbf{Calculate Impact Score ($I_{F,L}$):} We weigh the Z-Score by the language's overall Failure Rate.
    \begin{equation}
    \label{eq:impact}
      I_{F,L} = Z_{F,L} \times \text{FR}_L
    \end{equation}
\end{enumerate}

\noindent \textbf{Visualization.} The final output is a matrix of Impact Scores, with languages as rows and features as columns. This matrix is visualized as a 1D-stacked heatmap for each language. A diverging colormap (e.g., blue-white-red) is centered at zero and clipped to a fixed scale (e.g., -0.2 to +0.2) to ensure all heatmaps are directly comparable. This final visualization directly highlights which features are the most significant drivers of robustness failures for each language. \ref{fig:combined_perturbation_heatmaps}

\noindent \textbf{Result.}
Figure~\ref{fig:combined_perturbation_heatmaps} illustrates the correlation between features and robustness drop across programming languages. In each plot of the figure, the y-axis represents the programming languages, while the x-axis represents the evaluated features. Each cell displays an impact score, calculated using Equations \ref{eq:z} and \ref{eq:impact}. A higher positive value indicates a stronger positive correlation between the feature and the language, whereas a negative value suggests a negative correlation. A value of zero indicates no relationship.
Each subplot corresponds to one perturbation type (\texttt{Function Name}, \texttt{DocString}, \texttt{Syntax}, and \texttt{Format}). Rows represent programming languages (C++, Java, and JavaScript), while columns correspond to the features. The color and magnitude of each cell indicate the \textit{Impact Score}, which reflects the strength and direction of correlation between a given feature and the robustness drop for that language.

Overall, the heatmaps reveal language-specific patterns of sensitivity under different perturbations. JavaScript consistently shows the highest positive correlations across multiple features, aligning with its higher robustness drop reported in RQ1. In contrast, Java and C++ demonstrate more stable patterns, with lower overall correlation magnitudes and fewer dominant features.

Across the four perturbation types, the overall trend shows that Java tends to exhibit the strongest feature impact relationships under function name and docString perturbations, where several features (e.g., generated\_code\_change, perturbed\_code\_change) consistently reach impact scores around 0.10-0.15. In contrast, this pattern does not hold under syntax perturbations, where Java’s impact values drop noticeably, and most features fall close to zero, making Java the least sensitive language in this specific condition. C++ shows moderate but steady correlations across most perturbations, often around 0.05-0.11, without strong spikes, indicating a stable but weaker relationship between features and robustness drop. JavaScript generally exhibits low or near-zero correlations across all perturbation types, occasionally showing small positive effects (e.g., perturbed\_tokens, reference\_tokens) but overall demonstrating the flattest patterns. Taken together, the results indicate that Java is feature-sensitive under perturbations in function name and docString, while syntax and formatting perturbations suppress its feature–impact relationships, making correlations there more diffuse, whereas C++ remains moderate and steady, and JavaScript remains consistently low across all cases.

\noindent \textbf{Summary of Findings.} Across perturbations, several consistent trends emerge:
\begin{enumerate}
\item Changes in the DocString have a significant impact on the robustness drop for Java and JavaScript, and a moderate impact for C++. From the DocString perturbation, we observe that even small modifications to the DocString, caused indirectly by other name changes, can also lead to notable performance variations.
\item For the generated code from perturbed prompts, we observe substantial deviations compared to the code generated from nominal prompts and the original human-written solutions.
\item For function name and DocString perturbations, many features strongly affect robustness in Java and JavaScript but have a smaller influence on C++. In contrast, C++ is more affected by syntax-level perturbations than the other languages.
\end{enumerate}

\begin{figure}
\centering
\includegraphics[width=.5\textwidth]{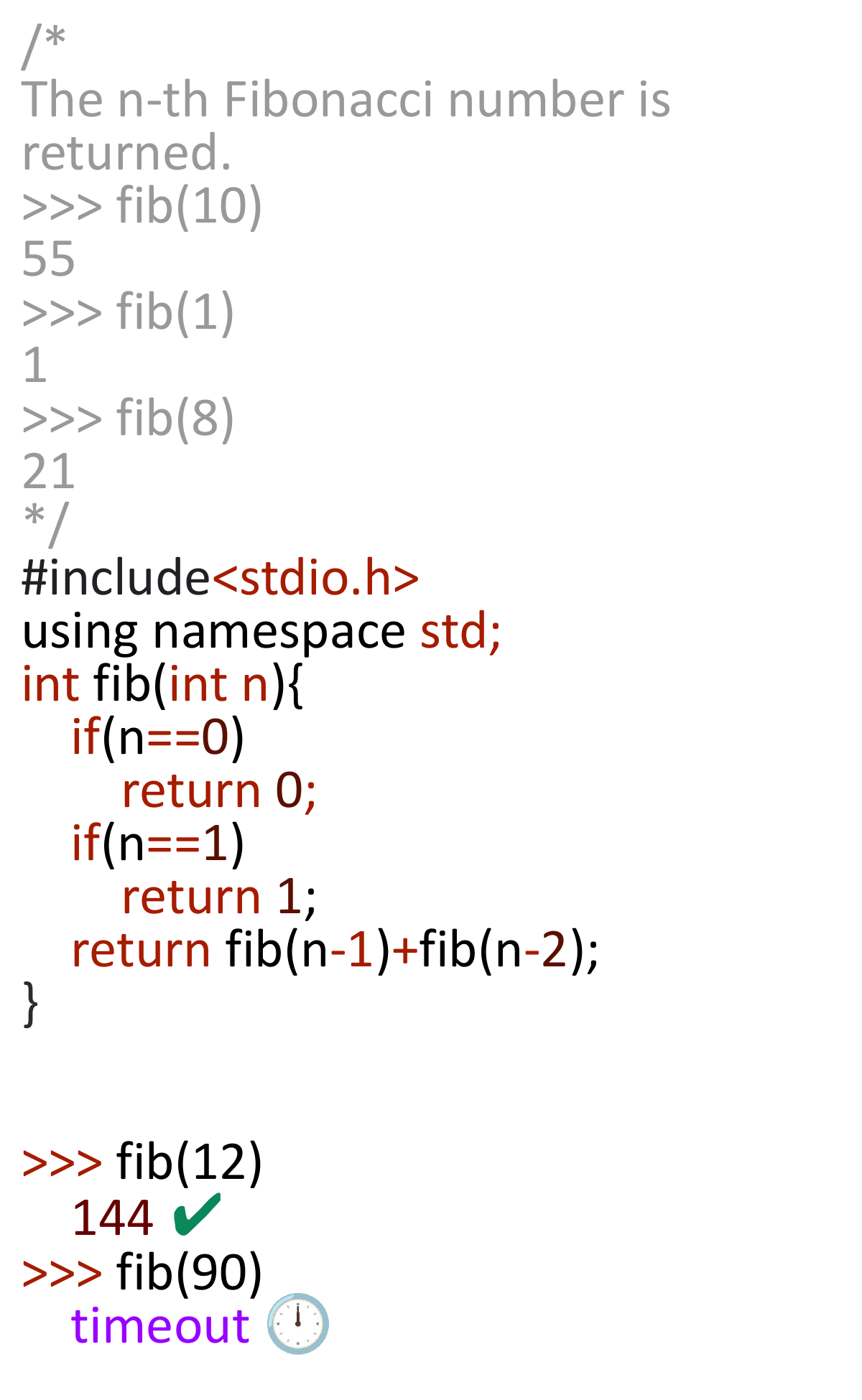}
\caption{A generated code by CodeGen-6B-Multi model which passes the test cases of HumanEval-X but fails against the evalplus translated test case}
\label{fig:humanevalnotenoughrobust}
\end{figure}

\begin{table}
\centering
\caption{Performance summary of perturbations against code generation models using EvalPlus test cases and worst case passes (5 out of 5). Up arrows indicate the extent of performance improvement, while down arrows indicate the extent of performance drop.}
\label{tab:ressummary_ep}
\resizebox{\textwidth}{!}{\large \begin{tabular}{|p{3cm}|p{2cm}|p{1cm}|p{1cm}|p{1cm}|p{1cm}|p{1cm}|p{1cm}|p{1cm}|p{1cm}|p{1cm}|p{1cm}|p{1cm}|p{1cm}|p{1cm}|p{1cm}|p{1cm}|p{1cm}|p{1cm}|p{1cm}|}
    \hline
    HumanEval-X & Model & \multicolumn{3}{|p{4cm}|}{\centering Incoder-1B} & \multicolumn{3}{|p{4cm}|}{\centering Incoder-6B} & \multicolumn{3}{|p{4cm}|}{\centering CodeGen-2B-multi} & \multicolumn{3}{|p{4cm}|}{\centering CodeGen-6B-multi} & \multicolumn{3}{|p{4cm}|}{\centering Magicoder7B} & \multicolumn{3}{|p{4cm}|}{\centering QwenCode} \\
    \hline
    Perturbation & Metric & Java & CPP & JS & Java & CPP & JS & Java & CPP & JS & Java & CPP & JS & Java & CPP & JS & Java & CPP & JS \\
    \hline
    Nominal & RP{\footnotesize5}@1 & 0.02 & 0.08 & 0.07 & 0.11 & 0.1 & 0.12 & 0.14 & 0.12 & 0.11 & 0.15 & 0.13 & 0.14 & 0.63 & 0.55 & 0.51 & 0.55 & 0.57 & 0.6 \\
    \hline
    Partial & RP{\footnotesize5}@1 & 0.32 & 0.09 & 0.37 & 0.45 & 0.14 & 0.44 & 0.56 & 0.13 & 0.39 & 0.55 & 0.2 & 0.44 & 0.86 & 0.61 & 0.74 & 0.77 & 0.63 & 0.68 \\
    \hline
    \multirow{4}{*}{\centering DocString} & Nominal$\uparrow$ & 0.02 & 0.08 & 0.07 & 0.11 & 0.1 & 0.12 & 0.14 & 0.12 & 0.11 & 0.15 & 0.13 & 0.14 & 0.63 & 0.55 & 0.51 & 0.55 & 0.57 & 0.6 \\
    & RP{\footnotesize5}@1$\uparrow$  & 0.0 & 0.02 & 0.03 & 0.02 & 0.05 & 0.05 & 0.06 & 0.05 & 0.04 & 0.05 & 0.06 & 0.05 & 0.42 & 0.37 & 0.3 & 0.27 & 0.38 & 0.35 \\
    & RD{\footnotesize5}@1$\downarrow$  & 0.88 & 0.71 & 0.6 & 0.78 & 0.48 & 0.6 & 0.56 & 0.57 & 0.64 & 0.66 & 0.5 & 0.63 & 0.33 & 0.32 & 0.41 & 0.52 & 0.34 & 0.42 \\
    & RR{\footnotesize5}@1$\downarrow$  & 0.02 & 0.07 & 0.05 & 0.1 & 0.05 & 0.07 & 0.08 & 0.07 & 0.07 & 0.1 & 0.07 & 0.09 & 0.24 & 0.19 & 0.24 & 0.33 & 0.21 & 0.29 \\
    \hline
    \multirow{4}{*}{\centering Syntax} & Nominal$\uparrow$ & 0.32 & 0.09 & 0.37 & 0.45 & 0.14 & 0.44 & 0.56 & 0.13 & 0.39 & 0.55 & 0.2 & 0.44 & 0.86 & 0.61 & 0.74 & 0.77 & 0.63 & 0.68 \\
    & RP{\footnotesize5}@1$\uparrow$  & 0.22 & 0.02 & 0.24 & 0.26 & 0.04 & 0.27 & 0.32 & 0.05 & 0.21 & 0.28 & 0.07 & 0.24 & 0.57 & 0.45 & 0.53 & 0.55 & 0.39 & 0.39 \\
    & RD{\footnotesize5}@1$\downarrow$  & 0.31 & 0.8 & 0.36 & 0.42 & 0.68 & 0.38 & 0.43 & 0.62 & 0.46 & 0.49 & 0.64 & 0.45 & 0.34 & 0.26 & 0.29 & 0.29 & 0.39 & 0.43 \\
    & RR{\footnotesize5}@1$\downarrow$  & 0.22 & 0.08 & 0.15 & 0.24 & 0.12 & 0.17 & 0.25 & 0.1 & 0.18 & 0.27 & 0.13 & 0.23 & 0.31 & 0.21 & 0.26 & 0.42 & 0.31 & 0.35 \\
    \hline
    \multirow{4}{*}{\centering Format} & Nominal$\uparrow$ & 0.32 & 0.09 & 0.37 & 0.45 & 0.14 & 0.44 & 0.56 & 0.13 & 0.39 & 0.55 & 0.2 & 0.44 & 0.86 & 0.61 & 0.74 & 0.77 & 0.63 & 0.68 \\
    & RP{\footnotesize5}@1$\uparrow$  & 0.15 & 0.02 & 0.25 & 0.29 & 0.06 & 0.3 & 0.37 & 0.08 & 0.28 & 0.34 & 0.1 & 0.28 & 0.79 & 0.51 & 0.67 & 0.69 & 0.51 & 0.55 \\
    & RD{\footnotesize5}@1$\downarrow$  & 0.53 & 0.73 & 0.33 & 0.36 & 0.57 & 0.31 & 0.35 & 0.38 & 0.28 & 0.37 & 0.46 & 0.37 & 0.09 & 0.17 & 0.1 & 0.1 & 0.19 & 0.19 \\
    & RR{\footnotesize5}@1$\downarrow$  & 0.19 & 0.07 & 0.14 & 0.18 & 0.09 & 0.15 & 0.2 & 0.06 & 0.15 & 0.21 & 0.1 & 0.2 & 0.09 & 0.14 & 0.12 & 0.21 & 0.15 & 0.23 \\
    \hline
    \multirow{4}{*}{\centering Function} & Nominal$\uparrow$ & 0.02 & 0.08 & 0.07 & 0.11 & 0.1 & 0.12 & 0.14 & 0.12 & 0.11 & 0.15 & 0.13 & 0.14 & 0.63 & 0.55 & 0.51 & 0.55 & 0.57 & 0.6 \\
    & RP{\footnotesize5}@1$\uparrow$  & 0.0 & 0.03 & 0.05 & 0.04 & 0.06 & 0.06 & 0.08 & 0.06 & 0.06 & 0.05 & 0.07 & 0.08 & 0.51 & 0.44 & 0.38 & 0.3 & 0.46 & 0.4 \\
    & RD{\footnotesize5}@1$\downarrow$  & 0.92 & 0.66 & 0.29 & 0.64 & 0.41 & 0.46 & 0.4 & 0.53 & 0.45 & 0.64 & 0.41 & 0.44 & 0.19 & 0.19 & 0.24 & 0.45 & 0.2 & 0.33 \\
    & RR{\footnotesize5}@1$\downarrow$  & 0.02 & 0.07 & 0.03 & 0.08 & 0.05 & 0.06 & 0.06 & 0.07 & 0.06 & 0.1 & 0.06 & 0.07 & 0.13 & 0.13 & 0.17 & 0.29 & 0.12 & 0.22 \\
    \hline
\end{tabular}}
\end{table} 

\begin{table*}
\centering
\caption{Performance summary of perturbations against code generation models using HumanEval-X test cases and worst case passes (5 out of 5).}
\label{tab:ressummary_he}
\resizebox{\textwidth}{!}{
\begin{tabular}{|p{3cm}|p{2cm}|p{1cm}|p{1cm}|p{1cm}|p{1cm}|p{1cm}|p{1cm}|p{1cm}|p{1cm}|p{1cm}|p{1cm}|p{1cm}|p{1cm}|p{1cm}|p{1cm}|p{1cm}|p{1cm}|p{1cm}|p{1cm}|}
    \hline
    HumanEval-X & Model & \multicolumn{3}{|p{4cm}|}{\centering Incoder-1B} & \multicolumn{3}{|p{4cm}|}{\centering Incoder-6B} & \multicolumn{3}{|p{4cm}|}{\centering CodeGen-2B-Multi} & \multicolumn{3}{|p{4cm}|}{\centering CodeGen-6B-Multi} & \multicolumn{3}{|p{4cm}|}{\centering Magicoder-7B} & \multicolumn{3}{|p{4cm}|}{\centering QwenCode} \\
    \hline
    Perturbation & Metric & Java & CPP & JS & Java & CPP & JS & Java & CPP & JS & Java & CPP & JS & Java & CPP & JS & Java & CPP & JS \\
    \hline
    Nominal & RP{\footnotesize5}@1 & 0.02 & 0.1 & 0.08 & 0.11 & 0.12 & 0.13 & 0.13 & 0.12 & 0.13 & 0.16 & 0.14 & 0.16 & 0.65 & 0.57 & 0.53 & 0.57 & 0.59 & 0.62 \\
    \hline
    Partial & RP{\footnotesize5}@1 & 0.33 & 0.1 & 0.38 & 0.45 & 0.17 & 0.44 & 0.56 & 0.13 & 0.39 & 0.54 & 0.22 & 0.44 & 0.88 & 0.63 & 0.76 & 0.79 & 0.65 & 0.70 \\
    \hline
    \multirow{4}{*}{\centering DocString} 
    & Nominal$\uparrow$ & 0.02 & 0.1 & 0.08 & 0.11 & 0.12 & 0.13 & 0.13 & 0.12 & 0.13 & 0.16 & 0.14 & 0.16 & 0.65 & 0.57 & 0.53 & 0.57 & 0.59 & 0.62 \\
	& RP{\footnotesize5}@1$\uparrow$  & 0.0 & 0.04 & 0.03 & 0.04 & 0.06 & 0.05 & 0.07 & 0.07 & 0.05 & 0.05 & 0.08 & 0.06 & 0.45 & 0.40 & 0.33 & 0.29 & 0.40 & 0.37 \\
	& RD{\footnotesize5}@1$\downarrow$  & 0.9 & 0.57 & 0.63 & 0.68 & 0.49 & 0.58 & 0.45 & 0.43 & 0.59 & 0.68 & 0.44 & 0.65 & 0.30 & 0.29 & 0.37 & 0.49 & 0.31 & 0.39 \\
	& RR{\footnotesize5}@1$\downarrow$  & 0.02 & 0.08 & 0.05 & 0.09 & 0.06 & 0.08 & 0.06 & 0.06 & 0.08 & 0.11 & 0.07 & 0.11 & 0.21 & 0.17 & 0.21 & 0.30 & 0.19 & 0.26 \\
	\hline
	\multirow{4}{*}{\centering Syntax} 
    & Nominal$\uparrow$ & 0.33 & 0.1 & 0.38 & 0.45 & 0.17 & 0.44 & 0.56 & 0.13 & 0.39 & 0.54 & 0.22 & 0.44 & 0.88 & 0.63 & 0.76 & 0.79 & 0.65 & 0.70 \\
	& RP{\footnotesize5}@1$\uparrow$  & 0.23 & 0.03 & 0.23 & 0.27 & 0.05 & 0.27 & 0.33 & 0.07 & 0.2 & 0.29 & 0.09 & 0.23 & 0.60 & 0.48 & 0.56 & 0.58 & 0.42 & 0.42 \\
	& RD{\footnotesize5}@1$\downarrow$  & 0.3 & 0.73 & 0.39 & 0.38 & 0.68 & 0.39 & 0.41 & 0.45 & 0.49 & 0.46 & 0.58 & 0.47 & 0.31 & 0.23 & 0.26 & 0.26 & 0.35 & 0.40 \\
	& RR{\footnotesize5}@1$\downarrow$  & 0.21 & 0.08 & 0.16 & 0.24 & 0.14 & 0.17 & 0.26 & 0.08 & 0.2 & 0.25 & 0.14 & 0.24 & 0.28 & 0.19 & 0.23 & 0.39 & 0.28 & 0.32 \\
	\hline
	\multirow{4}{*}{\centering Format} 
    & Nominal$\uparrow$ & 0.33 & 0.1 & 0.38 & 0.45 & 0.17 & 0.44 & 0.56 & 0.13 & 0.39 & 0.54 & 0.22 & 0.44 & 0.88 & 0.63 & 0.76 & 0.79 & 0.65 & 0.70 \\
	& RP{\footnotesize5}@1$\uparrow$  & 0.2 & 0.04 & 0.26 & 0.28 & 0.07 & 0.3 & 0.45 & 0.1 & 0.27 & 0.42 & 0.12 & 0.27 & 0.81 & 0.54 & 0.70 & 0.72 & 0.54 & 0.58 \\
	& RD{\footnotesize5}@1$\downarrow$  & 0.38 & 0.64 & 0.33 & 0.36 & 0.57 & 0.32 & 0.2 & 0.22 & 0.31 & 0.21 & 0.44 & 0.39 & 0.07 & 0.15 & 0.08 & 0.08 & 0.16 & 0.16 \\
	& RR{\footnotesize5}@1$\downarrow$  & 0.16 & 0.07 & 0.14 & 0.19 & 0.11 & 0.15 & 0.13 & 0.04 & 0.16 & 0.13 & 0.1 & 0.21 & 0.07 & 0.12 & 0.10 & 0.18 & 0.12 & 0.20 \\
	\hline
	\multirow{4}{*}{\centering Function} 
    & Nominal$\uparrow$ & 0.02 & 0.1 & 0.08 & 0.11 & 0.12 & 0.13 & 0.13 & 0.12 & 0.13 & 0.16 & 0.14 & 0.16 & 0.65 & 0.57 & 0.53 & 0.57 & 0.59 & 0.62 \\
	& RP{\footnotesize5}@1$\uparrow$  & 0.0 & 0.04 & 0.04 & 0.04 & 0.07 & 0.07 & 0.08 & 0.07 & 0.07 & 0.05 & 0.08 & 0.07 & 0.54 & 0.47 & 0.41 & 0.33 & 0.49 & 0.43 \\
	& RD{\footnotesize5}@1$\downarrow$  & 0.88 & 0.57 & 0.49 & 0.64 & 0.46 & 0.47 & 0.4 & 0.44 & 0.48 & 0.69 & 0.4 & 0.56 & 0.16 & 0.17 & 0.21 & 0.42 & 0.17 & 0.30 \\
	& RR{\footnotesize5}@1$\downarrow$  & 0.02 & 0.07 & 0.05 & 0.08 & 0.06 & 0.07 & 0.05 & 0.06 & 0.07 & 0.11 & 0.06 & 0.11 & 0.10 & 0.11 & 0.14 & 0.26 & 0.10 & 0.19 \\
	\hline

\end{tabular}}
\end{table*}

\begin{table}
\centering
\caption{Performance summary of perturbations using EvalPlus test cases and partial cases (3 out of 5)} 
\label{tab:ressummarypartial_ep}
\resizebox{\textwidth}{!}{\large \begin{tabular}{|p{3cm}|p{2cm}|p{1cm}|p{1cm}|p{1cm}|p{1cm}|p{1cm}|p{1cm}|p{1cm}|p{1cm}|p{1cm}|p{1cm}|p{1cm}|p{1cm}|p{1cm}|p{1cm}|p{1cm}|p{1cm}|p{1cm}|p{1cm}|}
    \hline
    HumanEval-X & Model & \multicolumn{3}{|p{4cm}|}{\centering Incoder-1B} & \multicolumn{3}{|p{4cm}|}{\centering Incoder-6B} & \multicolumn{3}{|p{4cm}|}{\centering CodeGen-2B-multi} & \multicolumn{3}{|p{4cm}|}{\centering CodeGen-6B-multi} & \multicolumn{3}{|p{4cm}|}{\centering Magicoder7B} & \multicolumn{3}{|p{4cm}|}{\centering QwenCode} \\
    \hline
    Perturbation & Metric & Java & CPP & JS & Java & CPP & JS & Java & CPP & JS & Java & CPP & JS & Java & CPP & JS & Java & CPP & JS \\
    \hline
    Nominal & RP{\footnotesize5}@1 & 0.02 & 0.08 & 0.07 & 0.11 & 0.1 & 0.12 & 0.14 & 0.12 & 0.11 & 0.15 & 0.13 & 0.14 & 0.63 & 0.55 & 0.51 & 0.55 & 0.57 & 0.6 \\
    \hline
    Partial & RP{\footnotesize5}@1 & 0.32 & 0.09 & 0.37 & 0.45 & 0.14 & 0.44 & 0.56 & 0.13 & 0.39 & 0.55 & 0.2 & 0.44 & 0.86 & 0.61 & 0.74 & 0.77 & 0.63 & 0.68 \\
    \hline
    \multirow{4}{*}{\centering DocString} & Nominal$\uparrow$ & 0.02 & 0.08 & 0.07 & 0.11 & 0.1 & 0.12 & 0.14 & 0.12 & 0.11 & 0.15 & 0.13 & 0.14 & 0.63 & 0.55 & 0.51 & 0.55 & 0.57 & 0.6 \\
    & RP{\footnotesize5}@1$\uparrow$  & 0.02 & 0.05 & 0.06 & 0.06 & 0.1 & 0.09 & 0.11 & 0.1 & 0.08 & 0.11 & 0.11 & 0.09 & 0.56 & 0.49 & 0.48 & 0.47 & 0.53 & 0.53 \\
    & RD{\footnotesize5}@1$\downarrow$  & 0.15 & 0.34 & 0.17 & 0.41 & 0.01 & 0.22 & 0.2 & 0.14 & 0.3 & 0.25 & 0.16 & 0.33 & 0.11 & 0.11 & 0.06 & 0.14 & 0.07 & 0.12 \\
    & RR{\footnotesize5}@1$\downarrow$  & 0.02 & 0.06 & 0.05 & 0.07 & 0.02 & 0.05 & 0.05 & 0.05 & 0.06 & 0.06 & 0.04 & 0.07 & 0.14 & 0.12 & 0.15 & 0.24 & 0.14 & 0.17 \\
    \hline
    \multirow{4}{*}{\centering Syntax} & Nominal$\uparrow$ & 0.32 & 0.09 & 0.37 & 0.45 & 0.14 & 0.44 & 0.56 & 0.13 & 0.39 & 0.55 & 0.2 & 0.44 & 0.86 & 0.61 & 0.74 & 0.77 & 0.63 & 0.68 \\
    & RP{\footnotesize5}@1$\uparrow$  & 0.27 & 0.05 & 0.29 & 0.35 & 0.1 & 0.32 & 0.38 & 0.09 & 0.29 & 0.39 & 0.12 & 0.33 & 0.62 & 0.53 & 0.6 & 0.64 & 0.5 & 0.55 \\
    & RD{\footnotesize5}@1$\downarrow$  & 0.17 & 0.4 & 0.22 & 0.22 & 0.29 & 0.28 & 0.32 & 0.31 & 0.26 & 0.3 & 0.4 & 0.24 & 0.28 & 0.13 & 0.19 & 0.16 & 0.21 & 0.19 \\
    & RR{\footnotesize5}@1$\downarrow$  & 0.22 & 0.08 & 0.11 & 0.19 & 0.14 & 0.14 & 0.21 & 0.1 & 0.15 & 0.19 & 0.12 & 0.17 & 0.27 & 0.18 & 0.22 & 0.38 & 0.25 & 0.27 \\
    \hline
    \multirow{4}{*}{\centering Format} & Nominal$\uparrow$ & 0.32 & 0.09 & 0.37 & 0.45 & 0.14 & 0.44 & 0.56 & 0.13 & 0.39 & 0.55 & 0.2 & 0.44 & 0.86 & 0.61 & 0.74 & 0.77 & 0.63 & 0.68 \\
    & RP{\footnotesize5}@1$\uparrow$  & 0.26 & 0.06 & 0.31 & 0.39 & 0.11 & 0.42 & 0.43 & 0.12 & 0.37 & 0.45 & 0.17 & 0.36 & 0.83 & 0.62 & 0.73 & 0.81 & 0.63 & 0.68 \\
    & RD{\footnotesize5}@1$\downarrow$  & 0.19 & 0.31 & 0.16 & 0.13 & 0.18 & 0.05 & 0.23 & 0.07 & 0.05 & 0.17 & 0.13 & 0.18 & 0.03 & -0.02 & 0.01 & -0.06 & 0.0 & 0.01 \\
    & RR{\footnotesize5}@1$\downarrow$  & 0.15 & 0.05 & 0.09 & 0.13 & 0.08 & 0.06 & 0.16 & 0.06 & 0.1 & 0.13 & 0.06 & 0.14 & 0.07 & 0.09 & 0.09 & 0.17 & 0.09 & 0.18 \\
    \hline
    \multirow{4}{*}{\centering Function} & Nominal$\uparrow$ & 0.02 & 0.08 & 0.07 & 0.11 & 0.1 & 0.12 & 0.14 & 0.12 & 0.11 & 0.15 & 0.13 & 0.14 & 0.63 & 0.55 & 0.51 & 0.55 & 0.57 & 0.6 \\
    & RP{\footnotesize5}@1$\uparrow$  & 0.02 & 0.07 & 0.08 & 0.08 & 0.09 & 0.11 & 0.13 & 0.1 & 0.11 & 0.12 & 0.11 & 0.12 & 0.62 & 0.54 & 0.52 & 0.51 & 0.56 & 0.56 \\
    & RD{\footnotesize5}@1$\downarrow$  & 0.21 & 0.1 & -0.08 & 0.31 & 0.1 & 0.08 & 0.09 & 0.15 & 0.03 & 0.23 & 0.12 & 0.12 & 0.01 & 0.01 & -0.03 & 0.08 & 0.01 & 0.07 \\
    & RR{\footnotesize5}@1$\downarrow$  & 0.03 & 0.06 & 0.03 & 0.05 & 0.03 & 0.04 & 0.05 & 0.05 & 0.05 & 0.06 & 0.03 & 0.05 & 0.09 & 0.09 & 0.12 & 0.18 & 0.09 & 0.13 \\
    \hline
\end{tabular}}
\end{table} 

\begin{table*}
\centering
\caption{Performance summary of perturbations using partial cases using HumanEval-X test cases and partial passes (3 out of 5).}
\label{tab:ressummarypartial_he}
\resizebox{\textwidth}{!}{
\begin{tabular}{|p{3cm}|p{2cm}|p{1cm}|p{1cm}|p{1cm}|p{1cm}|p{1cm}|p{1cm}|p{1cm}|p{1cm}|p{1cm}|p{1cm}|p{1cm}|p{1cm}|p{1cm}|p{1cm}|p{1cm}|p{1cm}|p{1cm}|p{1cm}|}
    \hline
    HumanEval-X & Model & \multicolumn{3}{|p{4cm}|}{\centering Incoder-1B} & \multicolumn{3}{|p{4cm}|}{\centering Incoder-6B} & \multicolumn{3}{|p{4cm}|}{\centering CodeGen-2B-Multi} & \multicolumn{3}{|p{4cm}|}{\centering CodeGen-6B-Multi} & \multicolumn{3}{|p{4cm}|}{\centering Magicoder-7B} & \multicolumn{3}{|p{4cm}|}{\centering QwenCode} \\
    \hline
    Perturbation & Metric & Java & CPP & JS & Java & CPP & JS & Java & CPP & JS & Java & CPP & JS & Java & CPP & JS & Java & CPP & JS \\
    \hline
    Nominal & RP{\footnotesize5}@1 & 0.02 & 0.1 & 0.08 & 0.11 & 0.12 & 0.13 & 0.13 & 0.12 & 0.13 & 0.16 & 0.14 & 0.16 & 0.65 & 0.57 & 0.53 & 0.57 & 0.59 & 0.62 \\
    \hline
    Partial & RP{\footnotesize5}@1 & 0.33 & 0.1 & 0.38 & 0.45 & 0.17 & 0.44 & 0.56 & 0.13 & 0.39 & 0.54 & 0.22 & 0.44 & 0.88 & 0.63 & 0.76 & 0.79 & 0.65 & 0.70 \\
    \hline
    \multirow{4}{*}{\centering DocString} 
    & Nominal$\uparrow$ & 0.02 & 0.1 & 0.08 & 0.11 & 0.12 & 0.13 & 0.13 & 0.12 & 0.13 & 0.16 & 0.14 & 0.16 & 0.65 & 0.57 & 0.53 & 0.57 & 0.59 & 0.62 \\
	& RP{\footnotesize5}@1$\uparrow$  & 0.03 & 0.07 & 0.07 & 0.07 & 0.11 & 0.11 & 0.12 & 0.12 & 0.1 & 0.12 & 0.13 & 0.12 & 0.58 & 0.51 & 0.50 & 0.50 & 0.55 & 0.55 \\
	& RD{\footnotesize5}@1$\downarrow$  & -0.07 & 0.31 & 0.16 & 0.35 & 0.11 & 0.16 & 0.09 & 0.05 & 0.23 & 0.22 & 0.07 & 0.28 & 0.09 & 0.09 & 0.04 & 0.12 & 0.05 & 0.10 \\
	& RR{\footnotesize5}@1$\downarrow$  & 0.02 & 0.06 & 0.05 & 0.06 & 0.04 & 0.05 & 0.04 & 0.05 & 0.05 & 0.08 & 0.04 & 0.09 & 0.12 & 0.10 & 0.13 & 0.22 & 0.12 & 0.15 \\
	\hline
	\multirow{4}{*}{\centering Syntax} 
    & Nominal$\uparrow$ & 0.33 & 0.1 & 0.38 & 0.45 & 0.17 & 0.44 & 0.56 & 0.13 & 0.39 & 0.54 & 0.22 & 0.44 & 0.88 & 0.63 & 0.76 & 0.79 & 0.65 & 0.70 \\
	& RP{\footnotesize5}@1$\uparrow$  & 0.27 & 0.07 & 0.28 & 0.36 & 0.11 & 0.32 & 0.4 & 0.11 & 0.28 & 0.39 & 0.14 & 0.33 & 0.64 & 0.55 & 0.62 & 0.66 & 0.52 & 0.57 \\
	& RD{\footnotesize5}@1$\downarrow$  & 0.18 & 0.37 & 0.26 & 0.19 & 0.36 & 0.27 & 0.29 & 0.13 & 0.28 & 0.27 & 0.35 & 0.24 & 0.25 & 0.11 & 0.16 & 0.14 & 0.19 & 0.17 \\
	& RR{\footnotesize5}@1$\downarrow$  & 0.2 & 0.08 & 0.11 & 0.2 & 0.17 & 0.13 & 0.21 & 0.09 & 0.16 & 0.18 & 0.13 & 0.17 & 0.25 & 0.16 & 0.20 & 0.36 & 0.23 & 0.25 \\
	\hline
	\multirow{4}{*}{\centering Format} 
    & Nominal$\uparrow$ & 0.33 & 0.1 & 0.38 & 0.45 & 0.17 & 0.44 & 0.56 & 0.13 & 0.39 & 0.54 & 0.22 & 0.44 & 0.88 & 0.63 & 0.76 & 0.79 & 0.65 & 0.70 \\
	& RP{\footnotesize5}@1$\uparrow$  & 0.32 & 0.07 & 0.32 & 0.39 & 0.13 & 0.41 & 0.52 & 0.14 & 0.36 & 0.53 & 0.19 & 0.35 & 0.85 & 0.64 & 0.75 & 0.83 & 0.65 & 0.70 \\
	& RD{\footnotesize5}@1$\downarrow$  & 0.03 & 0.29 & 0.18 & 0.12 & 0.21 & 0.06 & 0.07 & -0.07 & 0.09 & 0.01 & 0.12 & 0.2 & 0.01 & -0.04 & -0.01 & -0.08 & -0.02 & -0.01 \\
	& RR{\footnotesize5}@1$\downarrow$  & 0.11 & 0.05 & 0.09 & 0.14 & 0.1 & 0.07 & 0.08 & 0.04 & 0.12 & 0.05 & 0.06 & 0.15 & 0.05 & 0.07 & 0.07 & 0.15 & 0.07 & 0.16 \\
	\hline
	\multirow{4}{*}{\centering Function} 
    & Nominal$\uparrow$ & 0.02 & 0.1 & 0.08 & 0.11 & 0.12 & 0.13 & 0.13 & 0.12 & 0.13 & 0.16 & 0.14 & 0.16 & 0.65 & 0.57 & 0.53 & 0.57 & 0.59 & 0.62 \\
	& RP{\footnotesize5}@1$\uparrow$  & 0.03 & 0.08 & 0.08 & 0.07 & 0.1 & 0.12 & 0.12 & 0.12 & 0.11 & 0.11 & 0.12 & 0.15 & 0.64 & 0.56 & 0.54 & 0.53 & 0.58 & 0.58 \\
	& RD{\footnotesize5}@1$\downarrow$  & -0.04 & 0.19 & -0.01 & 0.37 & 0.15 & 0.1 & 0.06 & 0.05 & 0.11 & 0.28 & 0.12 & 0.12 & -0.01 & -0.01 & -0.05 & 0.06 & -0.01 & 0.05 \\
	& RR{\footnotesize5}@1$\downarrow$  & 0.02 & 0.05 & 0.04 & 0.07 & 0.05 & 0.05 & 0.05 & 0.05 & 0.05 & 0.09 & 0.04 & 0.07 & 0.07 & 0.07 & 0.10 & 0.16 & 0.07 & 0.11 \\
	\hline

\end{tabular}}
\end{table*}

\begin{table}
\centering
\caption{Performance of code generation models against docstring perturbations (worst case passes)} 
\label{tab:resdocs}
\resizebox{\textwidth}{!}{\large \begin{tabular}{|p{4.5cm}|p{2cm}|p{1cm}|p{1cm}|p{1cm}|p{1cm}|p{1cm}|p{1cm}|p{1cm}|p{1cm}|p{1cm}|p{1cm}|p{1cm}|p{1cm}|p{1cm}|p{1cm}|p{1cm}|p{1cm}|p{1cm}|p{1cm}|}
    \hline
    HumanEval-X & Model & \multicolumn{3}{|p{4cm}|}{\centering Incoder-1B} & \multicolumn{3}{|p{4cm}|}{\centering Incoder-6B} & \multicolumn{3}{|p{4cm}|}{\centering CodeGen-2B-multi} & \multicolumn{3}{|p{4cm}|}{\centering CodeGen-6B-multi} & \multicolumn{3}{|p{4cm}|}{\centering Magicoder7B} & \multicolumn{3}{|p{4cm}|}{\centering QwenCode} \\
    \hline
    Perturbation & Metric & Java & CPP & JS & Java & CPP & JS & Java & CPP & JS & Java & CPP & JS & Java & CPP & JS & Java & CPP & JS \\
    \hline
    Nominal & RP{\footnotesize5}@1 & 0.02 & 0.08 & 0.07 & 0.11 & 0.1 & 0.12 & 0.14 & 0.12 & 0.11 & 0.15 & 0.13 & 0.14 & 0.63 & 0.55 & 0.51 & 0.55 & 0.57 & 0.6 \\
    \hline
    \hline
	\multirow{3}{*}{\centering SynonymSubstitution} & RP{\footnotesize5}@1 & 0.01 & 0.01 & 0.02 & 0.02 & 0.04 & 0.02 & 0.05 & 0.04 & 0.02 & 0.04 & 0.06 & 0.05 & 0.36 & 0.33 & 0.26 & 0.16 & 0.31 & 0.31 \\
	& RD{\footnotesize5}@1  & 0.5 & 0.85 & 0.75 & 0.78 & 0.62 & 0.79 & 0.65 & 0.65 & 0.78 & 0.72 & 0.52 & 0.65 & 0.43 & 0.4 & 0.49 & 0.7 & 0.45 & 0.48 \\
	& RR{\footnotesize5}@1  & 0.02 & 0.07 & 0.05 & 0.11 & 0.06 & 0.09 & 0.09 & 0.08 & 0.09 & 0.11 & 0.07 & 0.1 & 0.29 & 0.23 & 0.27 & 0.41 & 0.27 & 0.32 \\
	\hline
	\multirow{3}{*}{\centering ButterFingersPerturbation} & RP{\footnotesize5}@1 & 0.0 & 0.02 & 0.04 & 0.02 & 0.05 & 0.05 & 0.04 & 0.05 & 0.04 & 0.04 & 0.06 & 0.05 & 0.38 & 0.34 & 0.25 & 0.23 & 0.37 & 0.29 \\
	& RD{\footnotesize5}@1  & 1.0 & 0.69 & 0.5 & 0.83 & 0.5 & 0.53 & 0.74 & 0.6 & 0.67 & 0.76 & 0.52 & 0.61 & 0.39 & 0.38 & 0.51 & 0.58 & 0.35 & 0.52 \\
	& RR{\footnotesize5}@1  & 0.02 & 0.07 & 0.05 & 0.1 & 0.05 & 0.06 & 0.1 & 0.09 & 0.07 & 0.12 & 0.07 & 0.09 & 0.26 & 0.22 & 0.3 & 0.34 & 0.23 & 0.35 \\
	\hline
	\multirow{3}{*}{\centering ChangeCharCase} & RP{\footnotesize5}@1 & 0.0 & 0.0 & 0.01 & 0.02 & 0.05 & 0.03 & 0.04 & 0.04 & 0.02 & 0.04 & 0.07 & 0.04 & 0.36 & 0.3 & 0.23 & 0.2 & 0.37 & 0.34 \\
	& RD{\footnotesize5}@1  & 1.0 & 1.0 & 0.83 & 0.83 & 0.5 & 0.74 & 0.74 & 0.7 & 0.83 & 0.72 & 0.43 & 0.74 & 0.43 & 0.44 & 0.54 & 0.64 & 0.34 & 0.43 \\
	& RR{\footnotesize5}@1  & 0.02 & 0.08 & 0.07 & 0.1 & 0.06 & 0.09 & 0.1 & 0.09 & 0.09 & 0.11 & 0.05 & 0.1 & 0.3 & 0.24 & 0.29 & 0.39 & 0.23 & 0.29 \\
	\hline
	\multirow{3}{*}{\centering SynonymInsertion} & RP{\footnotesize5}@1 & 0.01 & 0.03 & 0.02 & 0.02 & 0.05 & 0.03 & 0.08 & 0.05 & 0.02 & 0.07 & 0.06 & 0.02 & 0.42 & 0.38 & 0.34 & 0.22 & 0.35 & 0.35 \\
	& RD{\footnotesize5}@1  & 0.75 & 0.62 & 0.67 & 0.83 & 0.5 & 0.74 & 0.43 & 0.55 & 0.78 & 0.52 & 0.52 & 0.87 & 0.33 & 0.3 & 0.33 & 0.6 & 0.39 & 0.42 \\
	& RR{\footnotesize5}@1  & 0.02 & 0.06 & 0.06 & 0.1 & 0.05 & 0.09 & 0.07 & 0.07 & 0.09 & 0.09 & 0.07 & 0.12 & 0.21 & 0.18 & 0.23 & 0.37 & 0.22 & 0.3 \\
	\hline
	\multirow{3}{*}{\centering TenseTransformationFuture} & RP{\footnotesize5}@1 & 0.01 & 0.03 & 0.04 & 0.04 & 0.06 & 0.07 & 0.1 & 0.06 & 0.05 & 0.06 & 0.07 & 0.05 & 0.44 & 0.41 & 0.35 & 0.3 & 0.39 & 0.38 \\
	& RD{\footnotesize5}@1  & 0.5 & 0.62 & 0.5 & 0.67 & 0.38 & 0.42 & 0.3 & 0.5 & 0.5 & 0.6 & 0.48 & 0.61 & 0.3 & 0.24 & 0.31 & 0.46 & 0.31 & 0.37 \\
	& RR{\footnotesize5}@1  & 0.01 & 0.06 & 0.04 & 0.09 & 0.04 & 0.06 & 0.05 & 0.06 & 0.05 & 0.09 & 0.06 & 0.09 & 0.24 & 0.13 & 0.18 & 0.34 & 0.19 & 0.26 \\
	\hline
	\multirow{3}{*}{\centering BackTranslation} & RP{\footnotesize5}@1 & 0.0 & 0.02 & 0.05 & 0.04 & 0.07 & 0.05 & 0.09 & 0.07 & 0.07 & 0.08 & 0.07 & 0.08 & 0.46 & 0.42 & 0.29 & 0.29 & 0.43 & 0.4 \\
	& RD{\footnotesize5}@1  & 1.0 & 0.69 & 0.33 & 0.67 & 0.25 & 0.53 & 0.39 & 0.45 & 0.39 & 0.48 & 0.43 & 0.43 & 0.26 & 0.23 & 0.42 & 0.47 & 0.25 & 0.34 \\
	& RR{\footnotesize5}@1  & 0.02 & 0.07 & 0.02 & 0.09 & 0.04 & 0.07 & 0.07 & 0.05 & 0.04 & 0.07 & 0.07 & 0.06 & 0.2 & 0.14 & 0.24 & 0.29 & 0.15 & 0.23 \\
	\hline
	\multirow{3}{*}{\centering SwapCharactersPerturbation} & RP{\footnotesize5}@1 & 0.0 & 0.02 & 0.02 & 0.01 & 0.05 & 0.05 & 0.07 & 0.05 & 0.03 & 0.04 & 0.06 & 0.05 & 0.41 & 0.38 & 0.3 & 0.3 & 0.38 & 0.3 \\
	& RD{\footnotesize5}@1  & 1.0 & 0.77 & 0.75 & 0.89 & 0.44 & 0.58 & 0.52 & 0.6 & 0.72 & 0.72 & 0.52 & 0.61 & 0.35 & 0.31 & 0.41 & 0.46 & 0.32 & 0.51 \\
	& RR{\footnotesize5}@1  & 0.02 & 0.07 & 0.05 & 0.1 & 0.04 & 0.08 & 0.09 & 0.07 & 0.08 & 0.11 & 0.07 & 0.09 & 0.22 & 0.18 & 0.22 & 0.32 & 0.21 & 0.33 \\
	\hline
	\multirow{3}{*}{\centering WhitespacePerturbation} & RP{\footnotesize5}@1 & 0.0 & 0.03 & 0.02 & 0.02 & 0.05 & 0.03 & 0.04 & 0.05 & 0.03 & 0.04 & 0.05 & 0.04 & 0.37 & 0.32 & 0.25 & 0.27 & 0.33 & 0.37 \\
	& RD{\footnotesize5}@1  & 1.0 & 0.62 & 0.75 & 0.78 & 0.5 & 0.74 & 0.7 & 0.6 & 0.72 & 0.76 & 0.57 & 0.7 & 0.41 & 0.42 & 0.51 & 0.5 & 0.42 & 0.39 \\
	& RR{\footnotesize5}@1  & 0.02 & 0.07 & 0.05 & 0.1 & 0.05 & 0.1 & 0.1 & 0.07 & 0.08 & 0.12 & 0.07 & 0.1 & 0.27 & 0.23 & 0.27 & 0.34 & 0.25 & 0.27 \\
	\hline
	\multirow{3}{*}{\centering TenseTransformationPast} & RP{\footnotesize5}@1 & 0.0 & 0.04 & 0.04 & 0.04 & 0.05 & 0.06 & 0.08 & 0.07 & 0.07 & 0.04 & 0.07 & 0.06 & 0.51 & 0.43 & 0.37 & 0.32 & 0.43 & 0.38 \\
	& RD{\footnotesize5}@1  & 1.0 & 0.54 & 0.42 & 0.67 & 0.5 & 0.47 & 0.43 & 0.45 & 0.39 & 0.72 & 0.48 & 0.57 & 0.19 & 0.21 & 0.27 & 0.41 & 0.25 & 0.37 \\
	& RR{\footnotesize5}@1  & 0.02 & 0.05 & 0.03 & 0.09 & 0.05 & 0.05 & 0.06 & 0.07 & 0.04 & 0.11 & 0.06 & 0.08 & 0.16 & 0.14 & 0.16 & 0.27 & 0.16 & 0.25 \\
	\hline
	\multirow{3}{*}{\centering EnglishInflectionalVariation} & RP{\footnotesize5}@1 & 0.0 & 0.02 & 0.04 & 0.02 & 0.04 & 0.06 & 0.04 & 0.05 & 0.04 & 0.06 & 0.05 & 0.06 & 0.48 & 0.41 & 0.34 & 0.35 & 0.4 & 0.37 \\
	& RD{\footnotesize5}@1  & 1.0 & 0.69 & 0.5 & 0.83 & 0.56 & 0.47 & 0.7 & 0.55 & 0.61 & 0.6 & 0.57 & 0.57 & 0.24 & 0.24 & 0.34 & 0.36 & 0.3 & 0.39 \\
	& RR{\footnotesize5}@1  & 0.02 & 0.07 & 0.04 & 0.09 & 0.05 & 0.05 & 0.1 & 0.07 & 0.07 & 0.1 & 0.07 & 0.08 & 0.21 & 0.17 & 0.2 & 0.27 & 0.17 & 0.25 \\
	\hline
\end{tabular}}
\end{table} 

\begin{table}
\centering
\caption{Performance of code generation models against function name perturbations. (worst case passes)} 
\label{tab:resfunc}
\resizebox{\textwidth}{!}{\large \begin{tabular}{|p{5.5cm}|p{2cm}|p{1cm}|p{1cm}|p{1cm}|p{1cm}|p{1cm}|p{1cm}|p{1cm}|p{1cm}|p{1cm}|p{1cm}|p{1cm}|p{1cm}|p{1cm}|p{1cm}|p{1cm}|p{1cm}|p{1cm}|p{1cm}|}
    \hline
    HumanEval-X & Model & \multicolumn{3}{|p{4cm}|}{\centering Incoder-1B} & \multicolumn{3}{|p{4cm}|}{\centering Incoder-6B} & \multicolumn{3}{|p{4cm}|}{\centering CodeGen-2B-multi} & \multicolumn{3}{|p{4cm}|}{\centering CodeGen-6B-multi} & \multicolumn{3}{|p{4cm}|}{\centering Magicoder7B} & \multicolumn{3}{|p{4cm}|}{\centering QwenCode} \\
    \hline
    Perturbation & Metric & Java & CPP & JS & Java & CPP & JS & Java & CPP & JS & Java & CPP & JS & Java & CPP & JS & Java & CPP & JS \\
    \hline
    Nominal & RP{\footnotesize5}@1 & 0.02 & 0.08 & 0.07 & 0.11 & 0.1 & 0.12 & 0.14 & 0.12 & 0.11 & 0.15 & 0.13 & 0.14 & 0.63 & 0.55 & 0.51 & 0.55 & 0.57 & 0.6 \\
    \hline
    \multirow{3}{*}{\centering FuncRenameSwapChar} & RP{\footnotesize5}@1 & 0.01 & 0.04 & 0.05 & 0.05 & 0.06 & 0.05 & 0.09 & 0.05 & 0.06 & 0.05 & 0.09 & 0.08 & 0.52 & 0.41 & 0.39 & 0.3 & 0.47 & 0.39 \\
	& RD{\footnotesize5}@1  & 0.75 & 0.54 & 0.25 & 0.56 & 0.38 & 0.53 & 0.39 & 0.55 & 0.44 & 0.64 & 0.33 & 0.43 & 0.17 & 0.24 & 0.23 & 0.46 & 0.17 & 0.35 \\
	& RR{\footnotesize5}@1  & 0.02 & 0.05 & 0.03 & 0.07 & 0.04 & 0.06 & 0.05 & 0.08 & 0.05 & 0.1 & 0.05 & 0.09 & 0.11 & 0.15 & 0.16 & 0.29 & 0.1 & 0.23 \\
	\hline
	\multirow{3}{*}{\centering FuncRenameChangeChar} & RP{\footnotesize5}@1 & 0.0 & 0.01 & 0.07 & 0.04 & 0.06 & 0.07 & 0.07 & 0.03 & 0.07 & 0.05 & 0.06 & 0.08 & 0.53 & 0.44 & 0.41 & 0.29 & 0.49 & 0.4 \\
	& RD{\footnotesize5}@1  & 1.0 & 0.85 & 0.08 & 0.61 & 0.38 & 0.42 & 0.52 & 0.75 & 0.39 & 0.68 & 0.52 & 0.43 & 0.16 & 0.2 & 0.18 & 0.47 & 0.13 & 0.34 \\
	& RR{\footnotesize5}@1  & 0.02 & 0.09 & 0.03 & 0.08 & 0.04 & 0.06 & 0.07 & 0.09 & 0.05 & 0.12 & 0.08 & 0.07 & 0.12 & 0.16 & 0.15 & 0.27 & 0.12 & 0.22 \\
	\hline
	\multirow{3}{*}{\centering FuncRenameSynonymSub} & RP{\footnotesize5}@1 & 0.0 & 0.02 & 0.04 & 0.04 & 0.05 & 0.07 & 0.08 & 0.06 & 0.04 & 0.07 & 0.07 & 0.08 & 0.48 & 0.42 & 0.37 & 0.3 & 0.46 & 0.42 \\
	& RD{\footnotesize5}@1  & 1.0 & 0.69 & 0.42 & 0.67 & 0.44 & 0.42 & 0.43 & 0.5 & 0.67 & 0.56 & 0.48 & 0.43 & 0.24 & 0.23 & 0.28 & 0.46 & 0.19 & 0.3 \\
	& RR{\footnotesize5}@1  & 0.02 & 0.07 & 0.04 & 0.07 & 0.04 & 0.05 & 0.06 & 0.07 & 0.07 & 0.09 & 0.06 & 0.07 & 0.18 & 0.14 & 0.18 & 0.3 & 0.13 & 0.23 \\
	\hline
	\multirow{3}{*}{\centering FuncRenameInflectionalVariation} & RP{\footnotesize5}@1 & 0.0 & 0.01 & 0.05 & 0.04 & 0.06 & 0.06 & 0.1 & 0.05 & 0.07 & 0.04 & 0.07 & 0.08 & 0.49 & 0.45 & 0.37 & 0.32 & 0.45 & 0.41 \\
	& RD{\footnotesize5}@1  & 1.0 & 0.85 & 0.33 & 0.61 & 0.38 & 0.47 & 0.26 & 0.6 & 0.39 & 0.72 & 0.43 & 0.43 & 0.22 & 0.19 & 0.27 & 0.42 & 0.2 & 0.32 \\
	& RR{\footnotesize5}@1  & 0.02 & 0.07 & 0.02 & 0.07 & 0.05 & 0.07 & 0.05 & 0.07 & 0.05 & 0.11 & 0.05 & 0.06 & 0.14 & 0.13 & 0.17 & 0.27 & 0.12 & 0.22 \\
	\hline
	\multirow{3}{*}{\centering FuncRenameButterFinger} & RP{\footnotesize5}@1 & 0.0 & 0.02 & 0.04 & 0.02 & 0.04 & 0.06 & 0.09 & 0.07 & 0.07 & 0.04 & 0.07 & 0.07 & 0.51 & 0.47 & 0.37 & 0.29 & 0.43 & 0.41 \\
	& RD{\footnotesize5}@1  & 1.0 & 0.69 & 0.42 & 0.78 & 0.62 & 0.47 & 0.39 & 0.4 & 0.39 & 0.72 & 0.48 & 0.48 & 0.18 & 0.14 & 0.27 & 0.47 & 0.25 & 0.32 \\
	& RR{\footnotesize5}@1  & 0.02 & 0.08 & 0.03 & 0.09 & 0.06 & 0.07 & 0.05 & 0.05 & 0.05 & 0.11 & 0.06 & 0.08 & 0.12 & 0.13 & 0.2 & 0.3 & 0.15 & 0.23 \\
	\hline
    \multirow{3}{*}{\centering FuncRenameSnake/CamelCase} & RP{\footnotesize5}@1 & 0.01 & 0.04 & 0.05 & 0.04 & 0.07 & 0.07 & 0.09 & 0.07 & 0.06 & 0.07 & 0.07 & 0.08 & 0.53 & 0.45 & 0.39 & 0.32 & 0.44 & 0.4 \\
	& RD{\footnotesize5}@1  & 0.75 & 0.54 & 0.25 & 0.61 & 0.31 & 0.42 & 0.39 & 0.4 & 0.44 & 0.52 & 0.43 & 0.43 & 0.16 & 0.18 & 0.23 & 0.41 & 0.23 & 0.33 \\
	& RR{\footnotesize5}@1  & 0.02 & 0.07 & 0.03 & 0.08 & 0.05 & 0.05 & 0.05 & 0.05 & 0.05 & 0.08 & 0.05 & 0.07 & 0.13 & 0.11 & 0.15 & 0.3 & 0.13 & 0.21 \\
    \hline
\end{tabular}}
\end{table} 

\begin{table}
\centering
\caption{Performance of code generation models against syntax perturbations. (worst case passes)} 
\label{tab:ressyntax}
\resizebox{\textwidth}{!}{\large \begin{tabular}{|p{4cm}|p{2cm}|p{1cm}|p{1cm}|p{1cm}|p{1cm}|p{1cm}|p{1cm}|p{1cm}|p{1cm}|p{1cm}|p{1cm}|p{1cm}|p{1cm}|p{1cm}|p{1cm}|p{1cm}|p{1cm}|p{1cm}|p{1cm}|}
    \hline
    HumanEval-X & Model & \multicolumn{3}{|p{4cm}|}{\centering Incoder-1B} & \multicolumn{3}{|p{4cm}|}{\centering Incoder-6B} & \multicolumn{3}{|p{4cm}|}{\centering CodeGen-2B-multi} & \multicolumn{3}{|p{4cm}|}{\centering CodeGen-6B-multi} & \multicolumn{3}{|p{4cm}|}{\centering Magicoder7B} & \multicolumn{3}{|p{4cm}|}{\centering QwenCode} \\
    \hline
    Perturbation & Metric & Java & CPP & JS & Java & CPP & JS & Java & CPP & JS & Java & CPP & JS & Java & CPP & JS & Java & CPP & JS \\
    \hline
    \hline
    Nominal (Partial) & RP{\footnotesize5}@1 & 0.32 & 0.09 & 0.37 & 0.45 & 0.14 & 0.44 & 0.56 & 0.13 & 0.39 & 0.55 & 0.2 & 0.44 & 0.86 & 0.61 & 0.74 & 0.77 & 0.63 & 0.68 \\
    \hline
	\multirow{3}{*}{\centering VarRenamerRN} & RP{\footnotesize5}@1 & 0.25 & 0.01 & 0.27 & 0.28 & 0.03 & 0.3 & 0.36 & 0.04 & 0.26 & 0.29 & 0.06 & 0.27 & 0.68 & 0.48 & 0.6 & 0.66 & 0.41 & 0.41 \\
	& RD{\footnotesize5}@1  & 0.23 & 0.86 & 0.28 & 0.37 & 0.78 & 0.31 & 0.36 & 0.67 & 0.34 & 0.47 & 0.69 & 0.39 & 0.21 & 0.21 & 0.19 & 0.14 & 0.35 & 0.39 \\
	& RR{\footnotesize5}@1  & 0.2 & 0.07 & 0.12 & 0.24 & 0.13 & 0.13 & 0.21 & 0.1 & 0.13 & 0.26 & 0.13 & 0.21 & 0.21 & 0.18 & 0.19 & 0.39 & 0.3 & 0.33 \\
	\hline
	\multirow{3}{*}{\centering VarRenamerNaive} & RP{\footnotesize5}@1 & 0.27 & 0.02 & 0.27 & 0.32 & 0.05 & 0.28 & 0.37 & 0.06 & 0.25 & 0.3 & 0.07 & 0.28 & 0.7 & 0.53 & 0.62 & 0.63 & 0.46 & 0.52 \\
	& RD{\footnotesize5}@1  & 0.17 & 0.79 & 0.26 & 0.29 & 0.61 & 0.36 & 0.35 & 0.52 & 0.36 & 0.44 & 0.62 & 0.36 & 0.19 & 0.13 & 0.17 & 0.17 & 0.28 & 0.23 \\
	& RR{\footnotesize5}@1  & 0.18 & 0.08 & 0.11 & 0.21 & 0.11 & 0.16 & 0.21 & 0.09 & 0.15 & 0.24 & 0.12 & 0.2 & 0.21 & 0.14 & 0.19 & 0.35 & 0.25 & 0.29 \\
	\hline
	\multirow{3}{*}{\centering ForWhileTransformer} & RP{\footnotesize5}@1 & 0.24 & 0.02 & 0.23 & 0.27 & 0.04 & 0.28 & 0.38 & 0.07 & 0.24 & 0.33 & 0.1 & 0.26 & 0.68 & 0.52 & 0.56 & 0.68 & 0.42 & 0.39 \\
	& RD{\footnotesize5}@1  & 0.25 & 0.71 & 0.38 & 0.38 & 0.7 & 0.36 & 0.33 & 0.43 & 0.39 & 0.4 & 0.47 & 0.42 & 0.21 & 0.15 & 0.25 & 0.11 & 0.34 & 0.43 \\
	& RR{\footnotesize5}@1  & 0.21 & 0.09 & 0.16 & 0.22 & 0.12 & 0.16 & 0.2 & 0.09 & 0.18 & 0.22 & 0.1 & 0.22 & 0.2 & 0.14 & 0.23 & 0.29 & 0.29 & 0.34 \\
	\hline
	\multirow{3}{*}{\centering OperandSwap} & RP{\footnotesize5}@1 & 0.28 & 0.02 & 0.27 & 0.32 & 0.07 & 0.33 & 0.38 & 0.07 & 0.23 & 0.34 & 0.09 & 0.28 & 0.64 & 0.54 & 0.65 & 0.62 & 0.49 & 0.5 \\
	& RD{\footnotesize5}@1  & 0.13 & 0.71 & 0.26 & 0.27 & 0.52 & 0.25 & 0.33 & 0.43 & 0.41 & 0.38 & 0.53 & 0.36 & 0.26 & 0.11 & 0.12 & 0.19 & 0.22 & 0.27 \\
	& RR{\footnotesize5}@1  & 0.2 & 0.09 & 0.11 & 0.17 & 0.11 & 0.11 & 0.2 & 0.09 & 0.16 & 0.21 & 0.13 & 0.2 & 0.22 & 0.15 & 0.14 & 0.37 & 0.21 & 0.27 \\
	\hline
	\multirow{3}{*}{\centering DeadCodeInserter} & RP{\footnotesize5}@1 & 0.04 & 0.01 & 0.09 & 0.05 & 0.01 & 0.1 & 0.07 & 0.01 & 0.07 & 0.07 & 0.02 & 0.1 & 0.04 & 0.13 & 0.09 & 0.03 & 0.12 & 0.07 \\
	& RD{\footnotesize5}@1  & 0.89 & 0.93 & 0.75 & 0.89 & 0.91 & 0.76 & 0.87 & 0.95 & 0.83 & 0.88 & 0.88 & 0.78 & 0.96 & 0.78 & 0.88 & 0.96 & 0.81 & 0.9 \\
	& RR{\footnotesize5}@1  & 0.3 & 0.08 & 0.28 & 0.4 & 0.13 & 0.34 & 0.49 & 0.12 & 0.32 & 0.48 & 0.17 & 0.37 & 0.82 & 0.48 & 0.66 & 0.74 & 0.51 & 0.62 \\
	\hline
	\multirow{3}{*}{\centering VarRenamerCB} & RP{\footnotesize5}@1 & 0.26 & 0.02 & 0.28 & 0.3 & 0.06 & 0.32 & 0.36 & 0.04 & 0.23 & 0.35 & 0.07 & 0.26 & 0.66 & 0.48 & 0.65 & 0.65 & 0.42 & 0.44 \\
	& RD{\footnotesize5}@1  & 0.19 & 0.79 & 0.25 & 0.33 & 0.57 & 0.26 & 0.36 & 0.71 & 0.42 & 0.36 & 0.66 & 0.4 & 0.23 & 0.21 & 0.13 & 0.16 & 0.34 & 0.36 \\
	& RR{\footnotesize5}@1  & 0.22 & 0.08 & 0.1 & 0.22 & 0.13 & 0.12 & 0.21 & 0.09 & 0.16 & 0.21 & 0.13 & 0.23 & 0.22 & 0.2 & 0.16 & 0.4 & 0.29 & 0.28 \\
	\hline
\end{tabular}}
\end{table} 

\begin{table}
\centering
\caption{Performance of code generation models against format perturbations (worst case passes).} 
\label{tab:resformat}
\resizebox{\textwidth}{!}{\large \begin{tabular}{|p{4cm}|p{2cm}|p{1cm}|p{1cm}|p{1cm}|p{1cm}|p{1cm}|p{1cm}|p{1cm}|p{1cm}|p{1cm}|p{1cm}|p{1cm}|p{1cm}|p{1cm}|p{1cm}|p{1cm}|p{1cm}|p{1cm}|p{1cm}|}
    \hline
    HumanEval-X & Model & \multicolumn{3}{|p{4cm}|}{\centering Incoder-1B} & \multicolumn{3}{|p{4cm}|}{\centering Incoder-6B} & \multicolumn{3}{|p{4cm}|}{\centering CodeGen-2B-multi} & \multicolumn{3}{|p{4cm}|}{\centering CodeGen-6B-multi} & \multicolumn{3}{|p{4cm}|}{\centering Magicoder7B} & \multicolumn{3}{|p{4cm}|}{\centering QwenCode} \\
    \hline
    Perturbation & Metric & Java & CPP & JS & Java & CPP & JS & Java & CPP & JS & Java & CPP & JS & Java & CPP & JS & Java & CPP & JS \\
    \hline
    \hline
    Nominal (Partial) & RP{\footnotesize5}@1 & 0.32 & 0.09 & 0.37 & 0.45 & 0.14 & 0.44 & 0.56 & 0.13 & 0.39 & 0.55 & 0.2 & 0.44 & 0.86 & 0.61 & 0.74 & 0.77 & 0.63 & 0.68 \\
    \hline
	\multirow{3}{*}{\centering split\_lines} & RP{\footnotesize5}@1 & 0.01 & 0.02 & 0.26 & 0.39 & 0.06 & 0.33 & 0.02 & 0.07 & 0.27 & 0.03 & 0.09 & 0.27 & 0.77 & 0.49 & 0.59 & 0.7 & 0.49 & 0.49 \\
	& RD{\footnotesize5}@1  & 0.98 & 0.71 & 0.3 & 0.12 & 0.57 & 0.25 & 0.97 & 0.43 & 0.31 & 0.94 & 0.56 & 0.37 & 0.1 & 0.19 & 0.21 & 0.09 & 0.22 & 0.29 \\
	& RR{\footnotesize5}@1  & 0.32 & 0.07 & 0.13 & 0.07 & 0.08 & 0.12 & 0.54 & 0.07 & 0.15 & 0.52 & 0.11 & 0.18 & 0.11 & 0.14 & 0.21 & 0.15 & 0.18 & 0.26 \\
	\hline
	\multirow{3}{*}{\centering doc2comments} & RP{\footnotesize5}@1 & 0.11 & 0.01 & 0.36 & 0.2 & 0.05 & 0.02 & 0.34 & 0.09 & 0.36 & 0.38 & 0.1 & 0.4 & 0.79 & 0.52 & 0.72 & 0.48 & 0.52 & 0.73 \\
	& RD{\footnotesize5}@1  & 0.66 & 0.86 & 0.03 & 0.55 & 0.65 & 0.94 & 0.4 & 0.33 & 0.08 & 0.3 & 0.47 & 0.1 & 0.08 & 0.14 & 0.03 & 0.37 & 0.17 & -0.06 \\
	& RR{\footnotesize5}@1  & 0.23 & 0.07 & 0.05 & 0.24 & 0.12 & 0.41 & 0.23 & 0.05 & 0.08 & 0.18 & 0.1 & 0.08 & 0.09 & 0.13 & 0.11 & 0.34 & 0.11 & 0.2 \\
	\hline
	\multirow{3}{*}{\centering new\_lines} & RP{\footnotesize5}@1 & 0.23 & 0.02 & 0.15 & 0.3 & 0.06 & 0.34 & 0.45 & 0.07 & 0.23 & 0.35 & 0.1 & 0.18 & 0.77 & 0.52 & 0.68 & 0.8 & 0.5 & 0.51 \\
	& RD{\footnotesize5}@1  & 0.3 & 0.71 & 0.61 & 0.32 & 0.57 & 0.22 & 0.2 & 0.43 & 0.41 & 0.36 & 0.47 & 0.6 & 0.11 & 0.15 & 0.09 & -0.04 & 0.21 & 0.26 \\
	& RR{\footnotesize5}@1  & 0.13 & 0.07 & 0.24 & 0.16 & 0.1 & 0.11 & 0.12 & 0.07 & 0.21 & 0.2 & 0.1 & 0.29 & 0.1 & 0.13 & 0.1 & 0.18 & 0.16 & 0.23 \\
	\hline
	\multirow{3}{*}{\centering new\_line\_afterdoc} & RP{\footnotesize5}@1 & 0.26 & 0.02 & 0.2 & 0.39 & 0.06 & 0.38 & 0.51 & 0.09 & 0.32 & 0.45 & 0.12 & 0.23 & 0.81 & 0.51 & 0.71 & 0.68 & 0.52 & 0.57 \\
	& RD{\footnotesize5}@1  & 0.21 & 0.71 & 0.46 & 0.12 & 0.57 & 0.14 & 0.1 & 0.29 & 0.17 & 0.18 & 0.38 & 0.47 & 0.06 & 0.16 & 0.05 & 0.12 & 0.17 & 0.17 \\
	& RR{\footnotesize5}@1  & 0.1 & 0.06 & 0.18 & 0.1 & 0.09 & 0.06 & 0.08 & 0.04 & 0.1 & 0.11 & 0.07 & 0.23 & 0.06 & 0.11 & 0.07 & 0.15 & 0.11 & 0.18 \\
	\hline
	\multirow{3}{*}{\centering new\_line\_aftercode} & RP{\footnotesize5}@1 & 0.19 & 0.02 & 0.18 & 0.35 & 0.05 & 0.34 & 0.41 & 0.07 & 0.2 & 0.42 & 0.1 & 0.2 & 0.76 & 0.52 & 0.62 & 0.77 & 0.54 & 0.45 \\
	& RD{\footnotesize5}@1  & 0.42 & 0.71 & 0.51 & 0.21 & 0.61 & 0.22 & 0.26 & 0.43 & 0.5 & 0.23 & 0.5 & 0.54 & 0.12 & 0.14 & 0.16 & -0.01 & 0.14 & 0.34 \\
	& RR{\footnotesize5}@1  & 0.17 & 0.07 & 0.2 & 0.14 & 0.09 & 0.11 & 0.16 & 0.05 & 0.23 & 0.14 & 0.11 & 0.27 & 0.12 & 0.13 & 0.15 & 0.2 & 0.15 & 0.26 \\
	\hline
	\multirow{3}{*}{\centering tab\_indent} & RP{\footnotesize5}@1 & 0.12 & 0.03 & 0.34 & 0.07 & 0.08 & 0.41 & 0.48 & 0.08 & 0.32 & 0.43 & 0.12 & 0.38 & 0.81 & 0.48 & 0.71 & 0.73 & 0.49 & 0.57 \\
	& RD{\footnotesize5}@1  & 0.64 & 0.64 & 0.08 & 0.84 & 0.43 & 0.07 & 0.15 & 0.38 & 0.19 & 0.22 & 0.41 & 0.12 & 0.06 & 0.21 & 0.05 & 0.05 & 0.22 & 0.17 \\
	& RR{\footnotesize5}@1  & 0.21 & 0.05 & 0.05 & 0.37 & 0.06 & 0.05 & 0.1 & 0.06 & 0.11 & 0.12 & 0.09 & 0.13 & 0.07 & 0.16 & 0.07 & 0.24 & 0.15 & 0.25 \\
	\hline
\end{tabular}}
\end{table}

\begin{table}
\centering
\caption{P Values Using Fisher's Exact Test} 
\label{tab:pvalues}
\resizebox{\textwidth}{!}{\large \begin{tabular}{|p{3cm}|p{1cm}|p{1cm}|p{1cm}|p{1cm}|p{1cm}|p{1cm}|p{1cm}|p{1cm}|p{1cm}|p{1cm}|p{1cm}|p{1cm}|}
    \hline
    HumanEval-X & \multicolumn{3}{|p{4cm}|}{\centering Incoder-1B} & \multicolumn{3}{|p{4cm}|}{\centering Incoder-6B} & \multicolumn{3}{|p{4cm}|}{\centering CodeGen-2B-Multi} & \multicolumn{3}{|p{4cm}|}{\centering CodeGen-6B-Multi} \\
    \hline
    Perturbation & Java & CPP & JS & Java & CPP & JS & Java & CPP & JS & Java & CPP & JS \\
    \hline
    DocString & 0.25 & 0.08 & 0.03 & 0.01 & 0.08 & 0.02 & 0.09 & 0.06 & 0.02 & 0.0 & 0.05 & 0.0 \\
    Syntax & 0.04 & 0.01 & 0.0 & 0.0 & 0.0 & 0.0 & 0.0 & 0.11 & 0.0 & 0.0 & 0.0 & 0.0 \\
    Format & 0.0 & 0.03 & 0.0 & 0.0 & 0.02 & 0.01 & 0.0 & 0.58 & 0.0 & 0.0 & 0.03 & 0.0 \\
    Function & 0.25 & 0.08 & 0.1 & 0.02 & 0.08 & 0.06 & 0.14 & 0.06 & 0.06 & 0.0 & 0.08 & 0.01 \\
    \hline
\end{tabular}}
\end{table} 

\subsection{\textbf{Can the robustness drop be mitigated by fixing the perturbed prompts?}}

\noindent\textbf{Motivation:} 
The results of R1 show that LLM performance drops significantly when prompts are perturbed, yet no strategies have been explored to mitigate this decline. While prior work has applied input smoothing techniques~\citep{ye2020safer,ji2024defending} to improve LLM performance on security-related instructions, these approaches are specific to security scenarios and do not address robustness issues arising from natural perturbations in code prompts. In particular, repairing or normalizing perturbed prompts, especially docstrings, to restore code generation performance remains largely unexplored. This research question investigates whether refining perturbed prompts can recover part of the lost performance, providing actionable insights for practitioners relying on LLMs for reliable code generation.



\noindent\textbf{Approach:} 
We focus on the \texttt{DocString} perturbation dataset for this experiment. Our approach is inspired by prior work in natural language processing, where LLMs are used to correct typographical and grammatical errors in English 
sentences~\citep{coyne2023analyzing}. Specifically, we select all ten  \texttt{DocString}-perturbed subsets from our dataset and employ the \texttt{Magicoder-7B} model to fix the prompts. { We choose \texttt{Magicoder} because it is the only instruction-tuned model among our six evaluated models capable of reliably following natural language repair instructions. The remaining five models: Incoder-1B, Incoder-6B, 
CodeGen-2B-Multi, CodeGen-6B-Multi, and QwenCode, are primarily optimized for code generation and tend to produce code snippets rather than corrected natural language text when given repair prompts, making them unsuitable for this task.}

For each sample, we first parse and extract the DocString from the perturbed prompt. The extracted DocString is then passed to the \texttt{Magicoder-7B} model using the instruction provided in Figure~\ref{fig:fixprompt} to correct grammatical errors, spelling mistakes, and textual inconsistencies. After obtaining the fixed docString, we reconstruct the prompt by reintegrating the corrected natural language portion with the original code context. Finally, the reconstructed fixed prompt is used to regenerate the code using the same \texttt{Magicoder-7B} model. The performance of this fixed-prompt generation is then compared against the baseline perturbed-prompt results to assess whether prompt correction helps mitigate the robustness drop observed in RQ1.

\begin{figure}
\centering
\fbox{%
    \begin{minipage}{0.95\linewidth}
        \vspace{0.5em}
        \small
        \texttt{Fix the grammatical errors, spelling, and typos in the following text and reply with the fixed text. Do not delete any information, and do not add any code snippets.}

        \vspace{0.5em}
        \texttt{Text: \{nl\}}

        \vspace{0.5em}
        \texttt{Fixed Text:}
    \end{minipage}
}
\caption{Prompt used for repairing perturbed docstrings.}
\label{fig:fixprompt}
\end{figure}
\begin{figure}
    \centering
    \begin{subfigure}{\textwidth}
        \centering
        \includegraphics[width=.9\linewidth]{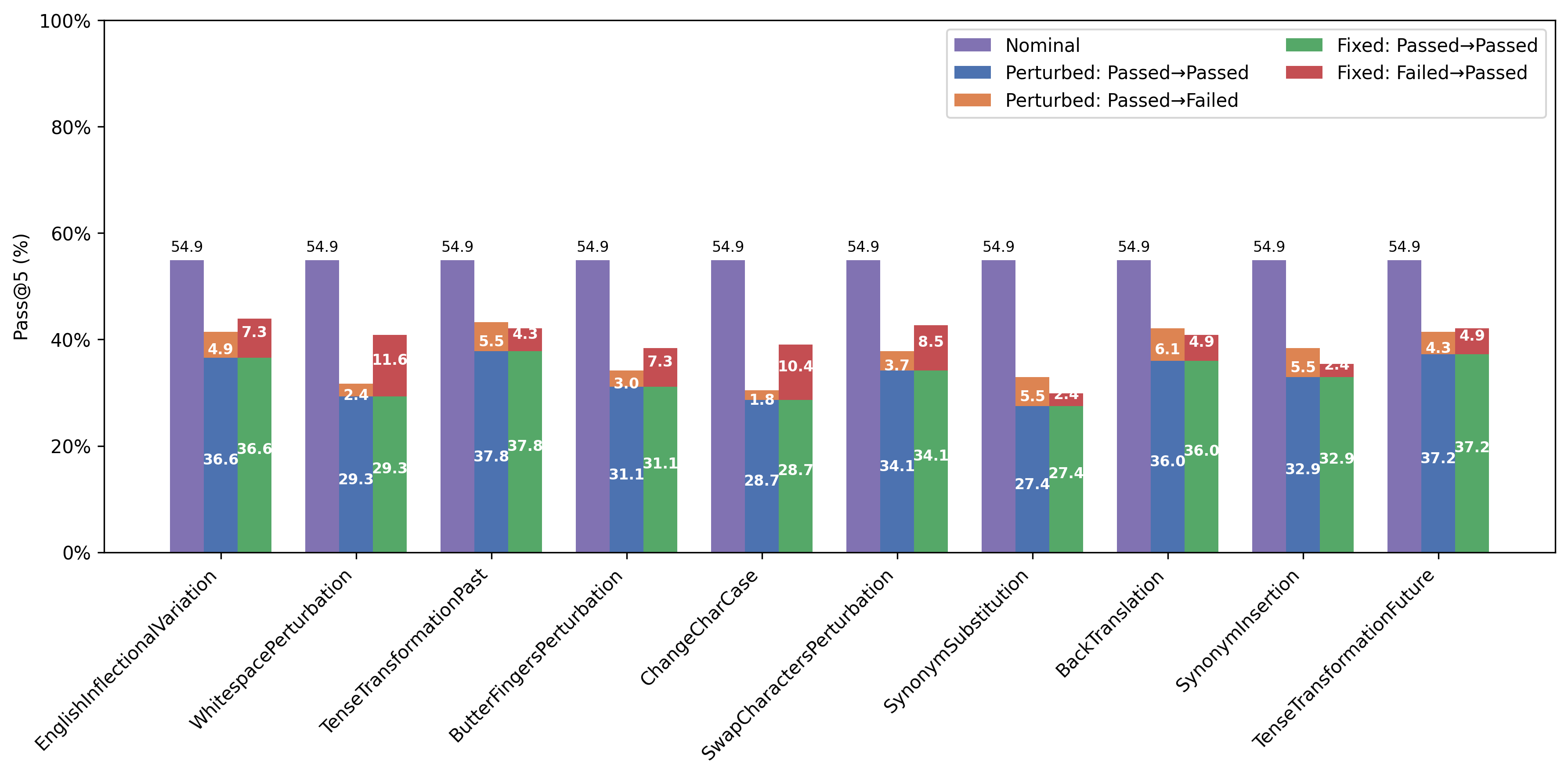}
        \caption{C++}
        \label{fig:cpp_results}
    \end{subfigure}
    \vspace{0.5em}
    
    \begin{subfigure}{\textwidth}
        \centering
        \includegraphics[width=.9\linewidth]{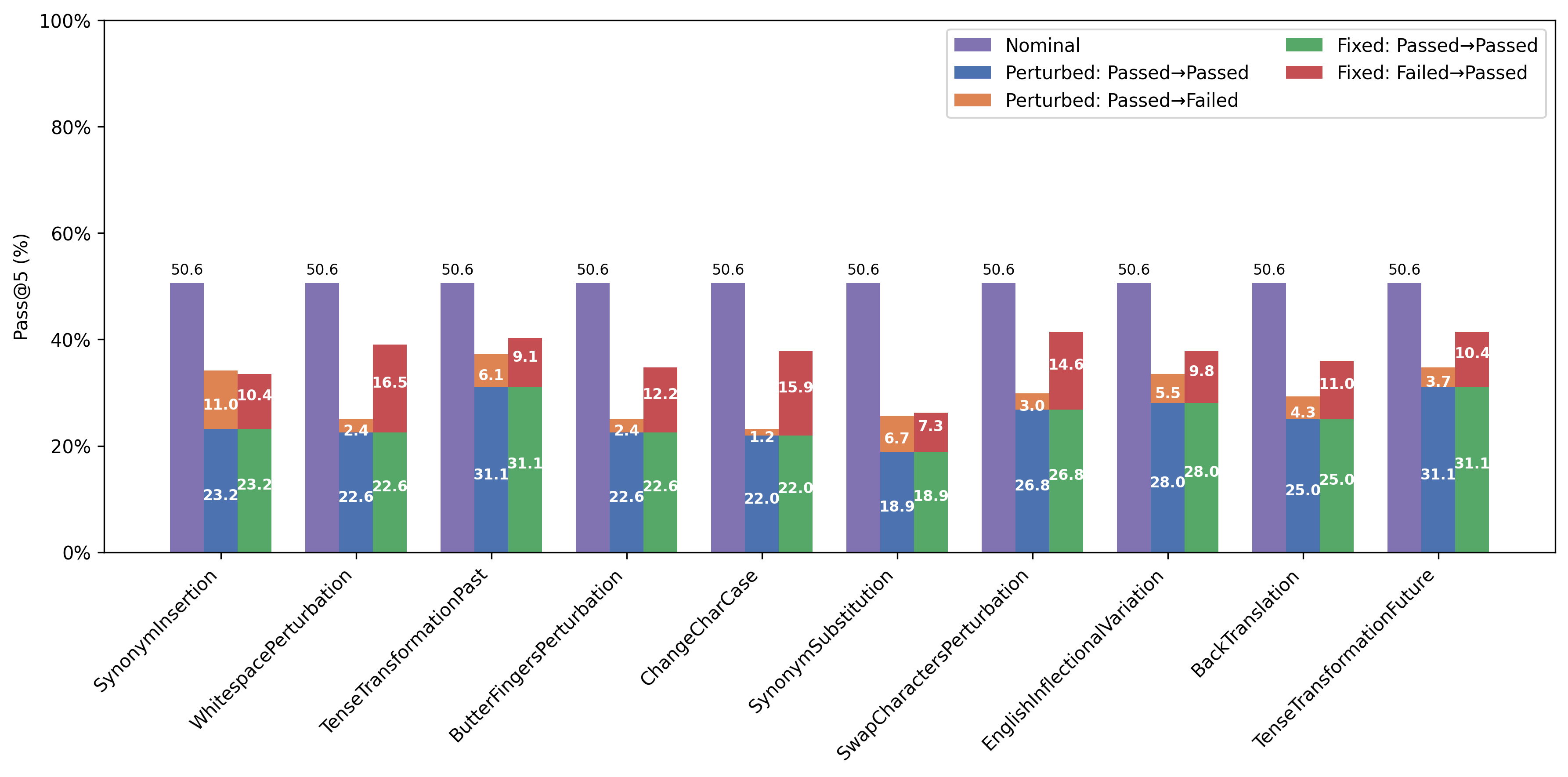}
        \caption{JavaScript}
        \label{fig:js_results}
    \end{subfigure}
    \vspace{0.5em}

    \begin{subfigure}{\textwidth}
        \centering
        \includegraphics[width=.9\linewidth]{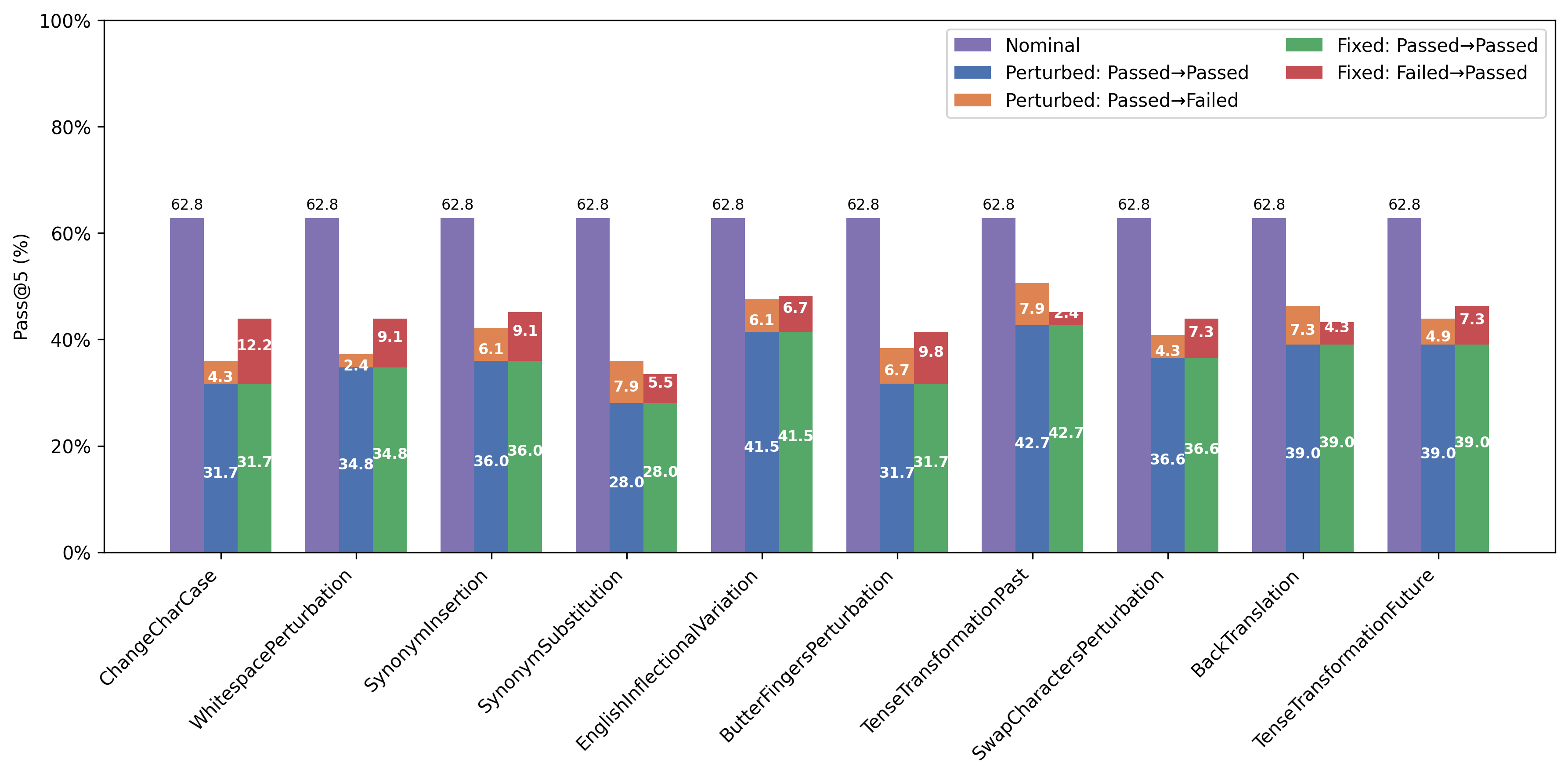}
        \caption{Java}
        \label{fig:java_results}
    \end{subfigure}
    
    \caption{Pass@5 scores for nominal, perturbed, and fixed prompts across three languages using the Magicoder-7B model.}
    \label{fig:pass_rate_vertical}
\end{figure}

\noindent\textbf{Results:}
Figure \ref{fig:pass_rate_vertical} summarizes the performance changes for each language under different perturbations. The three plots compare results across C++, Java, and JavaScript. In each plot, the y-axis lists the ten docString perturbations. For each perturbation, three bars represent the Pass@5 (\%) scores for the nominal, perturbed, and fixed prompts. Blue and green indicate the percentage of samples that passed under the perturbed and fixed prompts, respectively. Orange represents samples that passed under perturbation but failed after the fix, while red represents samples that failed under perturbation but passed after being fixed by the LLM.  

Overall, the results show that the pre-processing step helps recover part of the robustness loss, particularly for surface-level (syntactic) perturbations, while its effectiveness diminishes for deeper, semantic transformations.

\begin{itemize}
\item \textbf{C++.} The model achieves moderate recovery after perturbations, particularly for syntactic changes. Across perturbations, the mean Pass@5 for newly fixed cases (i.e., failed before but passed after pre-processing) is \textbf{6.4\%}, reaching up to \textbf{11.6\%} improvement in the best case. However, semantic perturbations continue to cause failures, with an average of \textbf{4.3\%} of cases newly failing. The stable proportion of already passed cases (\textbf{33.1\%}) indicates that rewriting neither significantly harms nor improves unaffected samples.

\item \textbf{Java.} The same overall trend is observed, but with smaller gains. Newly fixed cases average \textbf{7.4\%}, peaking at \textbf{12.2\%}. Meanwhile, newly failed cases average \textbf{5.8\%}, indicating that in some instances rewriting introduces additional errors. Overall, Java exhibits the narrowest recovery range, suggesting that the rewriting strategy is least effective for this language.

\item \textbf{JavaScript.} The strongest recovery is observed for JavaScript, where newly fixed cases average \textbf{11.7\%}, reaching up to \textbf{16.5\%}. Newly failed cases remain relatively low (\textbf{4.6\%} on average), and the rewriting process does not cause notable regressions. This implies that while the model cannot fully resolve semantic ambiguity, the rewriting step effectively restores syntactic robustness without destabilizing prior successes.
\end{itemize}

{ The limited recovery for semantic perturbations can be explained by the nature of the repair prompt, which instructs the model to fix grammar, spelling, and typos. Perturbations such as BackTranslation and Synonym-Substitution produce docstrings that are already grammatically correct and fluent, leaving the repair model with no clear signal to identify or correct. In some cases, the model rewrites the docString in a way that diverges further from the original intent, explaining the occasional performance degradation visible in Figure~\ref{fig:pass_rate_vertical}. Future approaches could explore context-aware repair strategies that use the function signature as an anchor to guide docString reconstruction, or employ retrieval-augmented generation to better align the repaired docString with the intended functionality. For other perturbation types, function name perturbations could, in principle, benefit from a similar word-level repair strategy, while syntax and format perturbations target the code structure directly and would require fundamentally different, code-specific repair 
strategies.\\}

\noindent \textbf{Summary of Findings.} From the results this following findings are ermerged.

\begin{enumerate}
    \item {Prompt repair yields only partial recovery, and its effectiveness is tied to perturbation type: it helps for surface-level changes such as whitespace or character modifications, but has little effect on deeper semantic edits where the perturbed text is already natural and fluent.}
    \item Effectiveness varies by language: prompt-fixing recovers \~11.7\% of failed cases in JavaScript on average (peaking at 16.5\%), versus \~7.4\% in Java (max 12.2\%) and \~6.4\% in C++ (max 11.6\%).
    \item New failures are rare: the rewriting step introduces additional errors in only about 4-6\% of cases (e.g. \~4.3\% for C++, \~5.8\% for Java, \~4.6\% for JavaScript), and it generally does not alter prompts that already passed.
    \item {The repair strategy is ineffective for semantic perturbations because they produce docstrings that are already grammatically valid, providing no signal for the repair model to act on. This is the key reason why BackTranslation and SynonymSubstitution show little or no recovery, and sometimes degrade performance. Future work should explore context-aware repair strategies, for instance, using the function signature as an anchor to reconstruct a semantically aligned docString.}
\end{enumerate}

{Extending the repair strategy to other perturbation types remains an open challenge. Syntax and format perturbations target code structure directly and would require code-specific repair strategies such as AST-based transformations. For function name perturbations, our natural language sentence repair approach does not directly apply, as function names consist of one or more words without sentence structure. Repairing them would require parsing the name into constituent words and replacing corrupted ones with the closest valid alternatives, a fundamentally different strategy that, given the limited context of a function name, remains an open direction for future work.}

%% file: refactoring.tex

    

    

%% file: lessons_learned.tex
\section{Lessons Learned}
\noindent Our multi-language analysis of LLM robustness provides several key takeaways for different stakeholders in the software engineering and machine learning communities.

\subsection*{\textbf{For Developers and Model Users (Practitioners)}}
\begin{itemize}[leftmargin=*]
    \item \textbf{Robustness is not guaranteed with a high correct rate of LLM-generated code.} \textit{RQ1 reveals: Even models with high nominal accuracy can fail catastrophically under small, semantics-preserving prompt changes.} Even when a model performs well on nominal prompts, it can still be highly vulnerable to small, semantics-preserving perturbations. For instance, in the stronger models such as Magicoder and QwenCode, Java achieves the highest success rates under nominal prompts, yet it also suffers the largest robustness drops compared to C++ and JavaScript, once minor prompt modifications are introduced. This pattern reveals a key cross-language issue: languages with higher nominal performance may in fact be more brittle, while with high correctness model can suffer robustness issue. These results show that high accuracy can conceal fragility, making robust testing across languages essential.

    \item \textbf{LLM-powered code generation is more fragile to semantic naming and documentation changes than formatting.} \textit{RQ1 reveals: Perturbing function names or docstrings often induces failures, indicating that LLMs rely heavily on these cues as semantic anchors.} This implies models rely heavily on these natural language cues as semantic anchors. Developers should be advised to use clear and meaningful function names and docstrings when leveraging LLMs for code generation, as unclear, mismatched, or slightly modified names or documentation can mislead the model into producing incorrect or non-functional code. or even slightly modified function names or docstrings with typos can mislead the model into generating incorrect or non-functional code.
 
\end{itemize}

\subsection*{\textbf{For LLM Designers (Model Builders)}}
\begin{itemize}[leftmargin=*]
    \item \textbf{Robustness requires more than just scaling.} \textit{RQ1 reveals: Increasing model size does not consistently improve robustness; in some cases, larger models were actually less robust to perturbations than smaller ones.} In several cases, larger models underperformed smaller ones when exposed to perturbed prompts. This suggests that robustness is not a byproduct of scale, but rather a distinct capability that requires targeted training strategies, diverse input exposure, and explicit design considerations.

    \item \textbf{Cross-language robustness must be evaluated independently.} \textit{RQ1 reveals: The same model’s robustness varies greatly by language: Java was usually the most resilient and C++ the most fragile under perturbations.} Our results show that robustness can vary substantially across programming languages, even when using the same model. This highlights the importance of assessing robustness on a per-language basis rather than assuming that performance in one language will generalize to others.

    \item \textbf{Each programming language has its own patterns of robustness drops.} \textit{RQ2 reveals: Semantically equivalent input perturbations produced very different failure patterns across languages.} Our findings from RQ2 shows that semantically equivalent perturbations cause very different failure patterns across languages. Because these perturbations change only surface form, not meaning, such language-dependent drops indicate that models are more sensitive to surface-level cues for some languages. This motivates the need for language-specific diagnostics to identify where and why models fail differently across programming languages.
\end{itemize}

\subsection*{\textbf{For Academia Researchers}}
\begin{itemize}[leftmargin=*]
    \item \textbf{Evaluation benchmarks must move beyond Python.} \textit{RQ1 reveals: Robust drop varies across languages, whereas prior Python-only work reported uniform patterns.} Our work shows that robustness varies substantially across programming languages, whereas prior work~\citep{wang2022recode} based solely on Python reported relatively consistent robustness patterns across perturbation types within a single language. In contrast, our multilingual results reveal that the magnitude and distribution of robustness drops differ across languages (e.g., Java is more resilient, while C++ degrades more sharply). This comparison indicates that Python-only evaluations may mask language-specific vulnerabilities. These findings motivate the need for community efforts to develop multilingual robustness benchmarks that capture the broader challenges faced by LLMs in real code-generation settings.

    \item \textbf{Robustness benchmarks must test diverse perturbation types} \textit{RQ1 reveals: Semantic and formatting perturbations are at least as effective as syntactic perturbations at causing failures} Our findings validate the approach of the Recode framework, showing that semantic and format perturbations are just as, if not more, effective at causing failures as standard syntactic noise across multiple languages.  While HumanEval and EvalPlus show similar robustness trends, EvalPlus consistently exposes more failures due to its stronger test cases. This underscores the need for robustness benchmarks that combine varied perturbations with comprehensive test suites. Future research should continue to build and refine these comprehensive robustness benchmarks

    \item \textbf{Failure analysis should go beyond pass/fail metrics.} \textit{RQ2 reveals: Analyzing only pass/fail rates misses why failures occur; feature-level metrics (e.g. prompt-change, code dissimilarity) provide deeper insight into failure causes.} Failure analysis should not rely only on overall pass/fail rates, because these metrics reveal whether a model fails but not why it fails. Our RQ2 analysis shows that examining additional measurable properties, such as how much a perturbation changes the prompt, how the generated code differs from the canonical solution, and which features correlate with failure provides deeper insight into the underlying causes of robustness breakdowns. By analyzing these attribute-level factors, we can identify specific weaknesses, such as sensitivity to surface-level changes or dependence on certain structures. This demonstrates the value of incorporating structured, feature-based analyses alongside pass/fail outcomes, and we encourage future work to adopt such methods to better explain the sources of model failures rather than only recording their occurrence.

     \item \textbf{Automated prompt-fixing is not a reliable solution.} \textit{RQ3 reveals: Using an LLM to auto-correct a perturbed prompt rarely restores original performance.} Although the NLP community has proposed prompt-repair techniques to address LLM robustness, our findings suggest that these techniques~\citep{ye2020safer,ji2024defending} do not transfer well to code generation because of their limitations to security vulnerabilities. In our RQ3 experiments, using an LLM to fix perturbed prompts offered only minor improvement for simple syntactic changes (e.g., whitespace) and failed for semantic perturbations such as back-translation, sometimes even degrading performance.
\end{itemize}

%% file: threats.tex
\section{Threats to Validity}
\label{threats}
\noindent \textbf{Internal Validity:} Threats to internal validity indicate the internal factors that may affect the result. For this work, generating the same codes from the same prompt each time by the code generation model can introduce biases into the result. To eliminate this, we generate 5 perturbed datasets from the original dataset, and all are used to mitigate the randomness of the code generation model. We run the whole process for all 28 perturbed datasets, 5 times each. Besides, the dataset we used is a public benchmark, and we have no control over it, which could impact the result. 

\noindent \textbf{External Validity:} Threats to external validity indicate outside effects other than the codes and used dataset. In this work as we use three code generation models and the human-eval-x dataset, which consists of datasets for multiple languages. The dataset and model, which could introduce bias for a limited number of models. However, we prepared our approach in a generic way that could be used for other languages and other datasets also to mitigate this threat.

{In addition, the benchmarks used in this study, HumanEval-X and EvalPlus, primarily consist of small, self-contained algorithmic tasks. While these benchmarks are widely adopted in the code generation community and enable controlled, reproducible evaluation, they may not fully represent the complexity of real-world software development tasks, which often involve larger codebases, complex dependencies, and domain-specific requirements. As a result, the robustness findings reported in this study may not directly generalize to more complex real-world scenarios. Evaluating robustness on more representative benchmarks covering diverse and complex programming tasks remains an important direction for future work.}

{Another external threat is that the six models evaluated in this study are all open-source and locally deployable. Widely used closed-source models such as GPT-4, CodeLlama, 
Claude and Gemini are not included in this study. The excluded models differ in architecture, training data, and scale, and may exhibit different robustness patterns under the same perturbations. In particular, larger instruction-tuned or RLHF-trained models may be more resilient to surface-level perturbations due to their stronger natural language understanding. As a result, the conclusions drawn from our six models may not fully generalize to closed-source or larger-scale models, and we acknowledge this as a limitation of our study.}

{In this study each prompt is perturbed by a single perturbation at a time, following the evaluation design of ReCode~\citep{wang2022recode}. In practice, a developer may inadvertently introduce multiple surface-level changes to a prompt simultaneously, for example, a typo in the function name combined with a paraphrased docString. Applying multiple perturbations simultaneously could compound the robustness drop beyond what is observed in single-perturbation settings, and the interaction effects between perturbation types remain an open question for future work.}

{\noindent \textbf{Construct Validity:} The perturbations used in this study target DocString, function name, syntax, and format, all of which are surface-level or natural language modifications (e.g., variable names). Our study does not include type-related perturbations, such as changes to type annotations, variable types, or type casts, which may affect statically typed languages like Java and C++ differently from dynamically typed ones like JavaScript. The absence of such perturbations may limit the completeness of our robustness evaluation with respect to language-specific type system characteristics. We consider the design and evaluation of type-aware 
perturbations as a direction for future work.}

%% file: conclusion.tex
\section{Conclusion and Future Work}
\label{conclusion} 
{\noindent This paper addresses the demand for a comparative study on the robustness of code generation models across multiple popular programming languages, including Java, C++, and JavaScript. Building upon the foundation of ReCode~\citep{wang2022recode}, which primarily focuses on Python, we extend the robustness evaluation to a multilingual setting by constructing EvalPlus-X, a new benchmark derived from HumanEval-X and EvalPlus with substantially more rigorous test cases. We evaluate six LLMs, Incoder-1B, Incoder-6B, CodeGen-2B-Multi, CodeGen-6B-Multi, Magicoder-7B, and QwenCode-2.5-7B, under 29 distinct semantic-preserving perturbations across four prompt scopes: DocString, function name, syntax, and format.} 

{Our investigation yields several key findings. All models consistently degrade under perturbations across all three languages, but the magnitude of robustness drops varies depending on the language and perturbation type, confirming that robustness is a language-dependent property that cannot be adequately assessed from a single language alone. Semantic and format perturbations prove at least as disruptive as syntactic ones, underscoring the need for diverse perturbation types in robustness benchmarks. Larger model size does not reliably predict robustness; in several cases, larger models are more brittle than smaller ones under semantics-preserving perturbations. Our feature-level analysis (RQ2) further reveals that the factors driving robustness drops differ across languages, with DocString and function name perturbations most strongly associated with failures in Java and JavaScript, while C++ is more sensitive to syntax-level perturbations. Our LLM-based docstring repair strategy (RQ3) yields only marginal gains for surface-level perturbations and is ineffective for semantic ones, highlighting the limits of lightweight prompt-level mitigation.} 

{Several directions remain open for future work. This study evaluates each perturbation type in isolation, following ReCode~\citep{wang2022recode}, which allows the individual effect of each perturbation to be isolated. However, real-world scenarios may involve multiple simultaneous changes to a single prompt, and prior NLP robustness work has shown that compound perturbations can reveal interaction effects beyond what single perturbations expose~\citep{dong2023revisit, gupta2024evaluating}. Extending our framework to support combined perturbations across multiple scopes represents a promising direction. Beyond that, evaluating robustness using white-box strategies, which provide access to model internals, could offer deeper insights beyond the black-box setting adopted here. Extending the evaluation to broader and more complex benchmarks, additional programming languages, and larger closed-source models would further strengthen the generalizability of the findings.}

%% file: funding.tex
\section*{Acknowledgement}
\label{acknowledgement}
\noindent This research was supported by the Fonds de recherche du Québec (Grant No.2024-NOVA346499)\citep{nova}, Natural Sciences and Engineering Research Council of Canada (NSERC) through the Alliance, Grant (Grant No.586838-23), the NSERC Discovery Grant (Grant No. RGPIN-2019-07007 and Grant No. DGECR-2019-00464), and NSERC CREATE Grant (Grant No.555406-2021). We gratefully acknowledge the support of all funding agencies.